\documentclass[usenatbib]{mnras}
\usepackage{natbib,amssymb,amsmath,times,graphicx,url,aas_macros}
\usepackage{setspace}      
\usepackage{epstopdf}       
\usepackage{verbatim}       
\usepackage{bm}		  
\usepackage{fleqn}		  

\newcommand{\tff}{$t_\text{ff}$}
\newcommand{\otftff}{$t = 1.34t_\text{ff}$}

\newcommand{\Bx}{$\bm{B}_\text{x}$}
\newcommand{\Bz}{$\bm{B}_\text{z}$}
\newcommand{\Bxm}{$-\bm{B}_\text{x}$}
\newcommand{\Bzm}{$-\bm{B}_\text{z}$}

\newcommand{\Bizm}{$\bm{B}_0 = -\bm{B}_\text{z}$}
\newcommand{\Bizpm}{$\bm{B}_0 = \pm\bm{B}_\text{z}$}
\newcommand{\Bixp}{$\bm{B}_0 = +\bm{B}_\text{x}$}
\newcommand{\Bixm}{$\bm{B}_0 = -\bm{B}_\text{x}$}
\newcommand{\Bixpm}{$\bm{B}_0 = \pm\bm{B}_\text{x}$}
\newcommand{\mufive}{$\mu_0=5$}
\newcommand{\muten}{$\mu_0=10$}
\defcitealias{PriceBate2007}{PB07}
\defcitealias{WPB2016}{WPB16}

\title[Non-ideal MHD and binary formation]{The impact of non-ideal magnetohydrodynamics on binary star formation}

\author[Wurster, Price \& Bate]{James Wurster$^{1,2}$\thanks{j.wurster@exeter.ac.uk}, Daniel J. Price$^{2}$ and Matthew R. Bate$^{1}$\thanks{mbate@astro.ex.ac.uk}  \\
$^{1}$School of Physics, University of Exeter, Stocker Rd, Exeter EX4 4QL, UK \\
$^{2}$Monash Centre for Astrophysics and School of Physics and Astronomy, Monash University, Vic 3800, Australia \\
}
\date{Submitted: Revised: Accepted: }
\pagerange{\pageref{firstpage}--\pageref{lastpage}} \pubyear{2016}

\begin{document}
\label{firstpage}
\bibliographystyle{mnras}
\maketitle

\begin{abstract}
We investigate the effect of non-ideal magnetohydrodynamics (MHD) on the formation of binary stars using a suite of three-dimensional smoothed particle magnetohydrodynamics simulations of the gravitational collapse of one solar mass, rotating, perturbed molecular cloud cores.  Alongside the role of Ohmic resistivity, ambipolar diffusion and the Hall effect, we also examine the effects of magnetic field strength, orientation and amplitude of the density perturbation. When modelling sub-critical cores, ideal MHD models do not collapse whereas non-ideal MHD models collapse to form single protostars.  In super-critical ideal MHD models, increasing the magnetic field strength or decreasing the initial density perturbation amplitude decreases the initial binary separation.  Strong magnetic fields initially perpendicular to the rotation axis suppress the formation of binaries and yield discs with magnetic fields $\sim$$10$ times stronger than if the magnetic field was initially aligned with the rotation axis.  When non-ideal MHD is included, the resulting discs are larger and more massive, and the binary forms on a wider orbit.   Small differences in the super-critical cores caused by non-ideal MHD effects are amplified by the binary interaction near periastron.  Overall, the non-ideal effects have only a small impact on binary formation and early evolution, with the initial conditions playing the dominant role.
\end{abstract}


\section{Introduction}
\label{intro}

Most observed stars are members of multiple systems \citep[e.g.][]{DuquennoyMayor1991,RaghavanEtAl2010}, so the formation of binaries is crucial to our understanding of the star formation process. The most common method for simulating binary formation numerically to date has been to model the gravitational collapse of a rotating, uniform sphere given an initial $m=2$ density perturbation either without \citep[e.g.][]{BossBodenheimer1979,Boss1986,BurkertBodenheimer1993,BateBonnellPrice1995,MachidaEtAl2008} or with \citep[e.g.][]{PriceBate2007,HennebelleTeyssier2008,BossKeiser2013} magnetic fields.

Using this approach, \citet{PriceBate2007} (hereafter \citetalias{PriceBate2007}) found that strong magnetic fields (mass-to-flux ratios $\lesssim 5$ times critical) could suppress binary formation --- leading to the formation of only one star where two would otherwise have formed --- influenced more by magnetic pressure than by magnetic tension or magnetic braking effects.  The effect was more pronounced with magnetic fields initially perpendicular to the rotation axis. A similar study by \citet{HennebelleTeyssier2008} considered magnetic fields parallel to the axis of rotation and an initial rotation that was $\sim$$4$ times slower than in \citetalias{PriceBate2007}. They also found that magnetic fields greatly inhibited binary formation, finding disc fragmentation and binary formation only for weak initial  magnetic fields (mass-to-flux ratios $\gtrsim 20$ times critical). When the initial density perturbation was increased, then binaries formed except when the initial cloud was close to being magnetically critical. They concluded that the suppression of binary formation was a result of the growth of the toroidal component of the magnetic field, and not a result of magnetic braking.  Further, large density perturbations were required to from binaries given the observed magnetic field strengths. 

\newpage 
However, these studies assumed ideal magnetohydrodynamics (MHD), which is a poor approximation for molecular clouds \citep{MestelSpitzer1956}, where ionisation fractions are of order $n_\text{e}/ n_{\text{H}_{2}} \sim 10^{-14}$ (\citealt{NakanoUmebayashi1986,UmebayashiNakano1990}).  Partial ionisation means that non-ideal MHD effects --- specifically Ohmic resistivity, the Hall effect, and ambipolar diffusion --- become important, with the relative importance of each depending, amongst other things, on the gas density and magnetic field strength \citep[e.g.][]{WardleNg1999,NakanoNishiUmebayashi2002,TassisMouschovias2007,Wardle2007,PandeyWardle2008,KeithWardle2014}. The Hall effect also depends on the direction of the magnetic field with respect to the axis of rotation \citep[e.g.][]{BraidingWardle2012accretion}.

Previous studies have examined the effects of non-ideal MHD on the formation of single stars (e.g. \citealp{NakanoUmebayashi1986b,FiedlerMouschovias1993,CiolekMouschovias1994,LiShu1996,Mouschovias1996,MouschoviasCiolek1999,ShuGalliLizanoCai2006,MellonLi2009,DuffinPudritz2009,DappBasu2010,MachidaInutsukaMatsumoto2011,LiKrasnopolskyShang2011,DappBasuKunz012,TomidaEtAl2013,TomidaOkuzumiMachida2015,TsukamotoEtAl2015,TsukamotoEtAl2015b,WPB2016}).  \citet{TsukamotoEtAl2015b} and \citet{WPB2016} (hereafter \citetalias{WPB2016}) found that, although the Hall effect was not the dominant non-ideal effect in numerical simulations of an isolated forming star, it was the controlling factor in disc formation, with a large disc forming when the initial magnetic field and rotation vector were anti-aligned but no disc forming when the initial magnetic field direction was reversed.

Here, we evaluate the influence of non-ideal MHD on the formation and early evolution of binary systems, following the original ideal MHD studies of \citetalias{PriceBate2007} and \citet{HennebelleTeyssier2008}.  We model 3D non-ideal self-gravitating smoothed particle magnetohydrodynamics simulations of collapsing, low mass cores, with the ionisation fractions calculated using the {\sc Nicil} library \citep{Wurster2016}.  We present the numerical formulation in Section~\ref{sec:numerics}, the initial conditions in Section~\ref{sec:ic}, the results in Section~\ref{sec:results} and the discussion and conclusions in Section~\ref{sec:discussion}.


\section{Numerical method}
\label{sec:numerics}
\subsection{Non-ideal Magnetohydrodynamics}
\label{sec:numerics:nimhd}
We solve the equations of self-gravitating, non-ideal magnetohydrodynamics in the form
\begin{eqnarray}
\frac{{\rm d}\rho}{{\rm d}t} & = & -\rho \nabla\cdot \bm{v}, \label{eq:cty} \\
\frac{{\rm d} \bm{v}}{\rm{d} t} & = & -\frac{1}{\rho}\bm{\nabla} \left[\left(P+\frac{1}{2}B^2\right)I - \bm{B}\bm{B}\right] - \nabla\Phi, \label{eq:mom} \\
\frac{{\rm d} \bm{B}}{\text{d} t} & = & \left(\bm{B}\cdot\bm{\nabla}\right)\bm{v}-\bm{B}\left(\bm{\nabla}\cdot\bm{v}\right) + \left.\frac{\text{d} \bm{B}}{\text{d} t}\right|_\text{non-ideal} \label{eq:ind}, \\
\nabla^{2}\Phi & = & 4\pi G\rho, \label{eq:grav}
\end{eqnarray}
where $\frac{\text{d}}{\text{d}t} \equiv \frac{\partial}{\partial t} + \bm{v}\cdot\bm{\nabla}$ is the Lagrangian derivative,  $\rho$ is the density, ${\bm  v}$ is the velocity, $P$ is the gas pressure, ${\bm B}$ is the magnetic field, $\Phi$ is the gravitational potential, $I$ is the identity matrix, and $\left.\frac{\text{d} \bm{B}}{\text{d} t}\right|_\text{non-ideal}$ is the non-ideal MHD term which is a sum of the Ohmic resistivity (OR), Hall effect (HE) and ambipolar diffusion (AD) terms, 
\begin{flalign}
\left.\frac{\text{d} \bm{B}}{\text{d} t}\right|_\text{OR} &= -\bm{\nabla} \times \left[  \eta_\text{OR}      \left(\bm{\nabla}\times\bm{B}\right)\right],                                                                     &\label{eq:ohm} \\
\left.\frac{\text{d} \bm{B}}{\text{d} t}\right|_\text{HE} &= -\bm{\nabla} \times \left[  \eta_\text{HE}       \left(\bm{\nabla}\times\bm{B}\right)\times\bm{\hat{B}}\right],                                         &\label{eq:hall} \\
\left.\frac{\text{d} \bm{B}}{\text{d} t}\right|_\text{AD} &=  \bm{\nabla} \times \left\{ \eta_\text{AD}\left[\left(\bm{\nabla}\times\bm{B}\right)\times\bm{\hat{B}}\right]\times\bm{\hat{B}}\right\}, &\label{eq:ambi}
\end{flalign}
where $\eta_\text{OR}$, $\eta_\text{HE}$ and $\eta_\text{AD}$ are the non-ideal MHD coefficients.  We assume units for the magnetic field such that the Alfv{\'e}n speed is $v_\text{A}\equiv B/\sqrt{\rho}$ (see \citealt{PriceMonaghan2004}).

We close the equation set using a barotropic equation of state,
\begin{equation}
\label{eq:eos}
P = \left\{ \begin{array}{l l} c_\text{s,0}^2\rho; 		    &  \rho < \rho_\text{c}, \\
                                           c_\text{s,0}^2\rho_\text{c}\left(\rho             /\rho_\text{c}\right)^{7/5};      &  \rho_\text{c} \leq \rho < \rho_\text{d}, \\
                                           c_\text{s,0}^2\rho_\text{c}\left(\rho_\text{d}/\rho_\text{c}\right)^{7/5} \left(\rho/\rho_\text{d}\right)^{11/10};      &  \rho \geq \rho_\text{d},
\end{array}\right.
\end{equation}
where $c_\text{s,0}$ is the initial isothermal sound speed and $\rho_\text{c} = 10^{-14}$ and $\rho_\text{d}~=~10^{-10}$ g~cm$^{-3}$.  These density thresholds are the same as in \citet{PriceTriccoBate2012}, \citet{LewisBatePrice2015} and \citetalias{WPB2016}.  Although we do not employ full radiation magnetohydrodynamics, the barotropic equation of state is designed to mimic the evolution of the equation of state in molecular clouds \citep{Larson1969,MasunagaInutsuka2000,MachidaInutsukaMatsumoto2008}.   Our value of $\rho_\text{c}$ is one order of magnitude lower than the typically chosen value, and is chosen to reduce fragmentation in the gas surrounding the protostars in the absence of radiation feedback.

We use Version 1.1 of the {\sc Nicil} library \citep{Wurster2016} to calculate the non-ideal MHD diffusion coefficients. \citetalias{WPB2016} used a precursor to {\sc Nicil}, and the main difference is that this library includes thermal ionisation and more detailed cosmic ray ionisation. The thermal ionisation processes can singly ionise hydrogen, and doubly ionise helium, sodium, magnesium and potassium; the mass fractions of the five elements are $0.747$, $0.252$, $2.96\times 10^{-5}$, $7.16\times 10^{-4}$ and $3.10\times 10^{-6}$, respectively  (e.g. \citealp{AsplundEtAl2009}; \citealp{KeithWardle2014}). Due to the cool temperatures in this study, we do not expect thermal ionisation to significantly contribute to the electron population. Cosmic rays have the ability to remove an electron to create an ion, which may be absorbed by a dust grain. We assume that two species of ions can be created: a heavy ion represented by magnesium \citep{AsplundEtAl2009} and a light ion representing hydrogen and helium compounds whose mass is calculated from the hydrogen and helium mass fractions (in \citetalias{WPB2016} we considered only the heavy ion species).  For most of the calculations, as in \citetalias{WPB2016}, we model a single grain species that can absorb the electrons, and further assume that these grains have radius, bulk density, and average electric charge of $a_\text{g} = 0.1 \mu$m, $\rho_\text{b} = 3$ g~cm$^{-3}$ \citep{PollackEtAl1994} and $\bar{Z}_\text{g} < 0$, respectively.  

\subsection{Smoothed Particle Magnetohydrodynamics}
\label{ssec:num:spmhd}
Our simulations are performed using the 3D smoothed particle magnetohydrodynamics (SPMHD) code {\sc Phantom} \citep{PriceFederrath2010,LodatoPrice2010}, which includes self-gravity. The discretised magnetohydrodynamic equations (see review by \citealt{Price2012}) are given in \citet{WPA2014} and  \citetalias{WPB2016}. We employ the constrained hyperbolic divergence cleaning algorithm described by \citet{TriccoPrice2012} and \citet{TriccoPriceBate2016} to control divergence errors in the magnetic field.

We adopt the usual cubic spline kernel, set such that the ratio of the smoothing length to the particle spacing is equivalent to $\sim$$58$ neighbours in three dimensions \citep{Price2012}.  We solve Poisson's equation, $\nabla^{2} \Phi_{a} = 4\pi G \rho_{a}$, following \citet{PriceMonaghan2007} at short range, and use a \emph{k}-d tree algorithm similar to that described in \citet{GaftonRosswog2011} to compute the long range gravitational interaction in an efficient manner.

Finally, the non-ideal MHD timestep is constrained by $\text{d}t_{a} < C_\text{non-ideal}h_a^2/\eta_a$, where $\eta_a = \max\left(\eta_{\text{OR},a},\left|\eta_{\text{HE},a}\right|,\eta_{\text{AD},a}\right)$ and $C_\text{non-ideal} = 1/2\pi < 1$ is a positive coefficient analogous to the Courant number.  Since this timestep can become prohibitively small due to ambipolar diffusion, we include super-timestepping \citep{AlexiadesEtAl96} for  $\text{d}t_\text{OR}$ and  $\text{d}t_\text{AD}$, using the implementation described in \citetalias{WPB2016}.

\section{Initial conditions}
\label{sec:ic} 
Our setup is a magnetised variant of the `standard isothermal test case' of \citet{BossBodenheimer1979}, similar to that used in \citetalias{PriceBate2007} and \citetalias{WPB2016}.  We use a spherical cloud of radius $R=4\times~10^{16}$~cm = 0.013~pc and mass $M=1$~M$_\odot$, yielding a mean density of $\rho_0=7.43\times~10^{-18}$~g~cm$^{-3}$.  The initial rotational velocity is $\Omega_0 = 1.006\times 10^{-12}$ rad s$^{-1}$, and the initial sound speed is $c_\text{s,0} = 1.87\times 10^4$~cm~s$^{-1}$.  The resultant thermal and rotational energy to gravitational potential energy ratios are $\alpha = 0.26$ and $\beta_\text{r} = 0.16$, respectively.  The free-fall time is $t_\text{ff}=2.4\times~10^4$~yr, which is the characteristic timescale for this study.

To facilitate the formation of a binary system, we perturb the initially uniform density sphere with a non-axisymmetric $m=2$ perturbation, such that
\begin{equation}
\label{eq:rhopert}
\rho = \rho_0\left[1+A_0\cos\left(2\phi\right)\right],
\end{equation}
where $A_0$ is the amplitude of the perturbation and $\phi$ is the azimuthal angle about the axis of rotation.  To achieve the new density profile with equal mass SPH particles we shift the particle positions using
\begin{equation}
\label{eq:rhopert_dphi}
\delta \phi = -\frac{A_0}{2}\sin\left(2\phi_0\right),
\end{equation}
where $\phi_0$ is the particle's unperturbed azimuthal angle \citepalias{PriceBate2007}.

We embed the initial cold spherical `cloud' in a uniform, low-density box of edge length $l = 4R = 0.052$~pc.  The cloud and surrounding medium are in pressure equilibrium and have a density contrast of 30:1.  We use quasi-periodic boundary conditions at the edge of the box, in which SPH particles interact hydrodynamically `across the box', but not gravitationally.

The densest gas particle is replaced by a sink particle \citep{BateBonnellPrice1995} when its density exceeds $\rho_\text{crit}~=~10^{-10}$~g~cm$^{-3}$ so long as the particle and its neighbours within $r_\text{acc} =$~3.35~AU meet the checks described in~\citet{BateBonnellPrice1995}; all the neighbours are immediately accreted onto the sink particle.  Gas which later enters this radius is checked against similar criteria to determine if it is also accreted onto the sink particle.  Sink particles interact with the gas only via gravity and accretion, thus, the magnetic field in the central regions is removed and not allowed to feed back on the surrounding material.

Given our initial conditions and our chosen equation of state, we require at least 30~000 particles to resolve the local Jeans mass for the entirety of the calculation (\citealt{BateBurkert1997}; \citetalias{PriceBate2007}).  Our simulations include 445~000 particles, with 302~000 in the sphere, thus the Jeans resolution condition is easily satisfied.
We also run selected models with $10^6$ particles in the sphere to test the effect of resolution, however the higher resolution models do not cover our entire suite nor are evolved as long; see Section~\ref{sec:results:res}. 
We set up particles initially on a regular close-packed lattice (e.g. \citealp{Morris1996}), and any undesirable effects initially introduced by the regularity of the lattice are transient and washed out long before the star formation occurs.  

We specify the magnetic field strength in terms of the mass-to-flux ratio expressed in units of the critical value, viz.,
\begin{equation}
\label{eq:masstofluxmu}
\mu \equiv \frac{M/\Phi_\text{B}}{\left(M/\Phi_\text{B}\right)_\text{crit}},
\end{equation}
where
\begin{equation}
\label{eq:masstoflux}
\frac{M}{\Phi_\text{B}} \equiv \frac{M}{\pi R^2 B},
\end{equation}
is the mass-to-flux ratio and
\begin{equation}
\label{eq:masstofluxcrit}
\left(\frac{M}{\Phi_\text{B}}\right)_\text{crit} = \frac{c_1}{3\pi}\sqrt{ \frac{5}{G} },
\end{equation}
is the critical value where magnetic fields prevent gravitational collapse altogether; here, $M$ is the total mass contained within the cloud, $\Phi_\text{B}$ is the magnetic flux threading the surface of the (spherical) cloud at radius $R$ assuming a uniform magnetic field of strength $B$, $G$ is the gravitational constant and $c_1 \simeq 0.53$ is a parameter determined numerically by \citet{MouschoviasSpitzer1976}. 

To study the effects of non-ideal MHD on binary formation we perform both ideal and non-ideal MHD simulations, where the non-ideal MHD models include Ohmic resistivity, the Hall effect and ambipolar diffusion.  We test `strong' and `weak' magnetic fields of $\mu_0 = 5$ and $10$, corresponding to $B = 163$ and $81.7 \ \mu G$, respectively; we also examine three different initial amplitudes of the density perturbation, $A_0= 0.2,\ 0.1 \text{ and } 0.05$, and four different initial magnetic field vectors, $\bm{B}_0 = \pm\bm{B}_\text{x} \text{ and } \pm\bm{B}_\text{z}$ where $\pm\bm{B}_\text{x} \equiv \pm B\hat{\bm{x}}$ and similarly for $\pm\bm{B}_\text{z}$.  The cloud is initially rotating counter-clockwise around the $z$-axis, so a positive $\bm{B}_\text{z}$ implies a magnetic field that is initially aligned with the rotation axis, while negative $\bm{B}_\text{z}$ implies an anti-aligned initial field.

In Section~\ref{sec:results:altP}, we briefly expand our parameter space to test the sub-critical mass-to-flux ratio of $\mu_0= 0.75 \left(B = 1090 \ \mu\text{G}\right)$, slower initial rotations, and the effect of using multiple grain populations in the non-ideal MHD algorithm.

We evolve the simulations that form binaries until at least first apoastron.

\section{Results}
\label{sec:results}
 In the following, we refer to gas densities $\rho > \rho_\text{disc,min} = 10^{-13}$~g~cm$^{-3}$ centred on a sink particle as a `disc', although these are not necessarily long-lived or rotationally supported.  The radius of the disc is defined as where the density drops to 10 per cent of the maximum density and the disc mass is defined as the gas mass with $\rho > \rho_\text{disc,min}$ enclosed within this radius.  The star+disc mass is the sum of the disc and sink particle masses.  This definition differs from the one used in \citetalias{PriceBate2007} and \citetalias{WPB2016} to avoid calculating artificially large radii and masses due to tidal bridges between two sink particles. 

When a sink particle is first formed, it represents a first hydrostatic core, which exists for $10^3 - 10^4$ yr \citep[e.g.][]{TomidaMachidaSaigoTomisaka2010,Bate2011} before collapsing to a stellar core \citep{Larson1969}.  Given that we follow the binary for $0.5$-$3\times10^4$ yr after the formation of the first sink particles, we refer to sink particles as protostars.

 Our full set of models is summarised in Table~\ref{table:results:all} of Appendix~\ref{app:results:all}, listing the time of first periastron $t_\text{peri}$, the initial period $T_0$ calculated using first periastron and first apoastron, and the separations at first periastron $R_\text{peri}$ and first apoastron $R_\text{apo}$.
 
\subsection{Ideal MHD with the magnetic field anti-aligned to the rotation axis}
\label{sec:results_IZ}
\begin{figure*}
\begin{center}
\includegraphics[width=\textwidth]{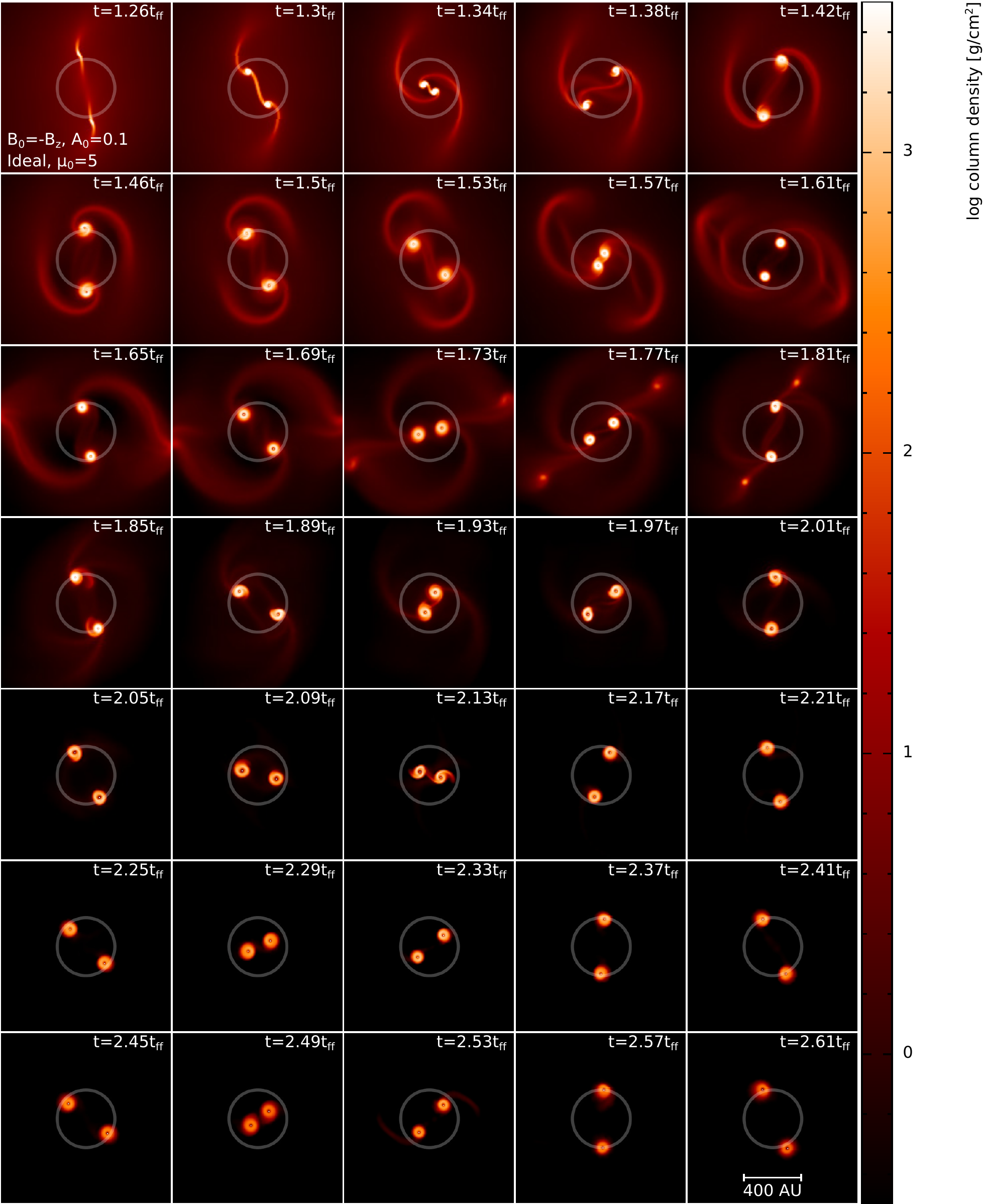}
\caption{Evolution of the face-on gas column density of the ideal MHD model with \mufive, $A_0=0.1$ and \Bizm, where $\mu_0$ is the mass-to-flux ratio in units of its critical value and $A_0$ is the amplitude of the initial $m=2$ density perturbation.  The frames are at intervals of $\text{d}t = 0.04$\tff, where the free-fall time is $t_\text{ff} = 2.4\times~10^4$~yr.  Each frame is (1200~AU)$^2$, and the grey circles of radius 200 au are included for reference.  The small solid white circles represent sink particles with the radius of the circle representing the accretion radius of the sink particle.  The binary is on a stable elliptical orbit, with first apoastron at $R_\text{apo} = 440$ au at $t = 1.46$\tff.  Over the first seven periods, the mean periastron and apoastron are $110$ and $400$ au, with an average period of $\text{d}t \approx 0.19$\tff.}
\label{fig:results:colden:evol}
\end{center}
\end{figure*}

 Our first set of simulations assume ideal MHD for comparison with previous studies. Given that the evolution of the magnetic field in ideal MHD is independent of the sign of $\bm{B}$, we consider only two initial magnetic field geometries: $\bm{B}_0 = -\bm{B}_\text{z}$ and $+\bm{B}_\text{x}$.  Fig.~\ref{fig:results:colden:evol} shows the evolution of the gas column density for the ideal MHD model with \mufive, $A_0=0.1$ and \Bizm, which is representative of our suite of models.

\begin{figure*}
\begin{center}
\includegraphics[width=0.75\textwidth]{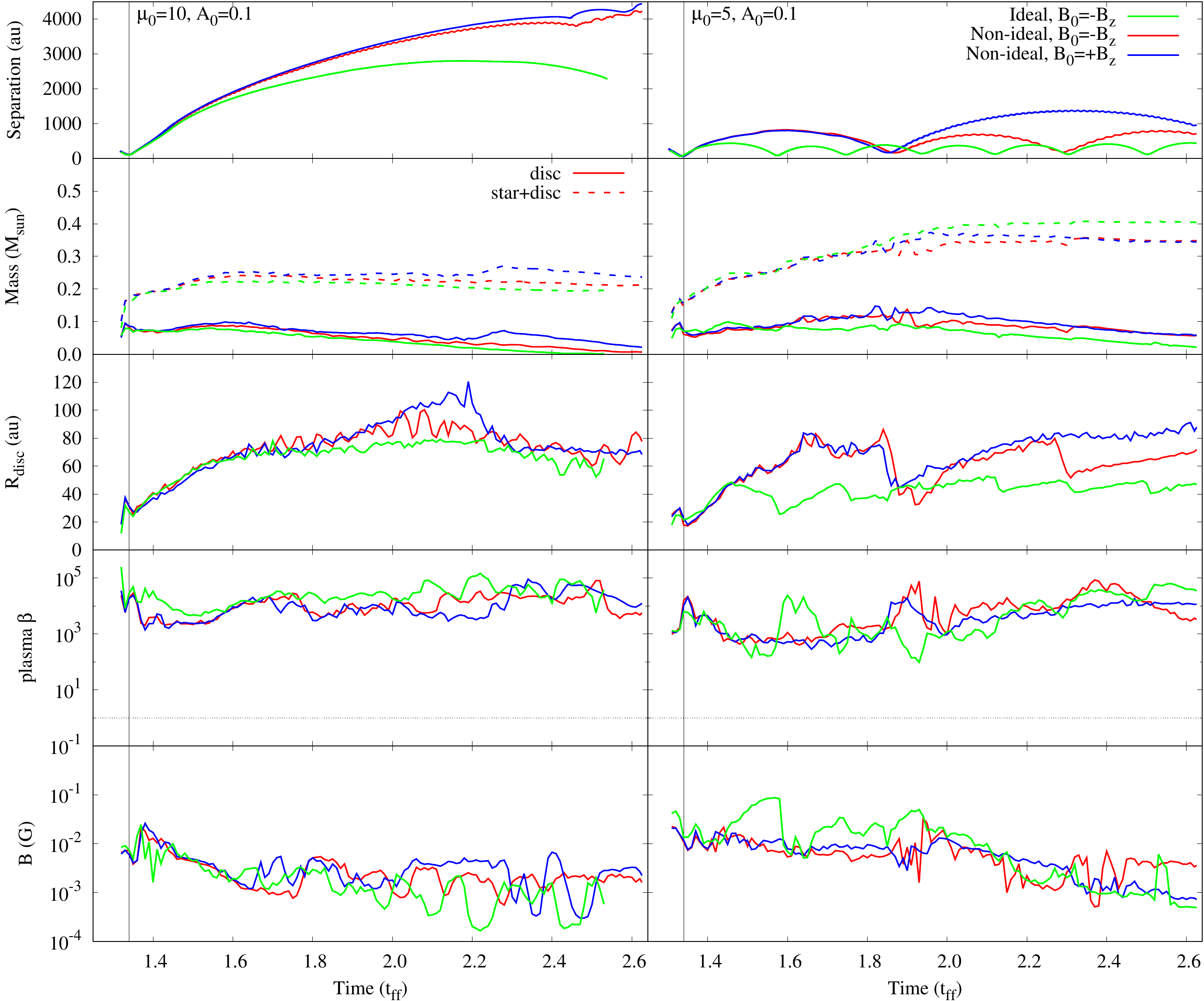}
\caption{Time evolution of selected models starting from the formation of the sink particles (first representing first hydrostatic cores which collapse to protostars; hereafter referred to as protostars).  Each model is initialised with an initial density perturbation of $A_0=0.1$ and $\bm{B}_0 = \pm\bm{B}_\text{z}$.  The left- and right-hand column shows three models with \muten \ and $5$, respectively.  \emph{Top to bottom}: the separation of the two protostars, the disc and star+disc masses, the disc radius, and the mass-weighted plasma $\beta$ and magnetic field strength in the disc around one protostar.  The vertical line at \otftff \ corresponds to the time of our early analysis.  The radius and mass of the disc in the non-ideal MHD models is for the non-fragmented disc.  For the ideal MHD model with \mufive, the mean period, periastron and apoastron are $0.19$\tff, $110$ and $400$ au, respectively.  The oscillations in the non-ideal MHD models with \mufive \ after second periastron are epicycles which are a result of one disc fragmenting and forming a well-behaved tight binary; the plotted binary separation is of the two initial protostars, and not to the barycentre of the newly formed tight binary.  The local minima and decreases in disc radii correspond to periastron. The protostar continues to accrete mass as the models evolve, while the mass of the discs generally decrease.  The increasing separation after $t\sim 2.4$\tff \ in the non-ideal models with \muten \ is a result of the primary protostars interacting with younger protostars that modify the orbit of the primaries.  The lines terminate at the end of the simulation.}
\label{fig:results:evolZ}
\end{center}
\end{figure*}

As the perturbed cloud collapses, each of the two over-densities collapse into a protostar.  Their first periastron and apoastron are $68$ and $440$ au, respectively.  Over the first seven periods, the mean periastron and apoastron are $110$ and $400$ au, with an average period of $0.19$\tff, indicating that the binary is dynamically evolving.  Fig. \ref{fig:results:evolZ} shows the evolution of the binary separation (top right panel, green curve).

\begin{figure}
\begin{center}
\includegraphics[width=\columnwidth]{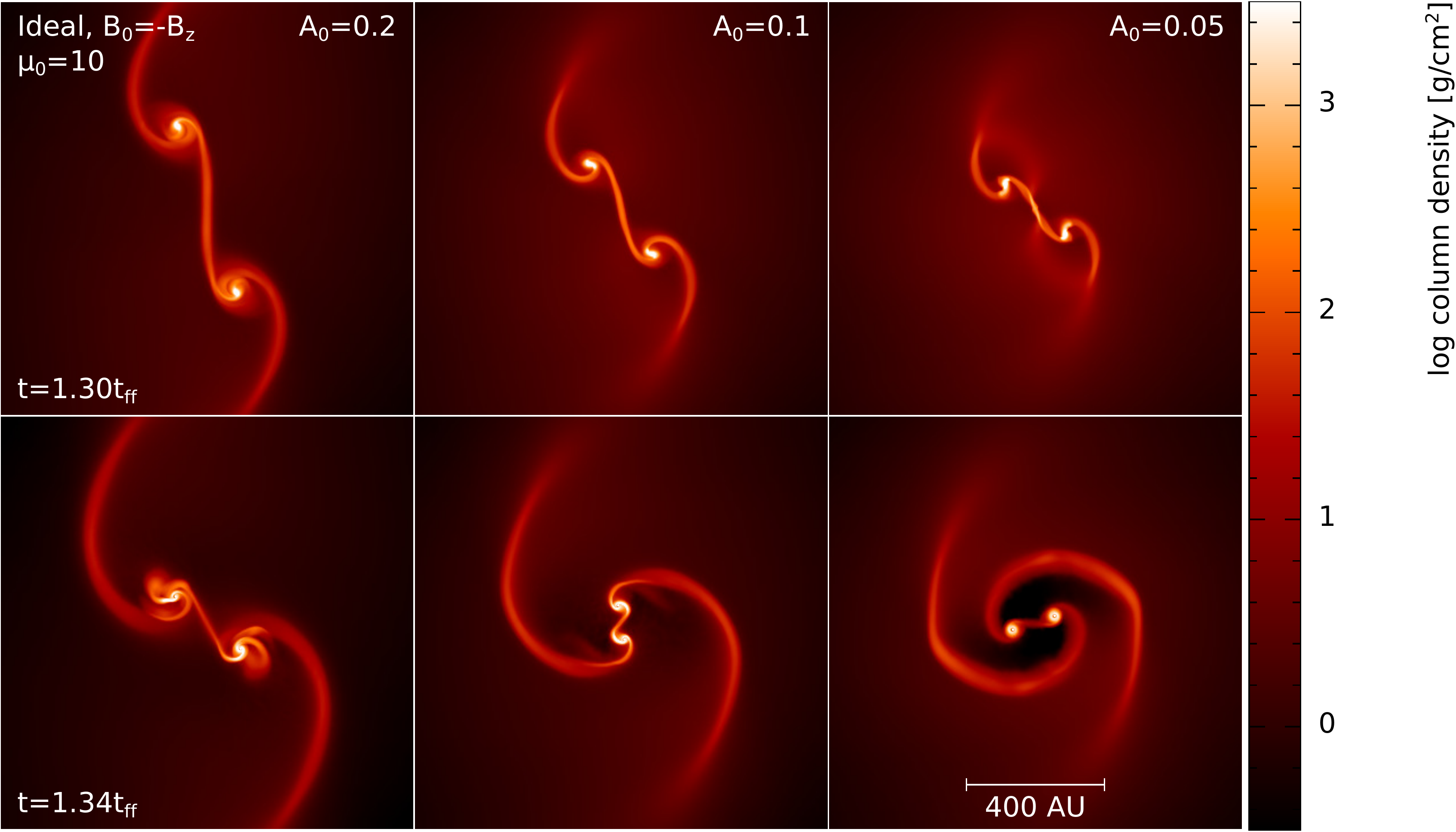}
\includegraphics[width=\columnwidth]{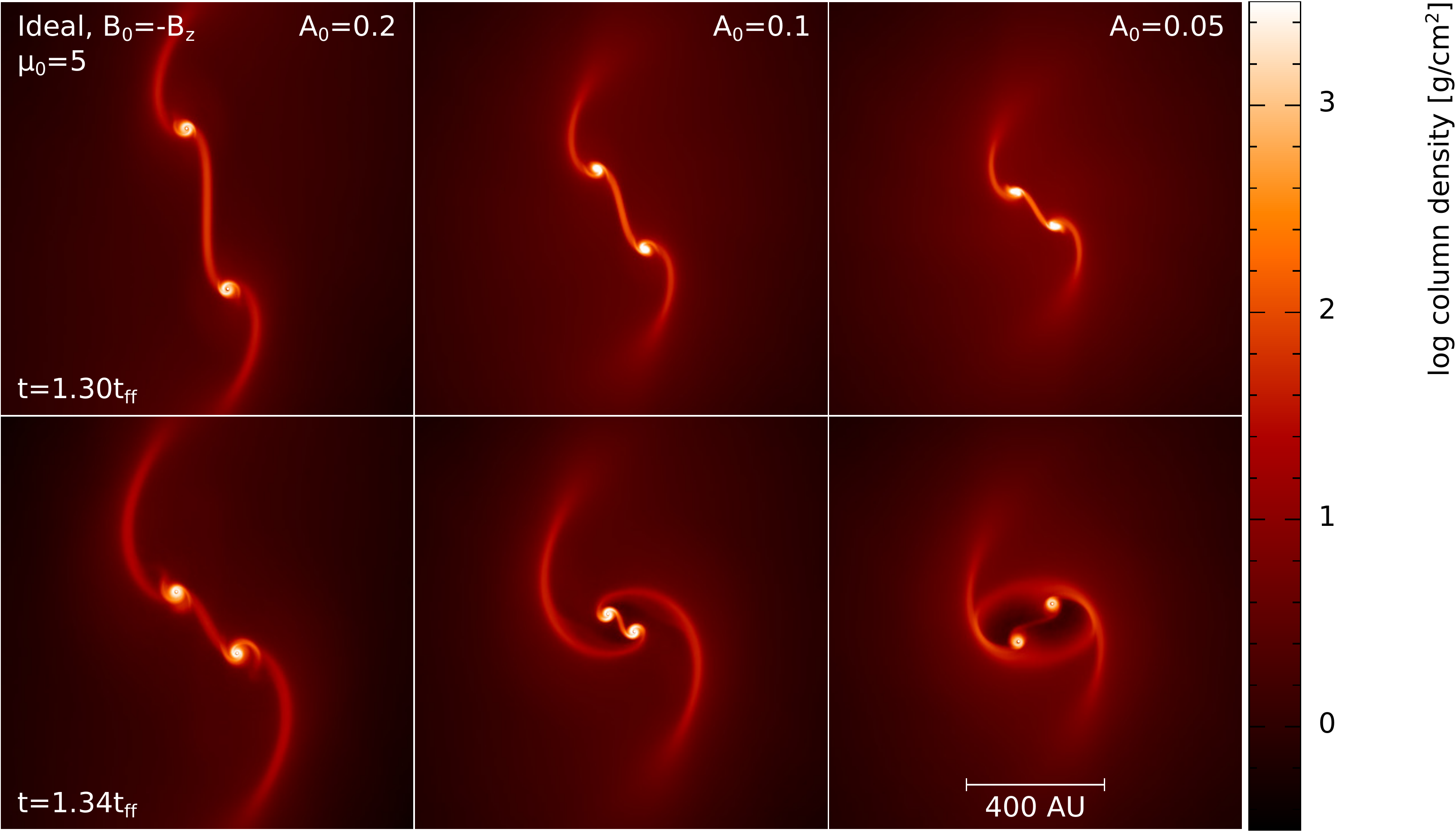}
\caption{Face-on gas column density snapshots from six ideal MHD models with $\bm{B}_0 = -\bm{B}_\text{z}$ at two different times (top and bottom row in each subfigure).  The top subfigure shows results with \muten, while the bottom subfigure shows the results with $\mu_{0}=5$. Decreasing the amplitude of the initial density perturbation $A_0$ (left to right) or increasing the magnetic field strength (i.e. decreasing $\mu_0$; top vs. bottom subfigure) decreases the first periastron separation as well as the disc mass and radius at first periastron.}
\label{fig:results:idealZ:130134}
\end{center}
\end{figure}

As found by \citetalias{PriceBate2007} and \citet{HennebelleTeyssier2008}, both the strength of the initial perturbation and the mass-to-flux ratio influence the formation of the binaries.  Fig.~\ref{fig:results:idealZ:130134} shows the snapshots for our models at $t = 1.30$ and 1.34\tff.
For smaller initial density perturbations, the time of first periastron, the first periastron and first apoastron separations all decrease.  As the magnetic field strength is decreased, $R_\text{peri}$ and $R_\text{apo}$ both increase since there is less magnetic braking.

Not all models yield stable orbits as shown in Fig.~\ref{fig:results:colden:evol}.  The model with \muten \ and $A_0=0.2$, for example, forms two massive, dense discs which fragment prior to first periastron.  Each pair of protostars forms a tight binary, and these pairs orbit one another on a long period; during their formation, the discs are totally disrupted.  Near apoastron, they interact with younger protostars which disrupts the orbit and makes the system chaotic.  Thus, this model, and many other models that have more than two protostars yield interactions that hinder a useful comparison.  

The green lines in Fig.~\ref{fig:results:evolZ} show the evolution of the disc and star+disc masses, the disc radius, and the mass-weighted plasma $\beta$ and magnetic field in the disc around one protostar for the ideal MHD models with $A_0=0.1$ and \muten \ (left-hand column) and \mufive \ (right-hand column).  Over the seven periods, starting at first periastron, the disc radius for the \mufive \ model varies between $22$ and $53$ au, and its mass varies between $0.02$ and $0.1$M$_\odot$, where the local minima corresponds to periastron.  After the initial rapid growth of the protostar, its subsequent growth is not dependent on orbital position, and the fluctuations in the star+disc mass correspond to fluctuations in the disc mass.  The disc is always dominated by gas pressure rather than magnetic pressure, with $\beta \gtrsim 10^3$.  As expected from the symmetry of the model, the properties around both protostars follow the same trends until one or both of the discs fragments.

As expected, the magnetic field strength is higher in the discs of the \mufive \ models compared to the \muten \ models. However, they are lower than in the discs produced during the collapse to form an isolated protostar \citepalias{WPB2016}, and hence have larger values of plasma $\beta$.  Thus, magnetic fields are less important in the evolution of the discs in these binary models than in the isolated protostar models of \citetalias{WPB2016}.

Weaker magnetic fields produce discs at larger separations, which have a larger gas reservoir than their strong field counterparts.  As shown in \citetalias{WPB2016}, larger discs form in weaker magnetic fields due to less efficient magnetic braking, independent of the gas reservoir.   Thus, these two complementary effects result in larger discs in weaker magnetic fields.  The largest discs form in the model with \muten \ and $A_0=0.2$ while the smallest discs form in the model with \mufive \ and $A=0.05$.  We would reach the same conclusions if the models were instead compared exactly at the time of first periastron.  

The weak field model exists on a long-period orbit, and does not have a second periastron by the end of the simulation (top left panel of Fig.~\ref{fig:results:evolZ}, green curve).  This allows the discs to essentially evolve in isolation, with the radius reaching a steady size of $r \sim 70$ au, even though the mass is continually decreasing.  The sharp and periodic decreases observed in the \mufive \ models do not occur.
 
\subsection{Ideal MHD with the magnetic field perpendicular to the rotation axis}
\label{sec:results_IX}

\begin{figure}
\begin{center}
\includegraphics[width=\columnwidth]{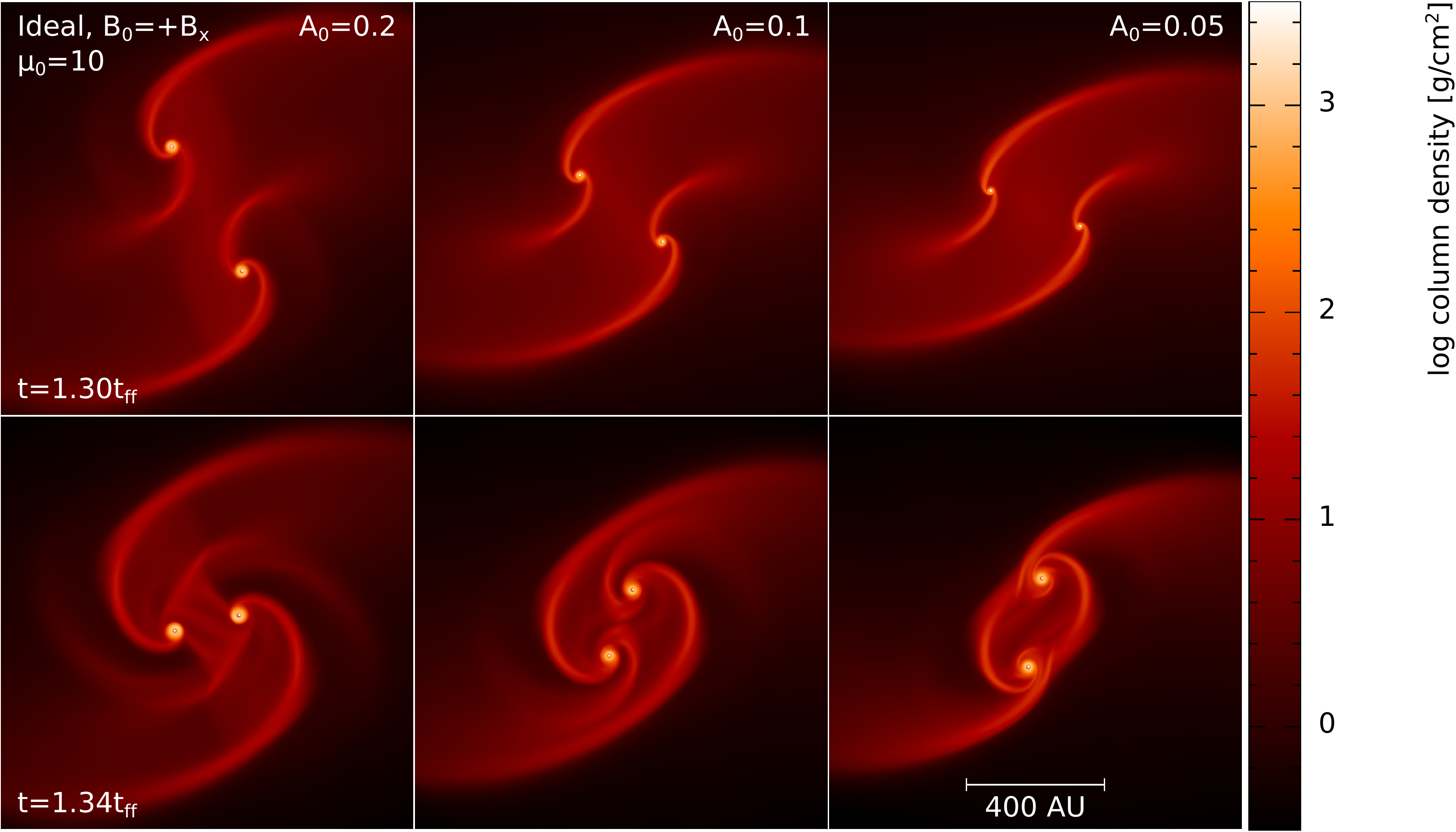}
\includegraphics[width=\columnwidth]{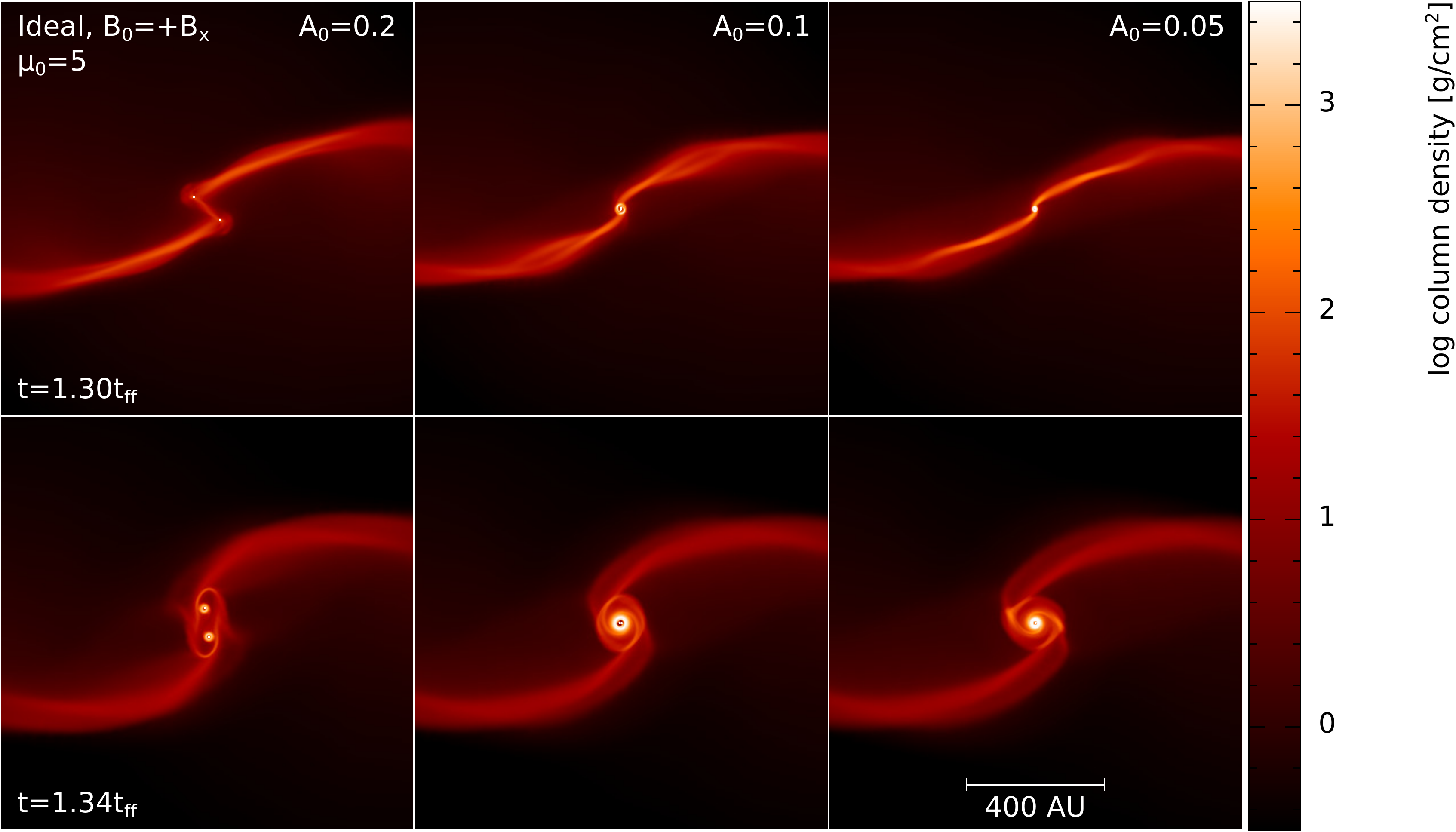}
\caption{As in Fig.~\ref{fig:results:idealZ:130134}, but with {\Bixp}.  With this magnetic field orientation, the initial magnetic field strength is the dominant parameter in determining the evolution. In the strong magnetic field case (bottom subfigure), either no binaries or tight binaries form.}
\label{fig:results:idealX:130134}
\end{center}
\end{figure}

\begin{figure}
\begin{center}
\includegraphics[width=0.9\columnwidth]{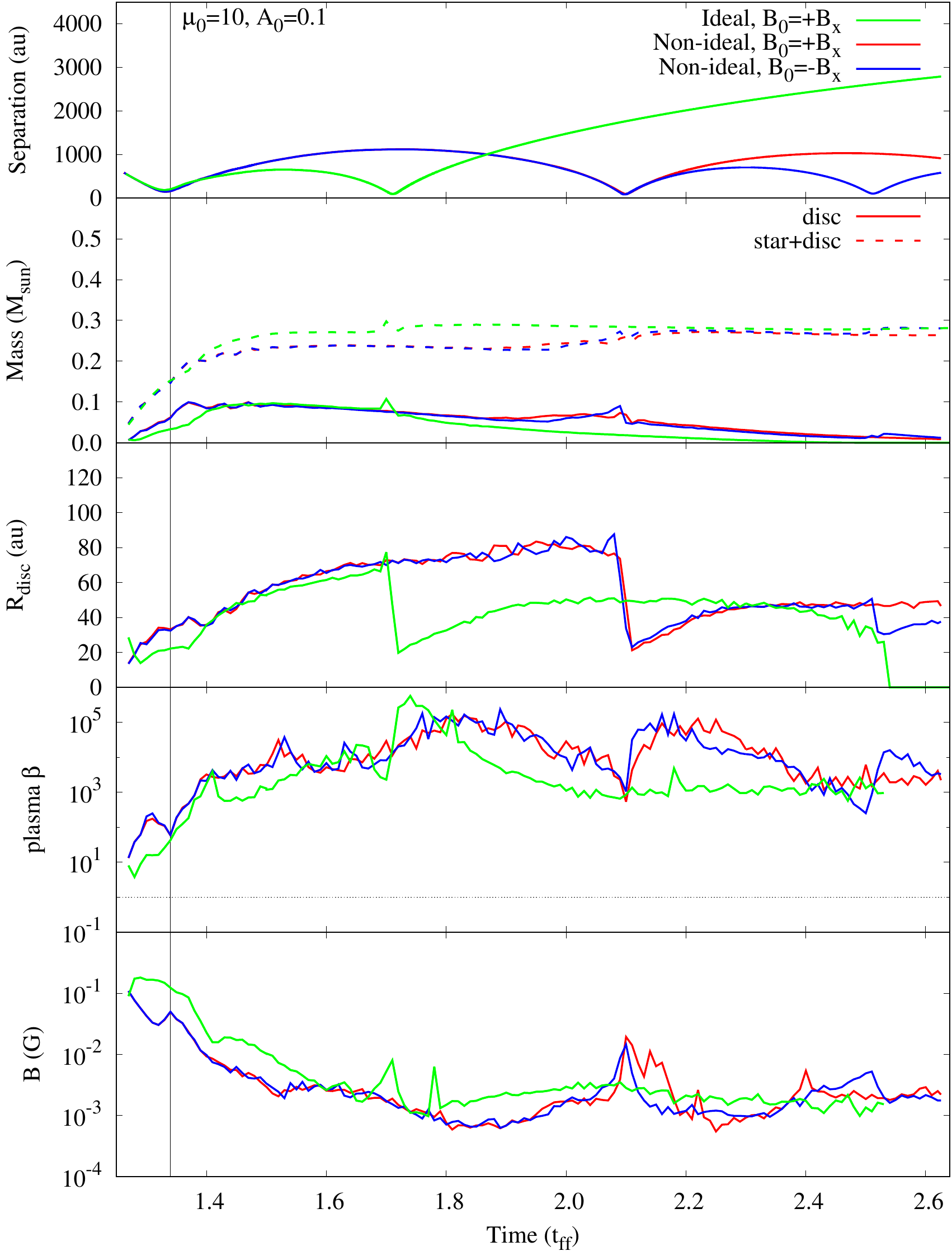}
\caption{Time evolution of selected models starting from the formation of the protostars as in Fig.~\ref{fig:results:evolZ}.  Each model is initialised with an initial density perturbation of $A_0=0.1$, \muten \ and $\bm{B}_0 = \pm\bm{B}_\text{x}$.  \emph{Top to bottom}: the separation of the two protostars, the disc and star+disc masses, the disc radius, and the mass-weighted plasma $\beta$ and magnetic field strength in the disc around one protostar.  The ideal MHD model has a larger first periastron, resulting in a shorter first period than the non-ideal models.  The subsequent periastron moves the ideal binary onto a long orbit, while the subsequent orbits of the non-ideal models decrease.}
\label{fig:results:evolX}
\end{center}
\end{figure}

Fig.~\ref{fig:results:idealX:130134} shows a repeat of the above calculations using an initial magnetic field perpendicular to the rotation axis (i.e. \Bixp); as previously shown, the gas column density is at $t = 1.30$ and 1.34\tff. The green lines in Fig.~\ref{fig:results:evolX} show the evolution of the separation of the two protostars and the evolution of the properties of one disc for the model with \muten \ and $A_0=0.1$.  As with the \Bizm \ models, after the initial growth of the protostar, the fluctuations in the star+disc mass correspond to the fluctuations in the disc mass, and the disc is always supported by gas pressure rather than magnetic pressure.

For weak magnetic fields ($\mu_0 = 10$; upper subfigure in Fig.~\ref{fig:results:idealX:130134}) initially perpendicular to the axis of rotation, the separation of the binary at first periastron is larger than its \Bizm \ counterpart, resulting in less interaction and a shorter period.  The corresponding disc radii and masses are smaller in the \Bixp \ models, discussed further in Section~\ref{sec:results_IZX} below.

Strong magnetic fields $(\mu_0 = 5$; lower subfigure in Fig.~\ref{fig:results:idealX:130134}) initially perpendicular to the axis of rotation suppress the formation of binaries.  For $A_0=0.2$, a binary forms early with first periastron occurring at $t\approx 1.29$\tff, compared to $t\approx 1.37$\tff \ for its \Bizm \ counterpart.  The apoastron distance is $R_\text{apo} < 120$ au, whereas this is a typical periastron distance for its $-\bm{B}_\text{z}$ counterpart.  The $A_0=0.1$ model forms a binary pair with a semi-major axis of 3 au and a common disc, while the $A_0=0.05$ model forms a single protostar and disc.  For the purposes of our analysis, the $A_0=0.1$ model is treated as single protostar.  

The magnetic field strength is larger and the plasma $\beta$ is smaller in the discs of the \Bixp \ models than their \Bizm \ counterparts, indicating that the magnetic field is more important in  the \Bixp \ models for the evolution of the disc.

\subsection{Ideal MHD: magnetic field evolution}
\label{sec:results_IZX}

Comparison of Fig.~\ref{fig:results:idealZ:130134} to Fig.~\ref{fig:results:idealX:130134} demonstrates that the evolution of the magnetic field depends on its initial orientation.  At \otftff, the net magnetic field in the discs of the \muten \ models are $\sim$$10$ times higher in the \Bixp \ models than their \Bzm \ counterparts with the same $A_0$, despite the same initial strength.  Since stronger magnetic fields enhance magnetic braking, the discs in the \Bixp \ models are smaller and less massive.

\begin{figure}
\begin{center}
\includegraphics[width=\columnwidth]{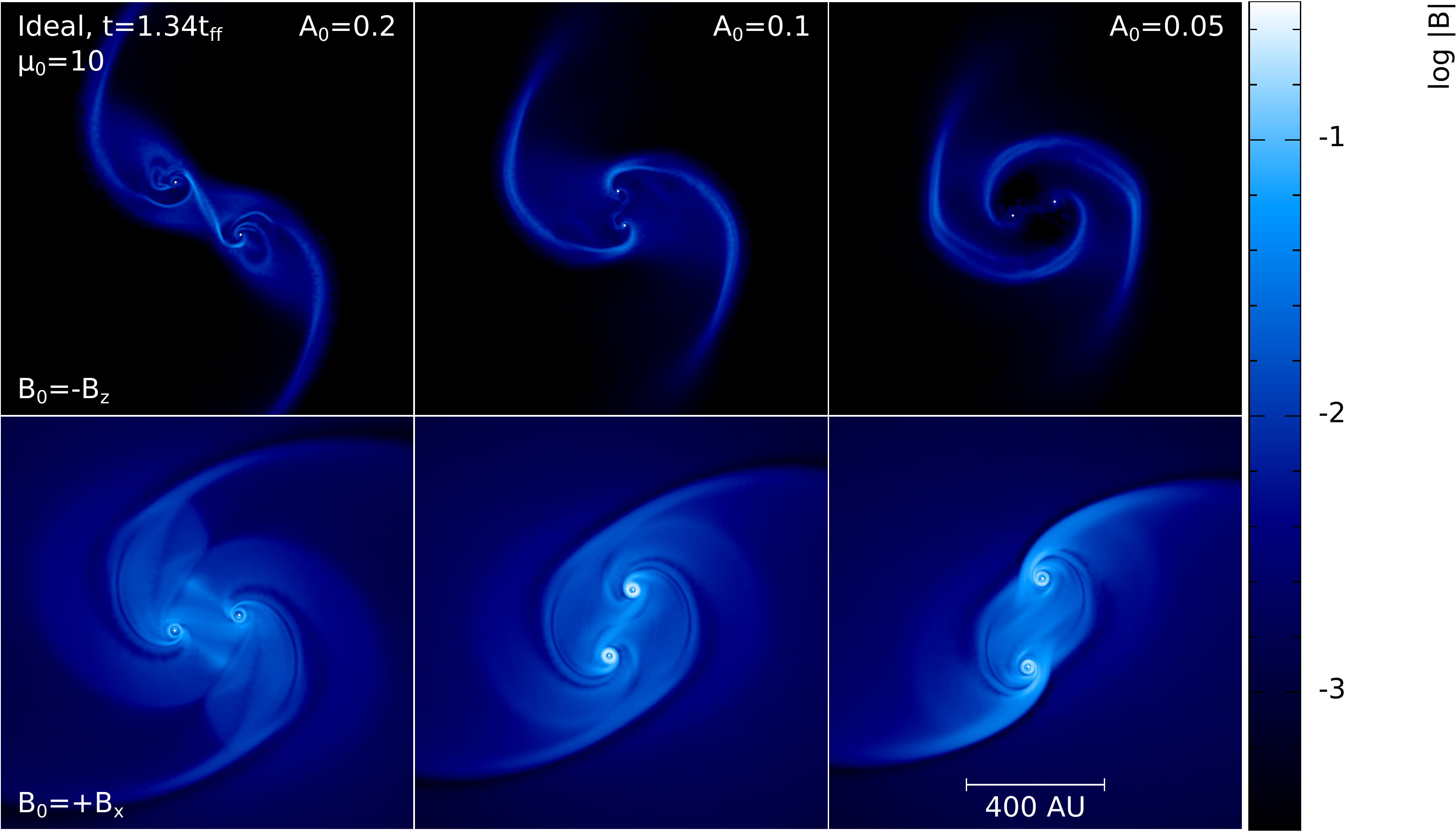}
\includegraphics[width=\columnwidth]{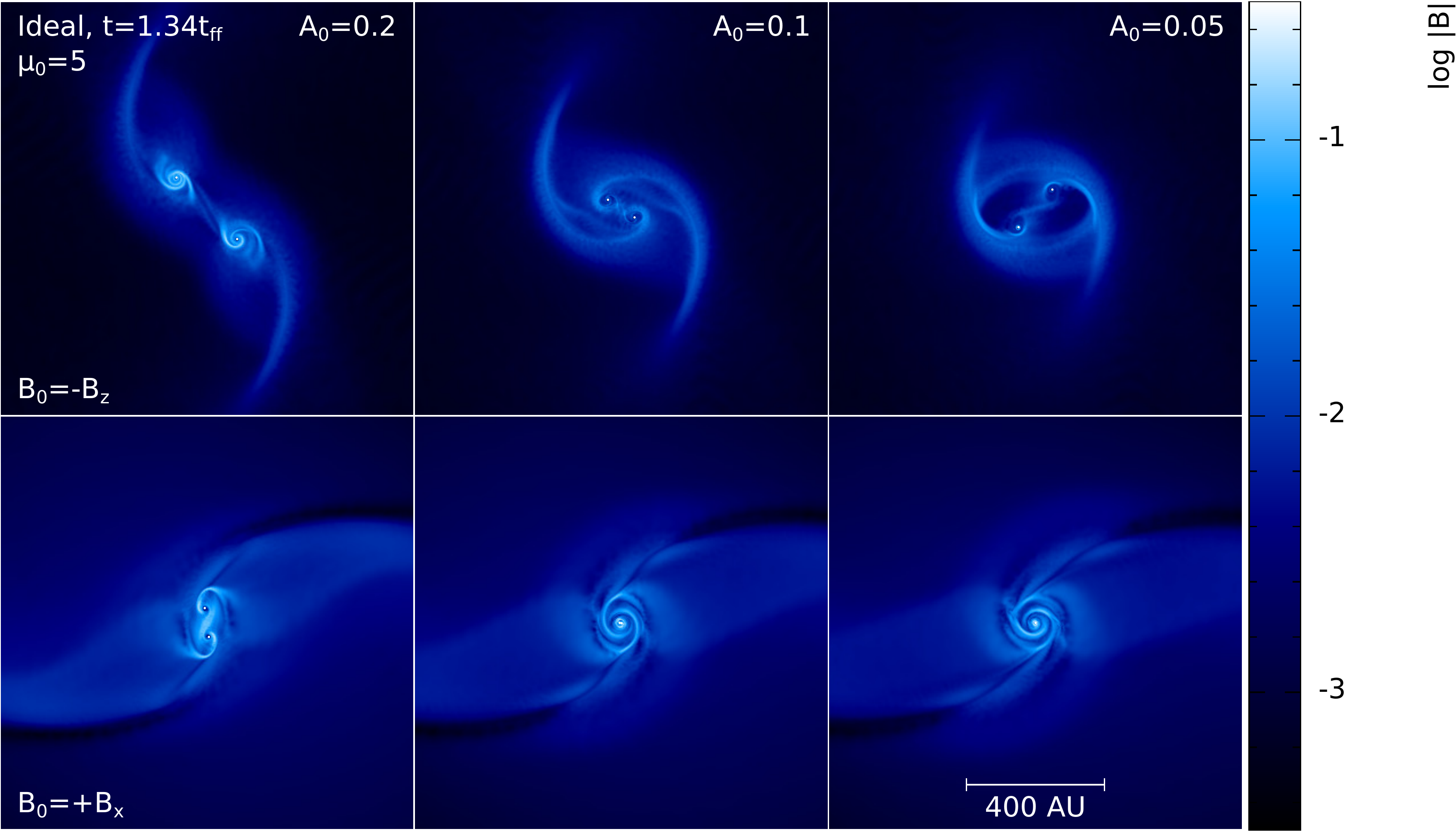}
\caption{Magnetic field strength in the mid-plane ($z=0$) at \otftff \ for the ideal MHD models with \muten \ (top subfigure) and \mufive \ (bottom subfigure).  At each initial magnetic field strength, the mid-plane magnetic field is always stronger in the models initialised with \Bixp, despite the same initial value.}
\label{fig:results:idealA01:B134:xy}
\end{center}
\end{figure}

To quantify this, Fig.~\ref{fig:results:idealA01:B134:xy} compares the magnetic field strength in the mid-plane ($z=0$) for the ideal MHD models at \otftff.
The magnetic field strength of the \Bixp \ models is higher throughout the mid-plane and in the discs than in their respective  \Bizm \ models.  As the vertical collapse proceeds in the \Bixp \ models, the field is dragged into the mid-plane.  When the discs form, the stronger magnetic field is wound into the disc, further enhancing its strength; in the \Bixp \ models, the azimuthal magnetic field, $B_\phi$, is the dominant component.  In the \Bizm \ models, the radial dragging of the magnetic fields enhances the field strength in the discs compared to the background, but not compared to their \Bx \ counterparts.

In all models, there is little conversion of horizontal magnetic fields into vertical fields or vice versa, hence only a weak vertical (horizontal) magnetic field develops in the \Bixp \ (\Bzm) models.  For example, on average at \otftff, the $\phi$-component of the magnetic field is $\sim$$9.0$ times stronger than the $z$-component for the model with \mufive, $A_0=0.1$ and \Bixp, while the $z$-component is $4.7$ times stronger than the $\phi$-component in its counterpart model with \Bixm. 

\subsection{Non-ideal MHD with the magnetic field aligned or anti-aligned to the rotation axis}
\label{sec:results_NI:aligned}

Previous studies have demonstrated that non-ideal MHD affects the formation and evolution of discs around protostars forming in isolation (e.g. \citealt{LiKrasnopolskyShang2011};  \citealt{TomidaOkuzumiMachida2015}; \citealt{TsukamotoEtAl2015b,TsukamotoEtAl2015}; \citetalias{WPB2016}).  Further, when the Hall effect is included, the direction of the magnetic field with respect to the axis of rotation affects the evolution, with larger discs forming for cases where the magnetic field is anti-aligned with the axis of rotation (\citealt{BraidingWardle2012accretion}; \citealt{TsukamotoEtAl2015b}; \citetalias{WPB2016}).  

 We thus perform a suite of non-ideal MHD models with the same parameters as the ideal MHD models discussed above, except that we also run models where the sign of the magnetic field is reversed.
 
\subsubsection{Early time evolution and disc properties}
\label{sec:results_NIZ:align:early}

\begin{figure}
\begin{center}
\includegraphics[width=\columnwidth]{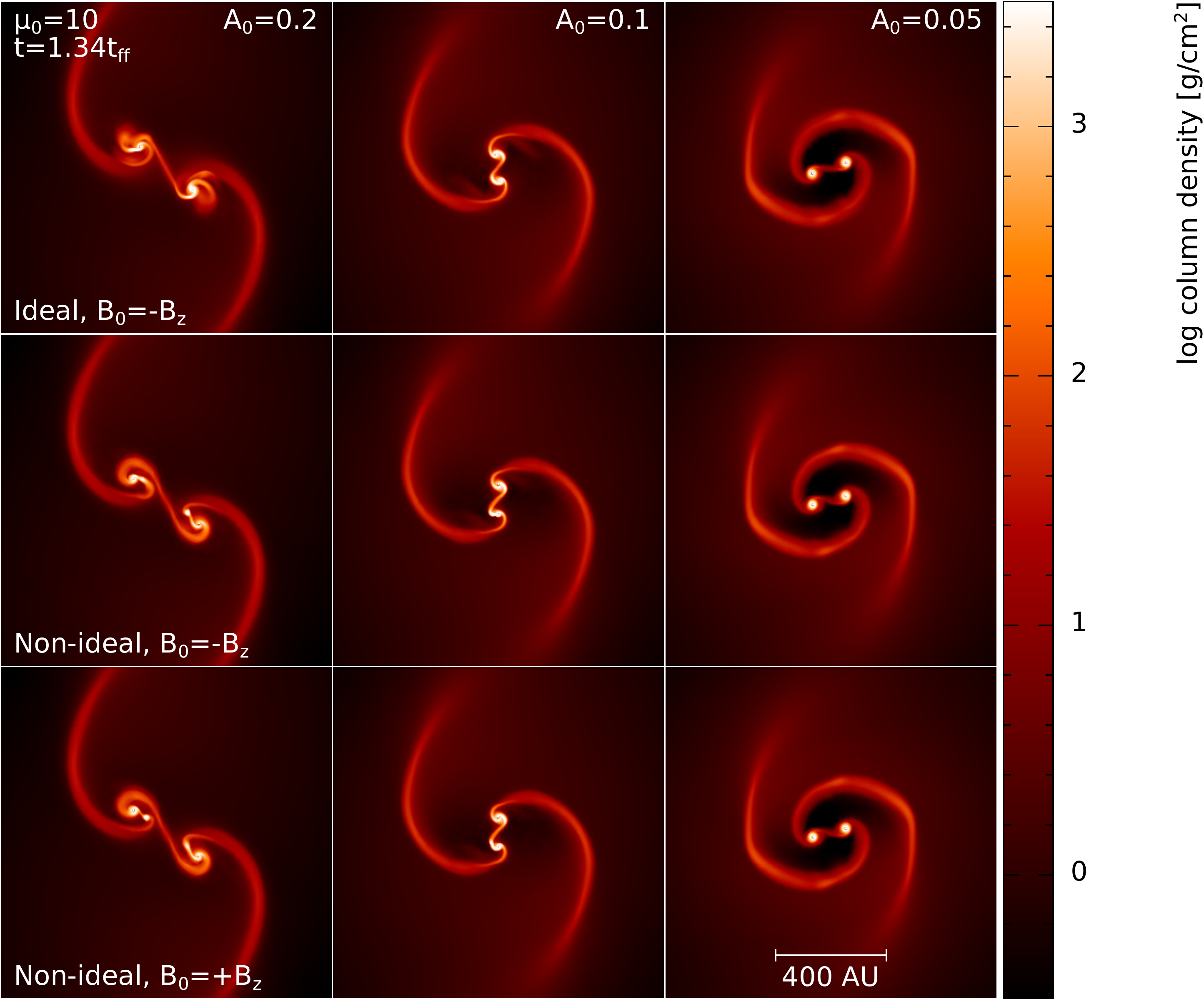}
\includegraphics[width=\columnwidth]{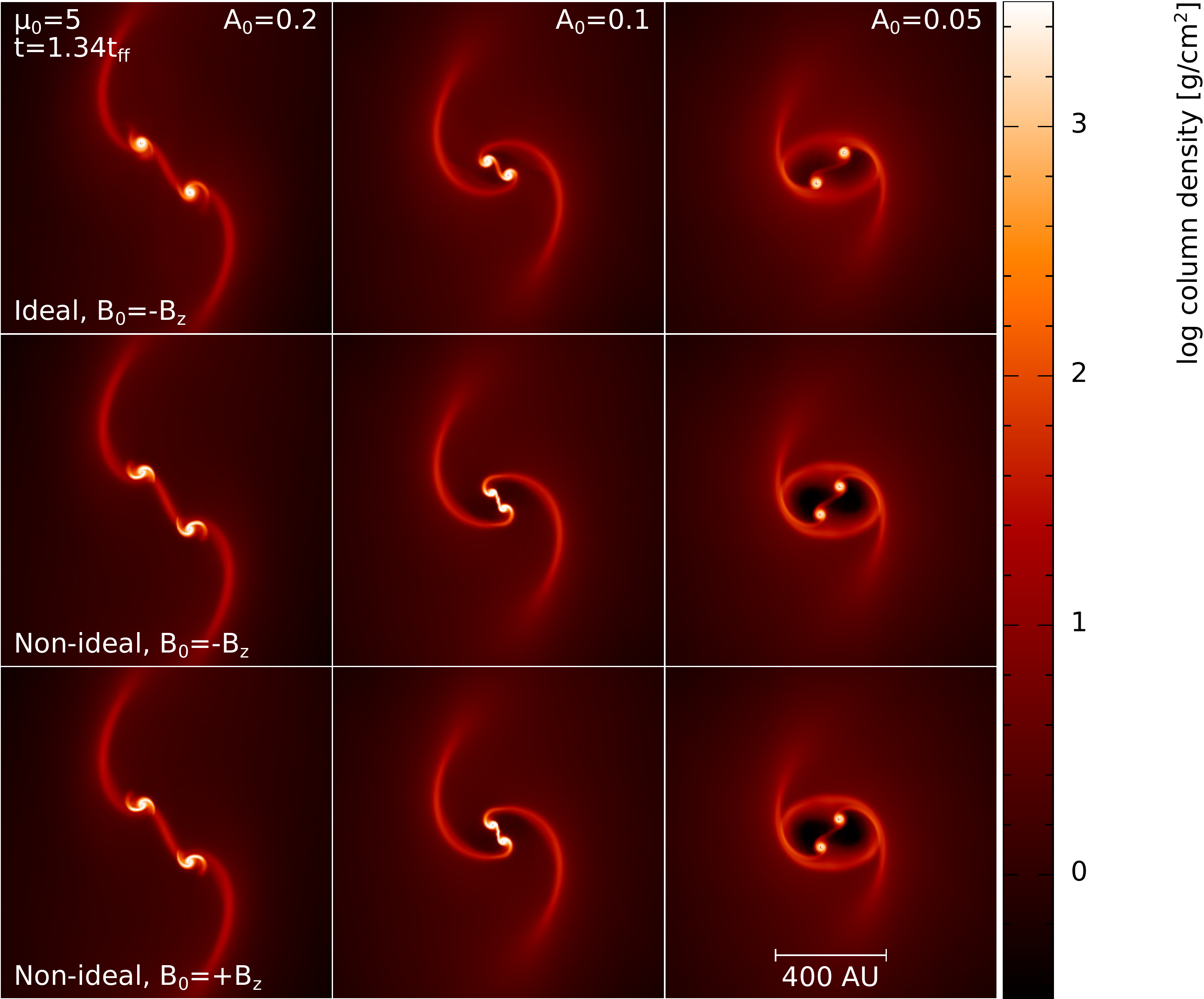}
\caption{Effect of non-ideal MHD on binary formation. Each panel shows the face-on column density of a particular model at \otftff.  \emph{Top to bottom in each subfigure}: ideal MHD, non-ideal MHD with \Bzm \ and with $+\bm{B}_\text{z}$, where the models in the top subfigure are initialised with \muten, while the bottom subfigure shows models with \mufive.  The addition of non-ideal MHD has only a small effect compared to changing the initial magnetic field strength (comparing top to bottom subfigures) or perturbation amplitude ($A_{0}$; comparing columns left to right).}
\label{fig:results_NI:Z:134}
\end{center}
\end{figure}

Fig.~\ref{fig:results_NI:Z:134} shows the gas column density at \otftff \ of the $\bm{B}_0 = \pm\bm{B}_\text{z}$ models.  The effect of non-ideal MHD on the results (comparing rows top to bottom in each subfigure) is small, with $t_\text{peri}$ differing by less than a percent, though with $R_\text{peri}$ differing by up to $18$ per cent, or a maximum of $23$ au. 
The disc radii and masses differ by up to $45$ and $65$ per cent, respectively.  The largest differences in disc mass between the ideal and non-ideal MHD calculations occurs in the calculations with \muten \ and $A_0=0.2$ (left-hand column; top subfigure of Fig.~\ref{fig:results_NI:Z:134}), which also show the largest difference in $R_\text{peri}$.  These discs are in the early stages of fragmentation, hence are irregularly shaped, which contributes to this difference in mass.  A third protostar forms at $t\approx 1.36$\tff, disrupting the host discs.  In the ideal MHD model, a fourth protostar is formed at approximately the same time to disrupt the second disc.

\begin{figure*}
\begin{center}
\includegraphics[width=\columnwidth]{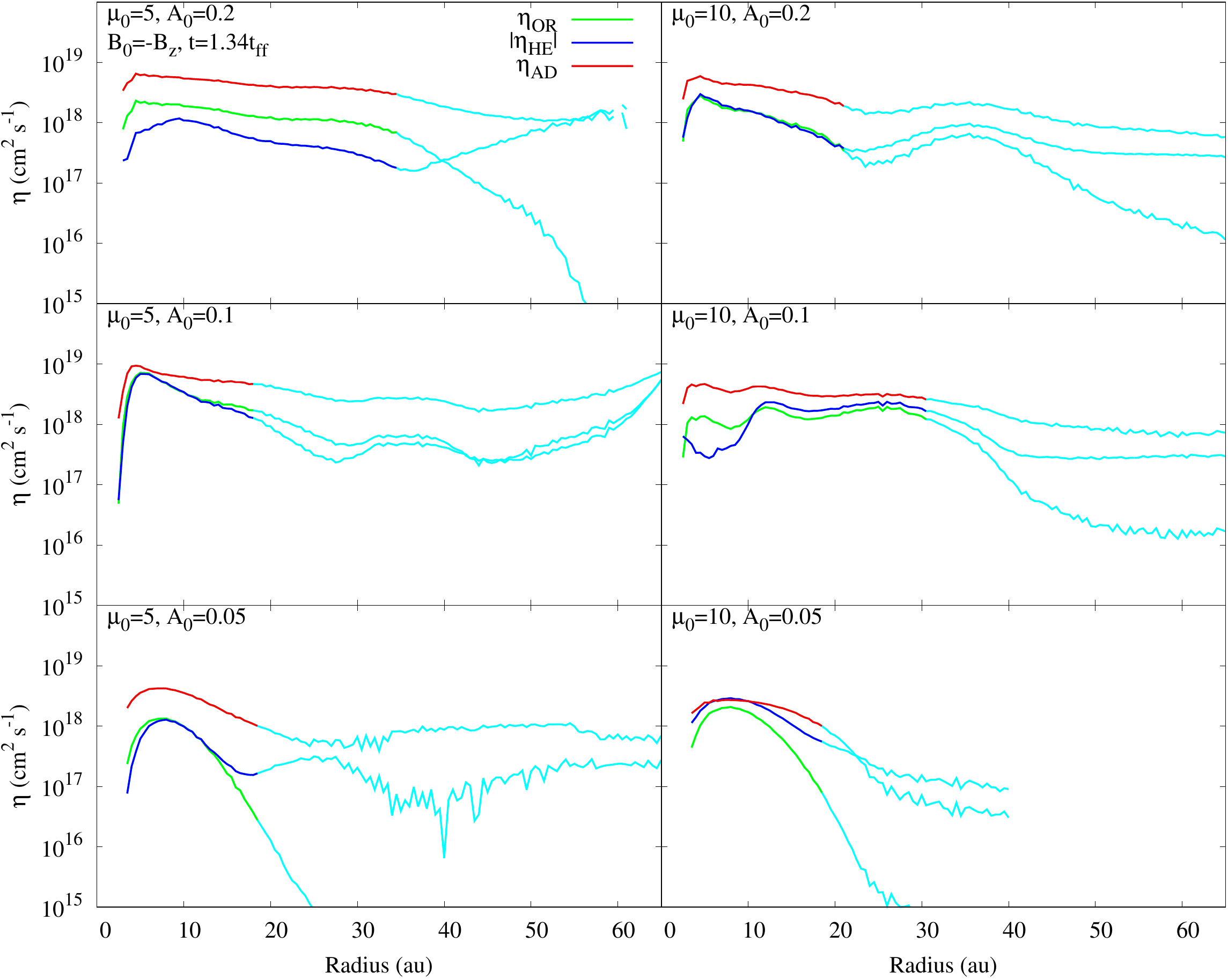}
\includegraphics[width=\columnwidth]{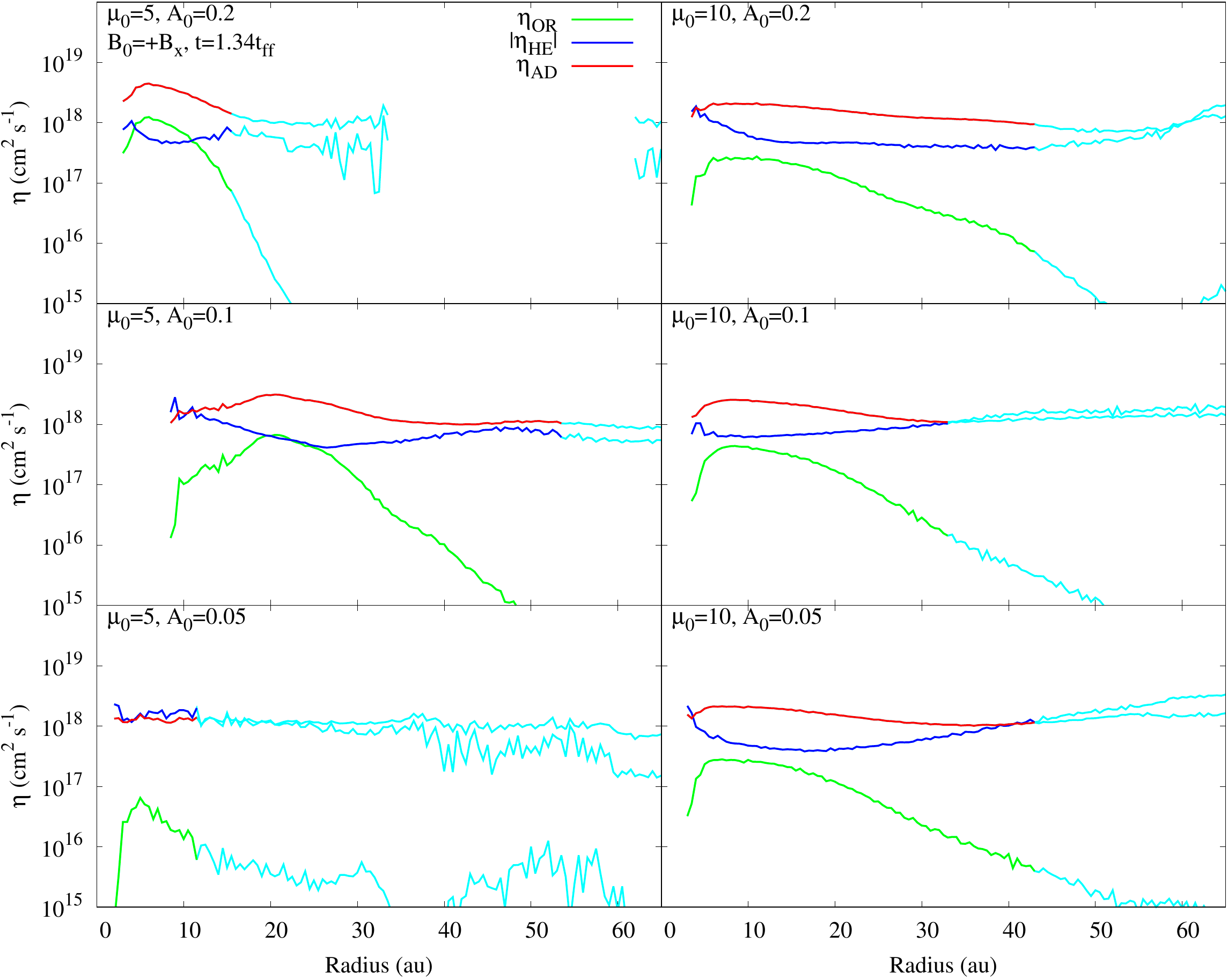}
\caption{Average values of Ohmic, ambipolar and Hall diffusion coefficients for the gas in the disc around one protostar at \otftff \ averaged over all gas particles with $\rho > \rho_\text{disc,min}$, for the models with \Bizm \ (left-hand subfigure) and \Bixp \ (right-hand subfigure).  The Hall coefficient is the average of its absolute values.  The lines switch to cyan at the defined edge of the disc.  Ambipolar diffusion is the dominant effect in the disc, although all three non-ideal coefficients typically differ by less than a factor of ten close to the protostar.  These coefficients are smaller than in models that form an isolated protostar presented in \citetalias{WPB2016}.}
\label{fig:results_NI:evol:eta}
\end{center}
\end{figure*}

At these early times, the main differences are caused by changes in $\mu_0$ and $A_0$.  This is expected since the density and magnetic field strengths are only starting to reach the limits where the non-ideal effects become important.  The left-hand subfigure in Fig.~\ref{fig:results_NI:evol:eta} show the coefficients for Ohmic, Hall and ambipolar diffusion in the discs of the non-ideal models with \Bizm \ at {\otftff} (we plot the average of the absolute value of the coefficients).
Ambipolar diffusion is the dominant effect in the disc, with the coefficients of the Hall effect and Ohmic resistivity lower by a factor of ten in the disc.  

For comparison, these values are $\sim$$1$ dex lower than in the discs formed in the isolated collapse simulations shown by \citetalias{WPB2016}.  Further, the plasma $\beta$ is smaller in the isolated collapse models, so we expect that non-ideal MHD will play a more minor role in the binary case, as we have already shown.

\subsubsection{Late time evolution and disc properties}
\label{sec:results_NIZ:align:late}

The $\mu_{0} = 10$ models shown in the upper subfigure of Fig.~\ref{fig:results_NI:Z:134} form large discs near first periastron.  These discs subsequently fragment, with the non-ideal MHD discs fragmenting before their ideal MHD counterparts.  This fragmentation hinders the analysis of the late time evolution of the weak field models, so in the rest of this section we focus on the {\mufive} models.
\begin{figure}
\begin{center}
\includegraphics[height=0.8\columnwidth]{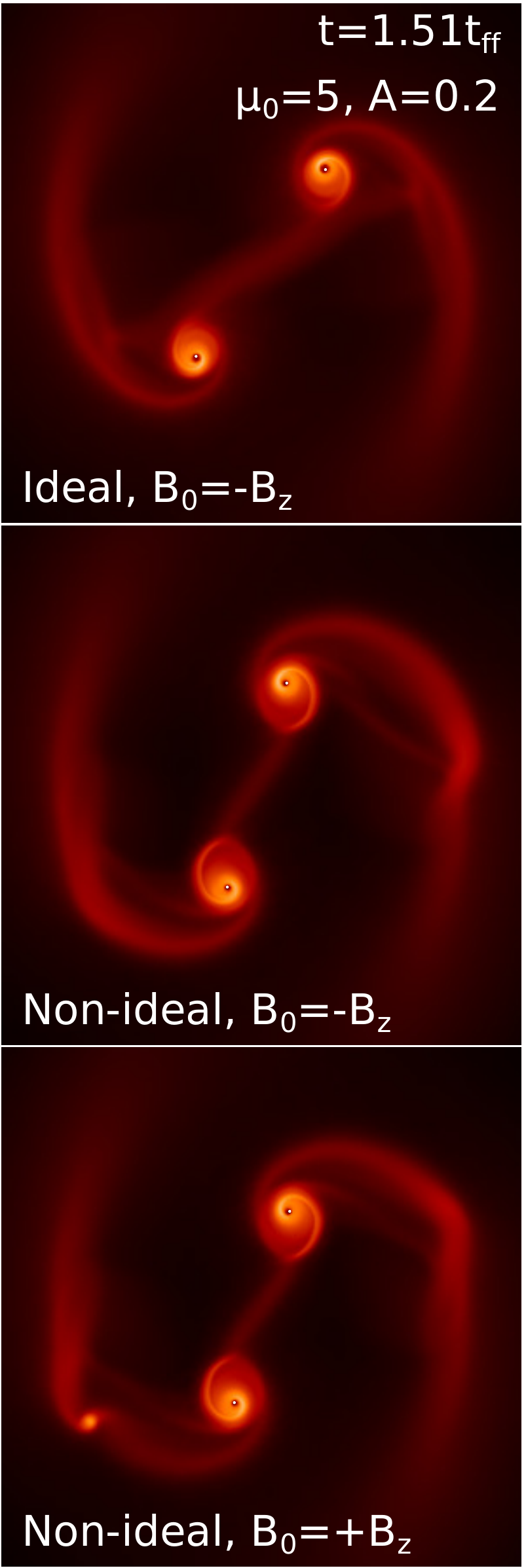}
\includegraphics[height=0.8\columnwidth]{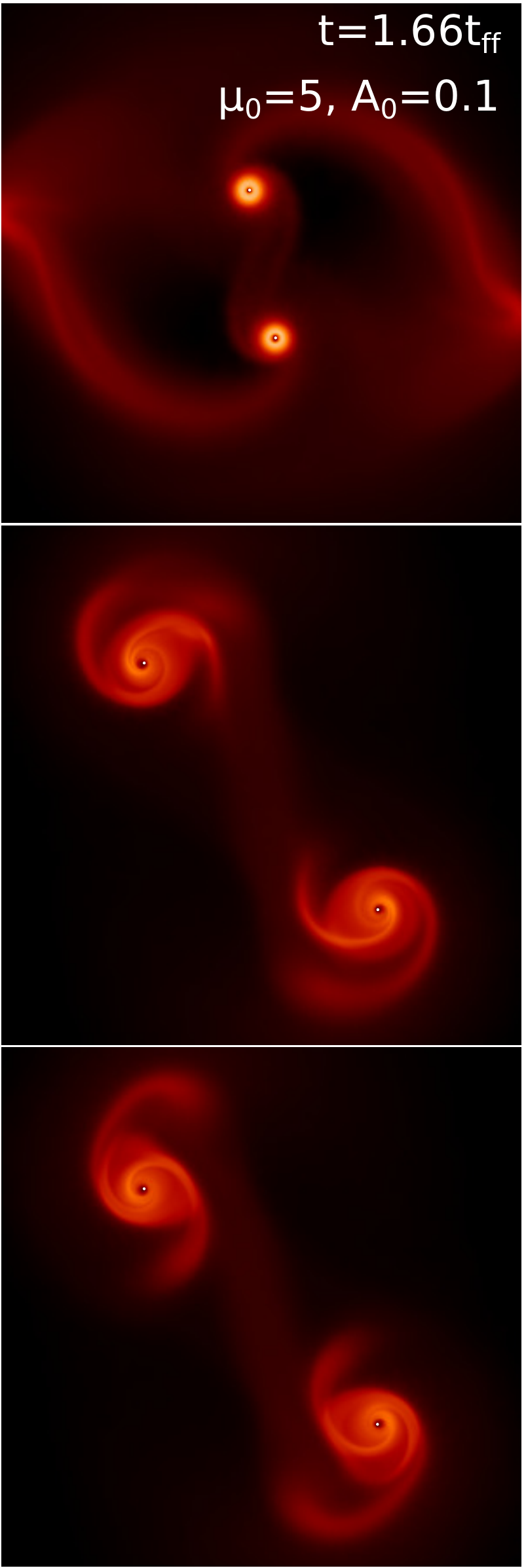}
\includegraphics[height=0.8\columnwidth]{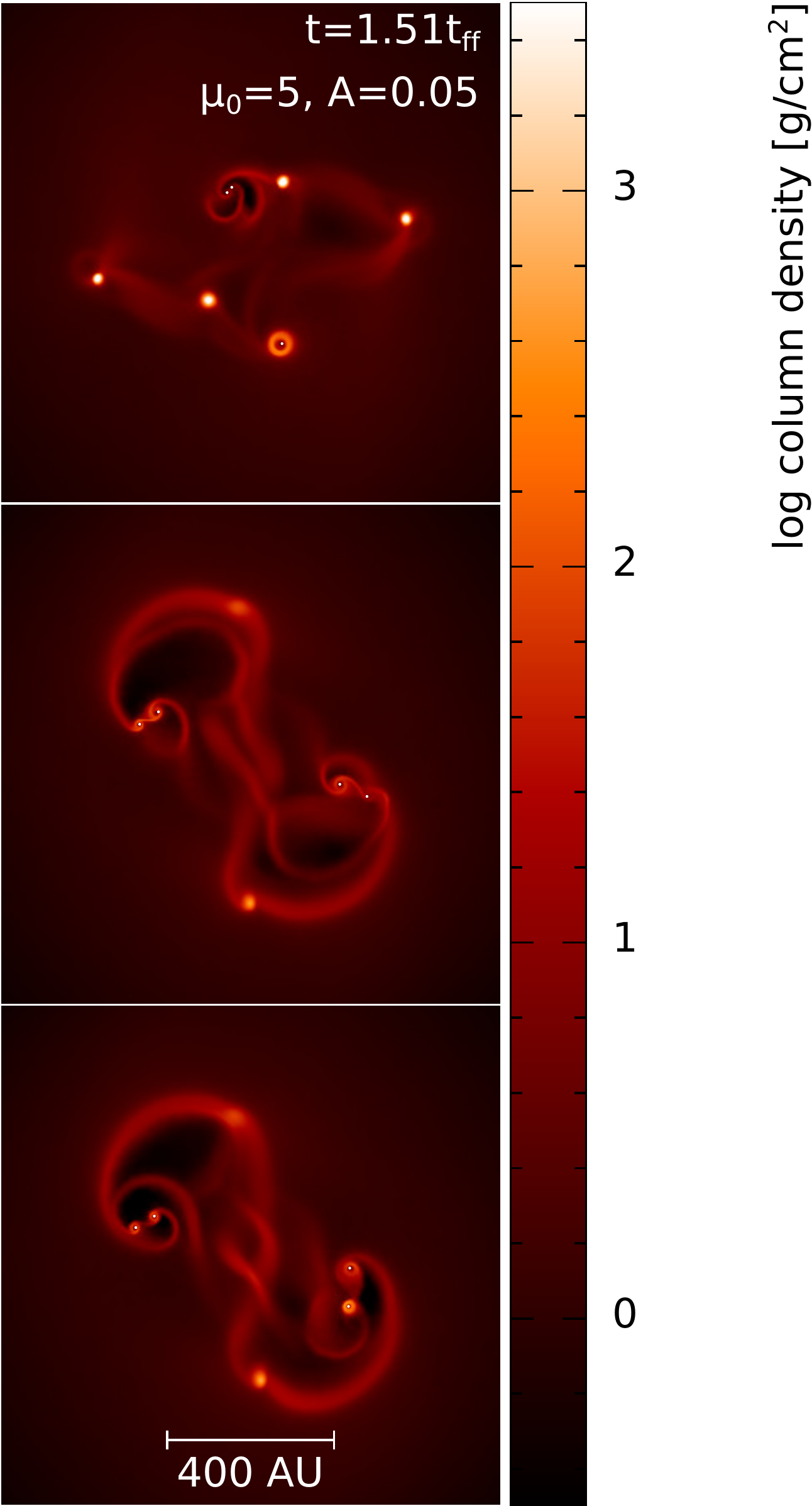}
\caption{Gas column density for nine models with \mufive \ and $\bm{B}_0=\pm\bm{B}_\text{z}$ at a later time.  
\emph{Left-hand column}: the models with $A_0=0.2$ at $t=1.51$\tff, which is first apoastron for all three models; the evolution is only weakly dependent on the non-ideal effects.  
\emph{Middle column}: the models with $A_0=0.1$ at $t=1.66$\tff, which is second apoastron for the ideal MHD model; at this time, the ideal MHD model has discs that are more massive and more concentrated near the protostar than in the non-ideal MHD models.  
\emph{Right-hand column}: the models with $A_0=0.05$ at $t=1.51$\tff; additional protostars form at $t \approx 1.4$\tff, which interact with the primary protostars starting at $t \approx 1.45$\tff \ to disrupt the disc by this time.}
\label{fig:results:inM05Z:late}
\end{center}
\end{figure}

The left-hand column of Fig.~\ref{fig:results:inM05Z:late} shows the gas column density at $t=1.51$\tff \ for the $A_0 = 0.2$ models, which is first apoastron for these three models.  The strong initial density perturbation yields an evolution that is very weakly dependent on the non-ideal MHD processes.  Amongst these three models the first period and first apoastron differ by less than $3$ and $13$ per cent, respectively, compared to $65$ and $56$ percent, respectively, for the models with $A_0 = 0.1$.  At this epoch, the difference between these three models are small, with the disc mass and magnetic field strengths at any given radius differing by less than a factor of $2$ and $3$, respectively.  

At second periastron, the discs in the non-ideal MHD models fragment and form more protostars, totally disrupting the discs.  This is mainly an artefact of our use of a barotropic equation of state to represent the thermodynamics.

The middle column of Fig.~\ref{fig:results:inM05Z:late} shows the gas column density of the $\mu_{0} = 5$, $A_0=0.1$ models at $t = 1.66$\tff, corresponding to the second apoastron in the ideal MHD model. At this epoch, the discs in the non-ideal MHD models are less massive and more extended than their ideal MHD counterpart (comparing middle and bottom panels to the top panel). This is a result of the different orbital histories, which diverge shortly after first periastron at $t_\text{peri} \approx 1.34$\tff. The blue and red lines in the right-hand column of Fig.~\ref{fig:results:evolZ} show the protostar separation and disc properties for the non-ideal models, which can be directly compared to their ideal MHD counterpart (green line).  
In the ideal MHD model, there have been two periastron approaches by $t = 1.66$\tff, keeping the discs small and concentrated around its host protostar; the non-ideal MHD discs have not interacted with one another again, thus effectively evolved in isolation for the previous $\text{d}t \approx 0.3$\tff.  Moreover, the values of the non-ideal MHD coefficients rapidly decrease after periastron, making the later evolution more ideal.  The coefficients increase briefly at second periastron when the close interaction increases the density of the disc.  However, the interaction also leads to an increase in the value of plasma $\beta$, counteracting any added effect. 

The right-hand column of Fig.~\ref{fig:results:inM05Z:late} shows the $A_0=0.05$ models at $t=1.51$\tff.  Additional clumps of gas form at $t\approx 1.4$\tff \ and $r \approx 270$ au from the centre of mass; at this time, the primary protostars are $r \approx 98$ au from the centre of mass.  The clumps spiral inwards and interact with the primary binary starting at $t\approx1.45$\tff, totally disrupting the primary binary. 

Throughout this paper, we have compared models at the same absolute times. However, this may be an unfair comparison in some cases due to different orbital dynamics.  For the above ideal and non-ideal MHD models with $A_0=0.1$ (middle column of Fig~\ref{fig:results:inM05Z:late}), $t_\text{apo} \approx1.46$ and $1.60$\tff, respectively; the non-ideal MHD models evolve very little between $t=1.60$ and $1.66$\tff, so the panels in Fig.~\ref{fig:results:inM05Z:late} are representative of both times.  The disc mass and radius of the ideal MHD model, however, decreases from $0.096$ M$_\odot$ and $48$ au to $0.081$ M$_\odot$ and $26$ au, respectively, between first and second apoastron.  The intervening periastron passages strip mass from the disc, concentrating the remaining disc mass closer to its host protostar.  This also results in a stronger magnetic field in the inner regions of the disc at  $t = 1.66$\tff.

\subsection{Non-ideal MHD with the magnetic field perpendicular to the rotation axis}
\label{sec:results_NI:perp}

We repeat the above study using a magnetic field initially perpendicular to the rotation axis. We consider both $\bm{B}_0=\pm\bm{B}_\text{x}$ since we expect a $\bm{B}_\text{z}$ component to be generated during the evolution.

\subsubsection{Early time evolution and disc properties}
\label{sec:results_NI:perp:early}

\begin{figure}
\begin{center}
\includegraphics[width=\columnwidth]{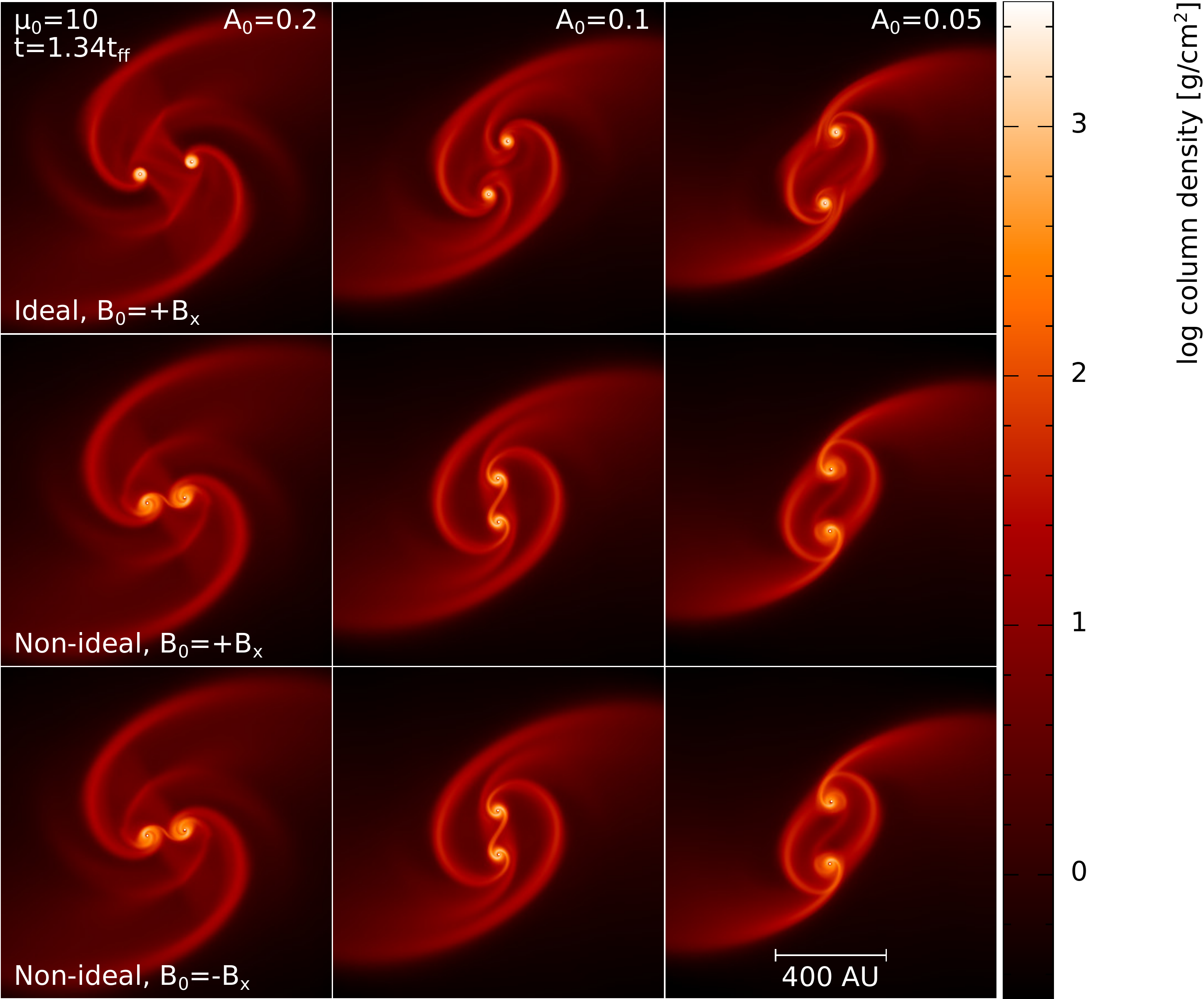}
\includegraphics[width=\columnwidth]{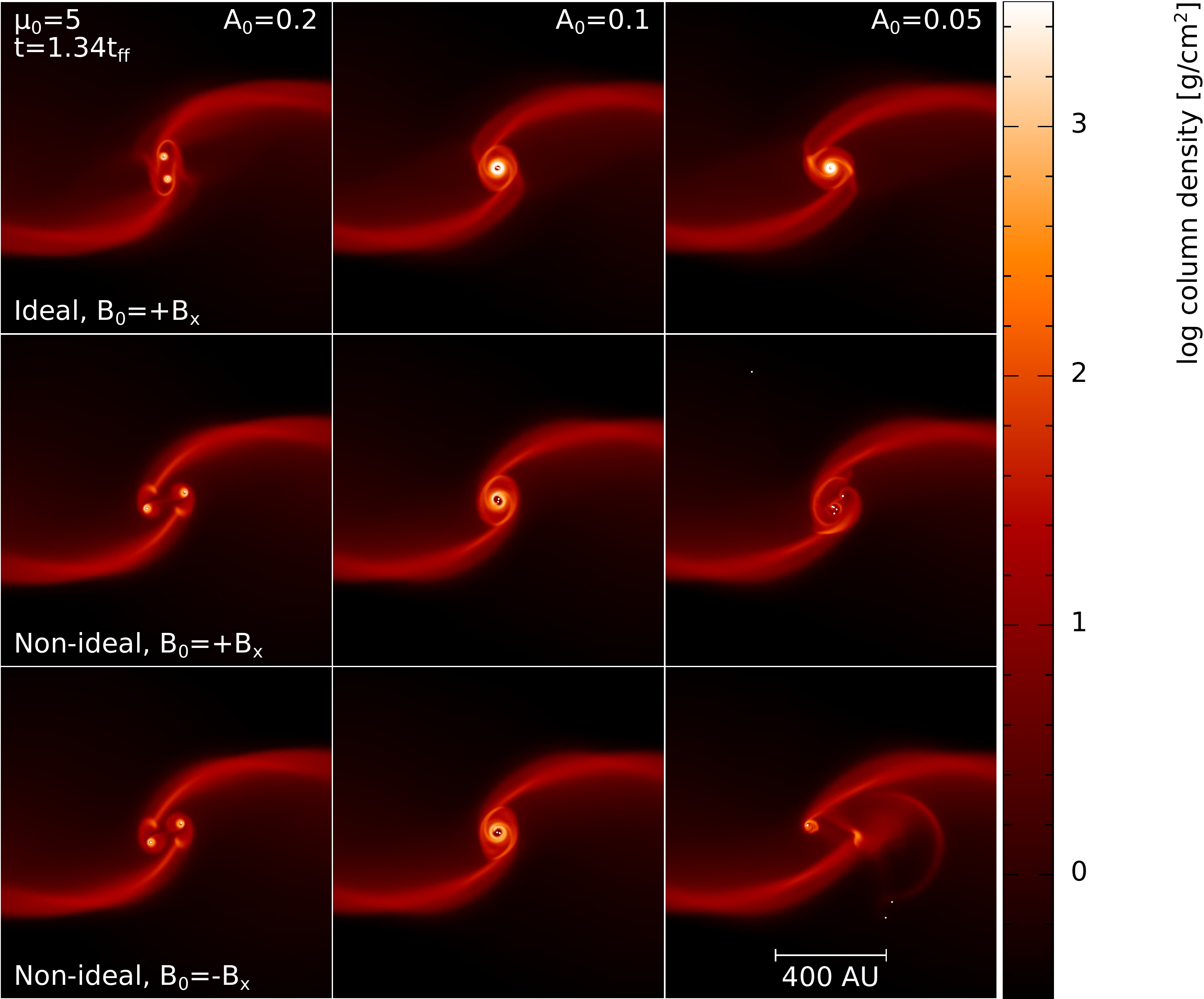}
\caption{Effect of non-ideal MHD on binary formation when the magnetic field is initially perpendicular to the axis of rotation. Each panel shows the face-on column density of a particular model \otftff.  \emph{Top to bottom in each subfigure}: ideal MHD, non-ideal MHD with \Bixp \ and with $-\bm{B}_\text{x}$, where the models in the top (bottom) subfigure are initialised with \muten \ (\mufive).  The non-ideal MHD models with \mufive \ and $A_0=0.1$ produce a tight binary with an orbit larger than their ideal MHD counterpart, yielding a larger central cavity and less massive disc.  The non-ideal MHD models with \mufive \ and $A_0=0.05$ produce multiple protostars, which immediately disrupt the system.  For weak magnetic fields, the non-ideal MHD models yield larger discs than their ideal counterparts.}
\label{fig:results_NI:inX:134}
\end{center}
\end{figure}

Fig.~\ref{fig:results_NI:inX:134} shows the gas column density of our suite of models with $\bm{B}_0=\pm\bm{B}_\text{x}$ at \otftff.
The weaker field models (\muten; top subfigure) yield well-separated binaries for all $A_0$, with the non-ideal MHD models forming larger and more massive discs.   The magnetic field in the non-ideal MHD discs is approximately constant at $\bm{B} \approx 0.05$ G, while in the ideal MHD model, it decreases by a factor of $\sim$$3$ between the maximum strength and the outer edge of the disc.

 By contrast, the \mufive \ and $A_0=0.2$ models (left-hand column, bottom subfigure) yield discs that are not significantly different from one another.  The \mufive \ and $A_0=0.1$ calculations (middle column, bottom subfigure) forms two protostars by $t=1.27$\tff, which all form a tight binary of semi-major axes $\sim$$3$ and $\sim$$5$ au for the ideal and non-ideal MHD models, respectively.  The larger semi-major axis in the non-ideal MHD models results in a larger central cavity in the circumbinary disc.

A single protostar forms in the ideal MHD model with $A_0 = 0.05$, while $5$ and $7$ protostars form by $t=1.305$\tff \ in the non-ideal MHD models with \Bixpm, respectively (right-hand column, bottom subfigure).  This plethora of protostars immediately disrupts the discs, and the remaining evolution is chaotic. 

As with the \Bizpm \ models, the dynamics are dominated by $\mu_0$ and $A_0$ rather than the effect of non-ideal MHD.  The right-hand subfigure in Fig.~\ref{fig:results_NI:evol:eta} shows the non-ideal MHD coefficients for the non-ideal models with \Bixp \ at \otftff.  As with their \Bizm \ counterparts, ambipolar diffusion is the dominant term, however, all the coefficients are lower despite the stronger magnetic field in the disc. 

\subsubsection{Late time evolution and disc properties}
\label{sec:results_NI:perp:late}
\begin{figure}
\begin{center}
\includegraphics[height=0.8\columnwidth]{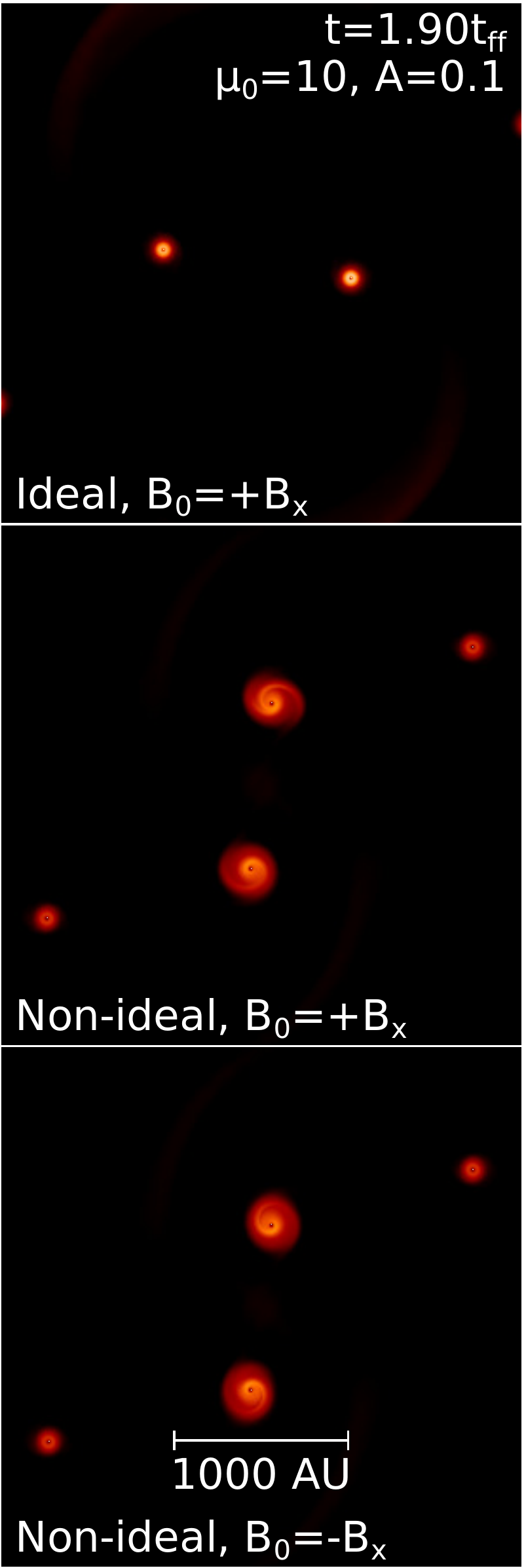}
\includegraphics[height=0.8\columnwidth]{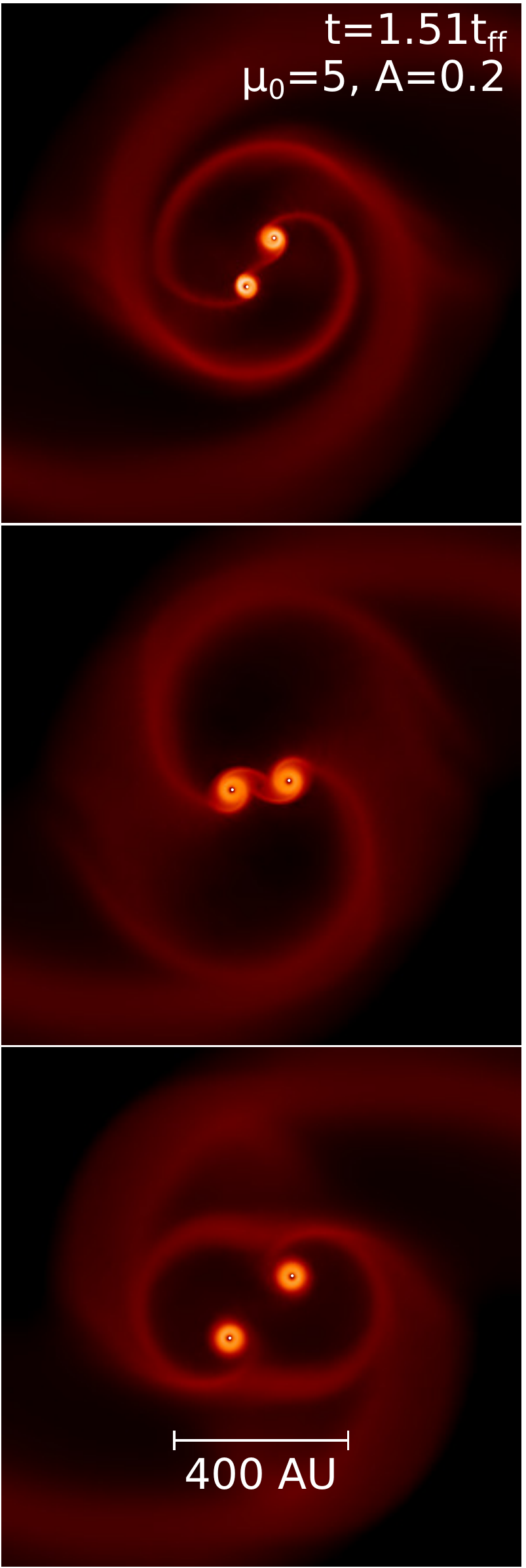}
\includegraphics[height=0.8\columnwidth]{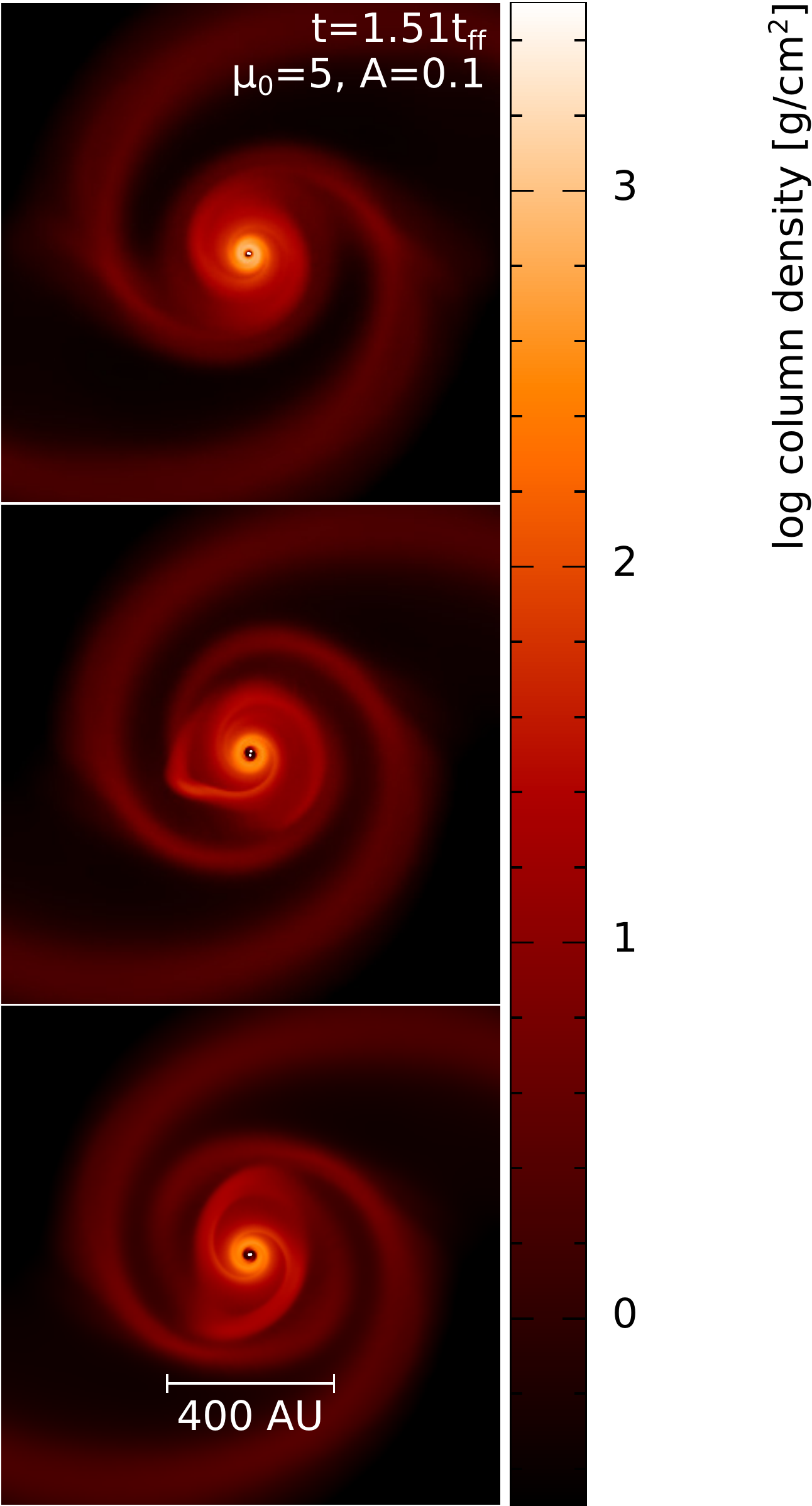}
\caption{Gas column density for nine models with $\bm{B}_0 = \pm\bm{B}_\text{x}$ at a later times.  
\emph{Left-hand column}: models with \muten \ and $A_0 = 0.1$ at $t = 1.90$\tff; the images in this column have frame sizes of (3000 au)$^2$ so that the four protostars in the non-ideal MHD models can be seen.  
\emph{Middle column}: models with \mufive \ and $A_0 = 0.2$ at $t = 1.51$\tff.  
\emph{Right-hand column}: models with \mufive \ and $A_0 = 0.1$ at $t = 1.51$\tff.  
In all the models presented, the ideal MHD models have more concentrated discs than their respective non-ideal MHD counterparts.}
\label{fig:results_NI:inX:late}
\end{center}
\end{figure}

The initial differences between the weak field models with $A_0=0.1$ caused by non-ideal effects trigger pronounced differences as the evolution continues.  For example, the larger $R_\text{peri}$ for the ideal model results in it reaching second periastron after $\text{d}t = 0.39$\tff, while the non-ideal models require $\text{d}t=0.79$\tff \ to reach second periastron; see the top panel of Fig.~\ref{fig:results:evolX} and left-hand column of Fig.~\ref{fig:results_NI:inX:late}.  Subtle differences in mass and radius of the non-ideal MHD models near second periastron causes their future evolution to diverge.

Shortly after first periastron, all nine weak field models produce an additional two protostars on orbits external to the primary binary, and their early evolution is independent of the primary binaries.  For $A_0=0.05$ and $0.1$, these external binaries do not interact with the primary prior to the end time of $t=2.64$\tff, but they interact near first apoastron in the non-ideal MHD models with $A_0=0.2$.

The \mufive\ models with $A_0=0.2$ and $0.1$ retain a binary until the end of the simulation at $t = 1.55$\tff, with an elliptical binary persisting in the former and a single, stable disc persisting in the latter.  The middle column of Fig.~\ref{fig:results_NI:inX:late} shows the $A_0=0.2$ models at $t=1.51$\tff.  At this time, the non-ideal MHD models have disc masses and radii that are 10 per cent larger and 4 per cent smaller, respectively, than their ideal counterpart.  From top to bottom in that column, each model has an increasing periastron and apoastron distance, and by $t = 1.51$\tff, the models have passed through periastron $6$, $3$ and $4$ times, respectively.  Non-ideal MHD effects contribute to these slight differences, but not enough to significantly change the overall evolution.  

The strong field models with $A_0=0.1$ form a single disc; see the right-hand column of Fig.~\ref{fig:results_NI:inX:late} for gas column densities at $t = 1.51$\tff.  The non-ideal MHD discs are $\sim$$3$ per cent larger but $\sim$$40$ per cent less massive as a result of the large central cavity.  For all three models, the mass and radius decrease with time.  The general trends amongst the three models are similar between the early and late epochs.

\subsection{Non-ideal MHD: magnetic field evolution}
\label{sec:results_NIZX}

\begin{figure}
\begin{center}
\includegraphics[width=\columnwidth]{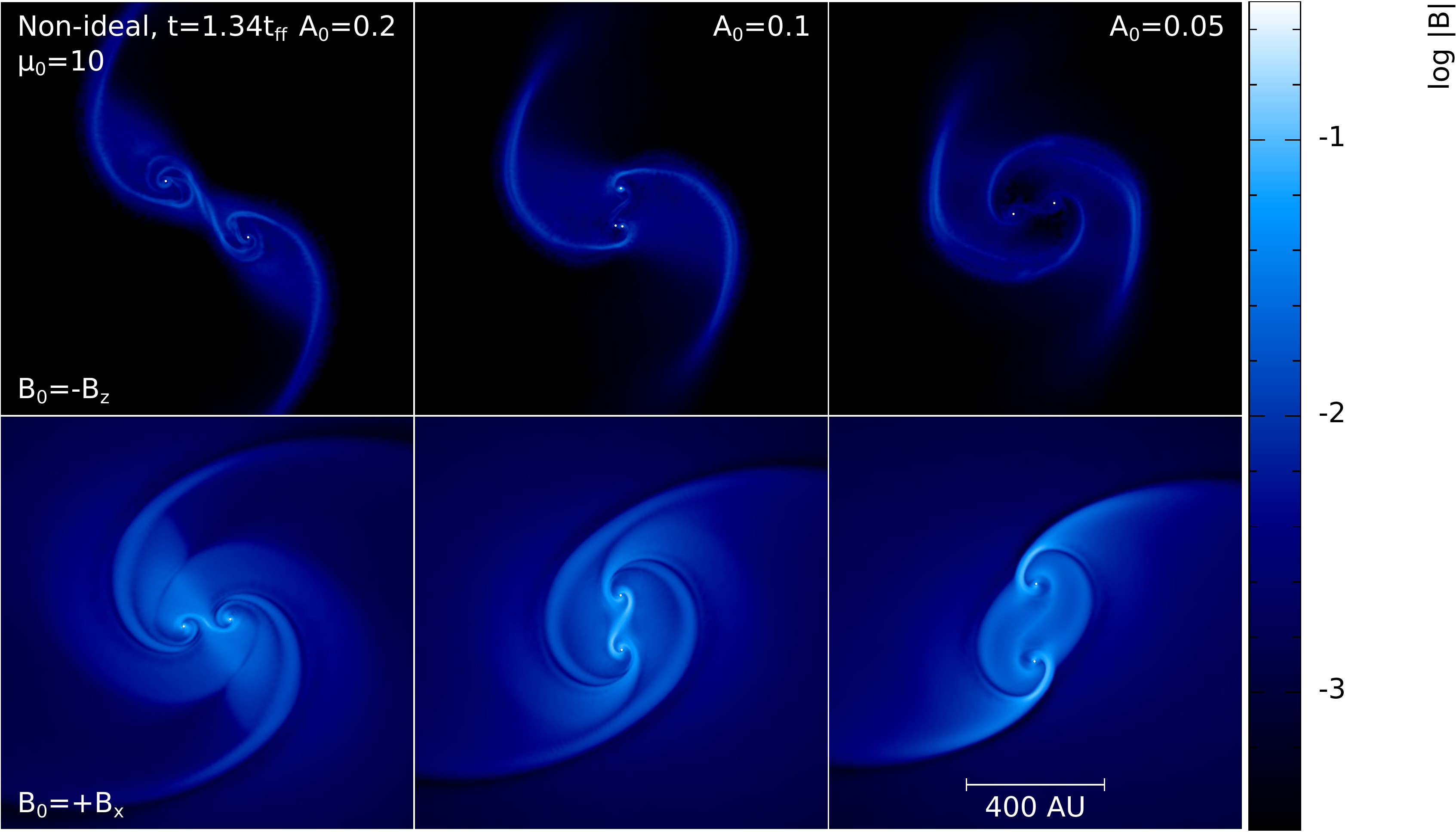}
\includegraphics[width=\columnwidth]{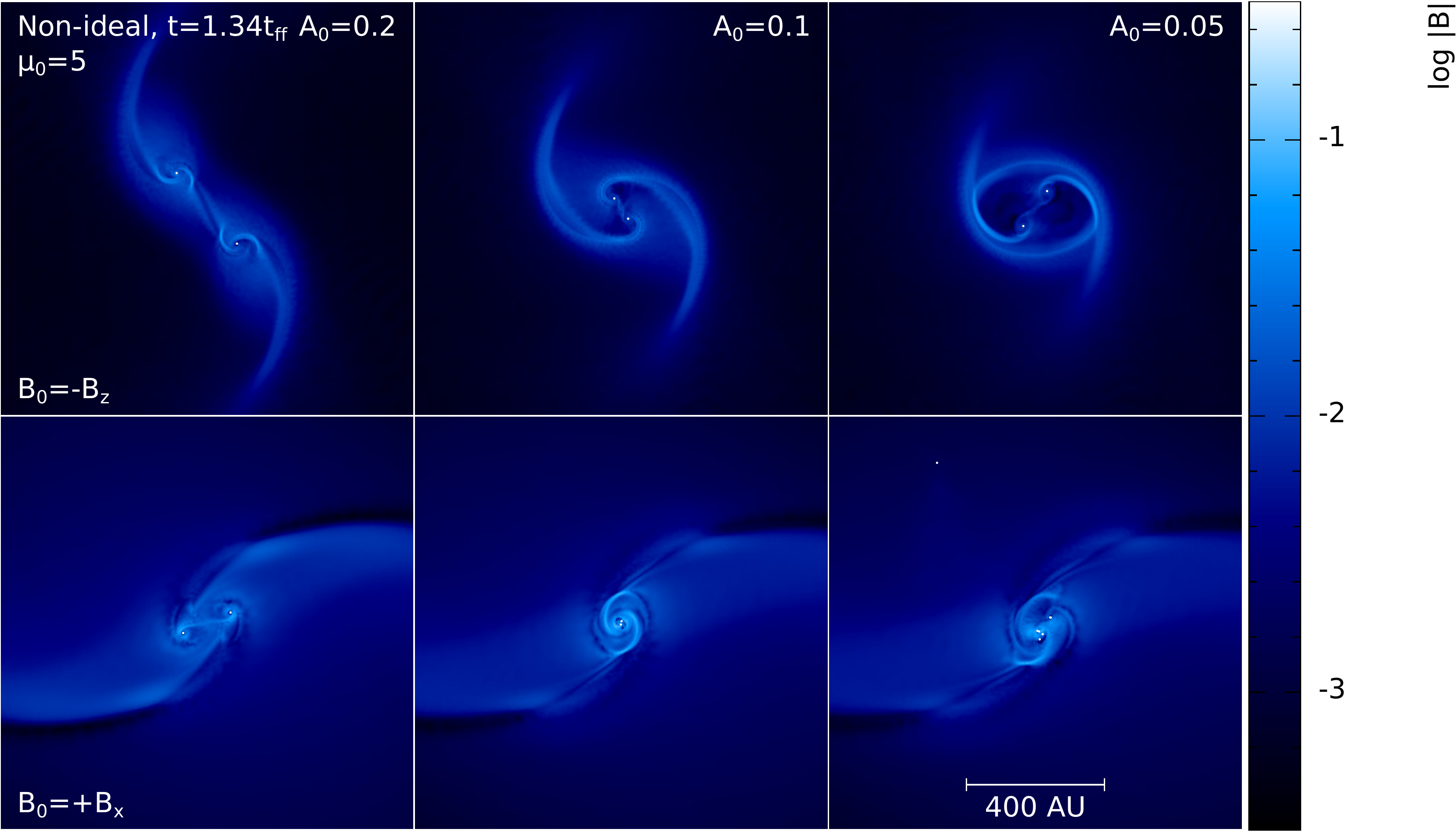}
\caption{As in Fig.~\ref{fig:results:idealA01:B134:xy}, the magnetic field strength in the mid-plane ($z=0$) at \otftff, but for the non-ideal MHD models with \muten \ (top subfigure) and \mufive \ (bottom subfigure).   As with the ideal MHD models, for each initial magnetic field strength, the mid-plane magnetic field is always stronger in the models initialised with \Bixp, despite the same initial value.  The magnetic field is weaker in the disc mid-plane when compared to their ideal MHD counterpart.}
\label{fig:results:IN:B134}
\end{center}
\end{figure}

Fig.~\ref{fig:results:IN:B134} compares the magnetic field strength in the mid-plane ($z=0$) for the non-ideal MHD models at \otftff; this figure is directly comparable to Fig.~\ref{fig:results:idealA01:B134:xy}.  Only the $\bm{B}_0 = -\bm{B}_\text{z}$ and $+\bm{B}_\text{x}$ models are shown, but the results are similar when the initial magnetic field direction is reversed.

Similar to the ideal MHD models, at \otftff, the net magnetic field in the discs are higher in the \Bixp \ models than their \Bzm \ counterparts with the same $\mu_0$ and same $A_0$.  When comparing a non-ideal MHD model to its ideal MHD counterpart, the magnetic field in the mid-plane, and specifically the disc, is weaker (see also Figs.~\ref{fig:results:evolZ} and \ref{fig:results:evolX}).

As in \citet{JoosEtAl2013}, we calculate the evolution of the mass-to-flux ratio inside a sphere of fixed radius, $R$, using  
\begin{equation}
\label{eq:masstoflux_calc}
\mu\left(R,t\right) = \frac{M(R)}{\pi R^2 \left<\bm{B}(R)\right>}\left(\frac{M}{\Phi_\text{B}}\right)_\text{crit}^{-1},
\end{equation}
where $M(R)$ is the enclosed mass including any protostars (i.e. sink particles), $\left<\bm{B}(R)\right>$ is the volume-averaged magnetic field within radius $R$, and $\left(M/\Phi_\text{B}\right)_\text{crit}$ is the critical mass-to-flux ratio that is independent of $M$, $R$ and $B$.  We have previous defined $\mu_0 \equiv \mu\left(R=0.013\text{pc},t=0\right)$.

\begin{figure*}
\begin{center}
\includegraphics[width=\columnwidth]{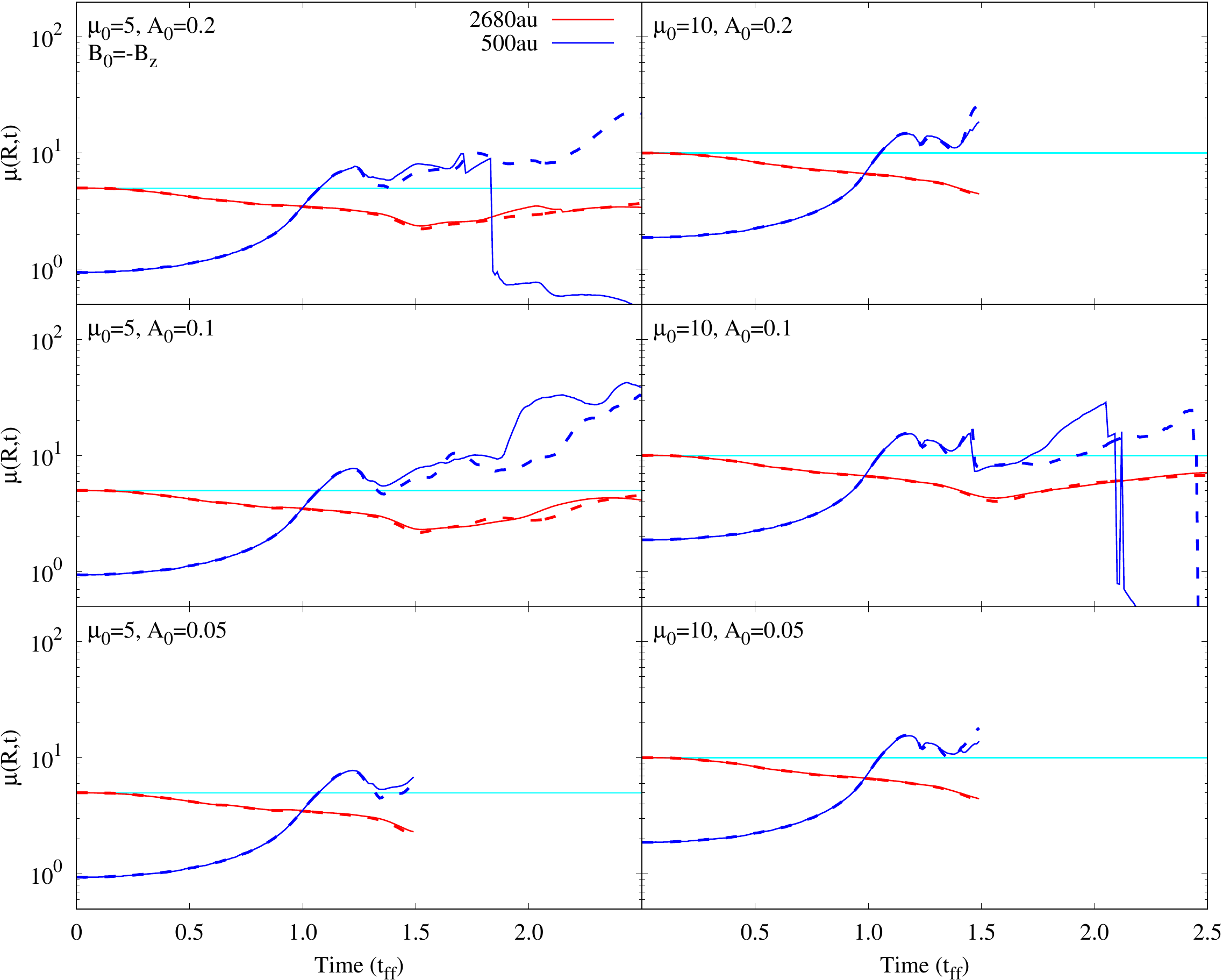}
\includegraphics[width=\columnwidth]{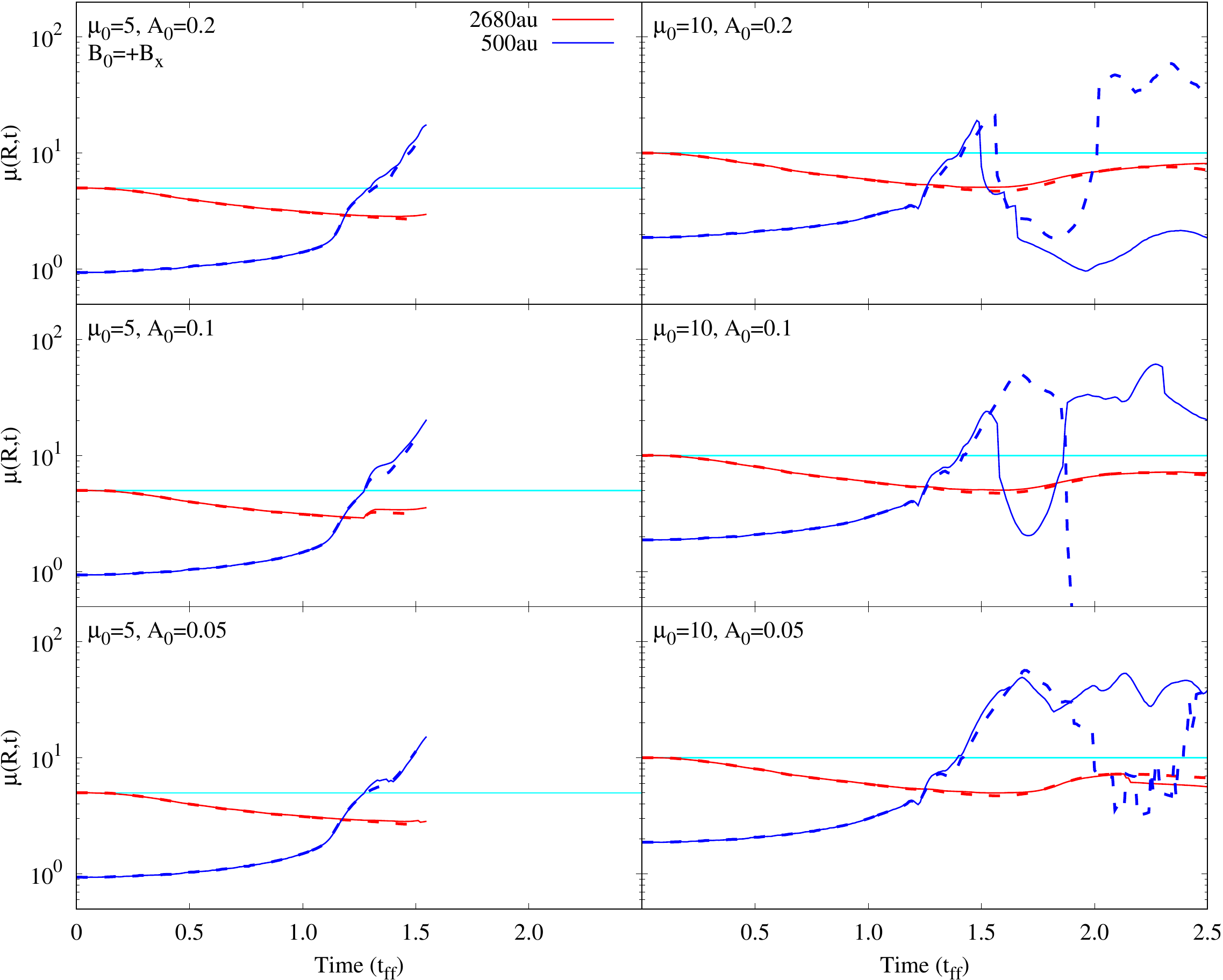}
\caption{Evolution of the mass-to-flux ratio, $\mu(R,t)$, for $R=2680$ au $= 0.013$ pc (i.e. the initial size of the gas cloud) and $500$ au for the ideal (dashed lines) and non-ideal (solid lines) MHD models.  The lines end when one of the models in the panel has reached its end time.  The cyan line represent the initial mass-to-flux ratio, $\mu_0$.  The ratio is $\mu\left(R,t\right) \propto M(R)/\left<\bm{B}(R)\right>$, where $M(R)$ is the mass enclosed within a sphere of radius $R$ centred on the origin, and $\left<\bm{B}(R)\right>$ is the volume-averaged magnetic field within the sphere.   In each panel, $\mu(R=2680\text{au},t)$ is similar for both models, whereas $\mu(R=500\text {au},t)$ may differ after $t \gtrsim 1.5$\tff \ due to difference in disc masses and magnetic fields (small deviations) or due to part or all of the discs leaving the sphere (big deviations and sudden drops).}
\label{fig:results:mu:global}
\end{center}
\end{figure*}

\begin{figure*}
\begin{center}
\includegraphics[width=\columnwidth]{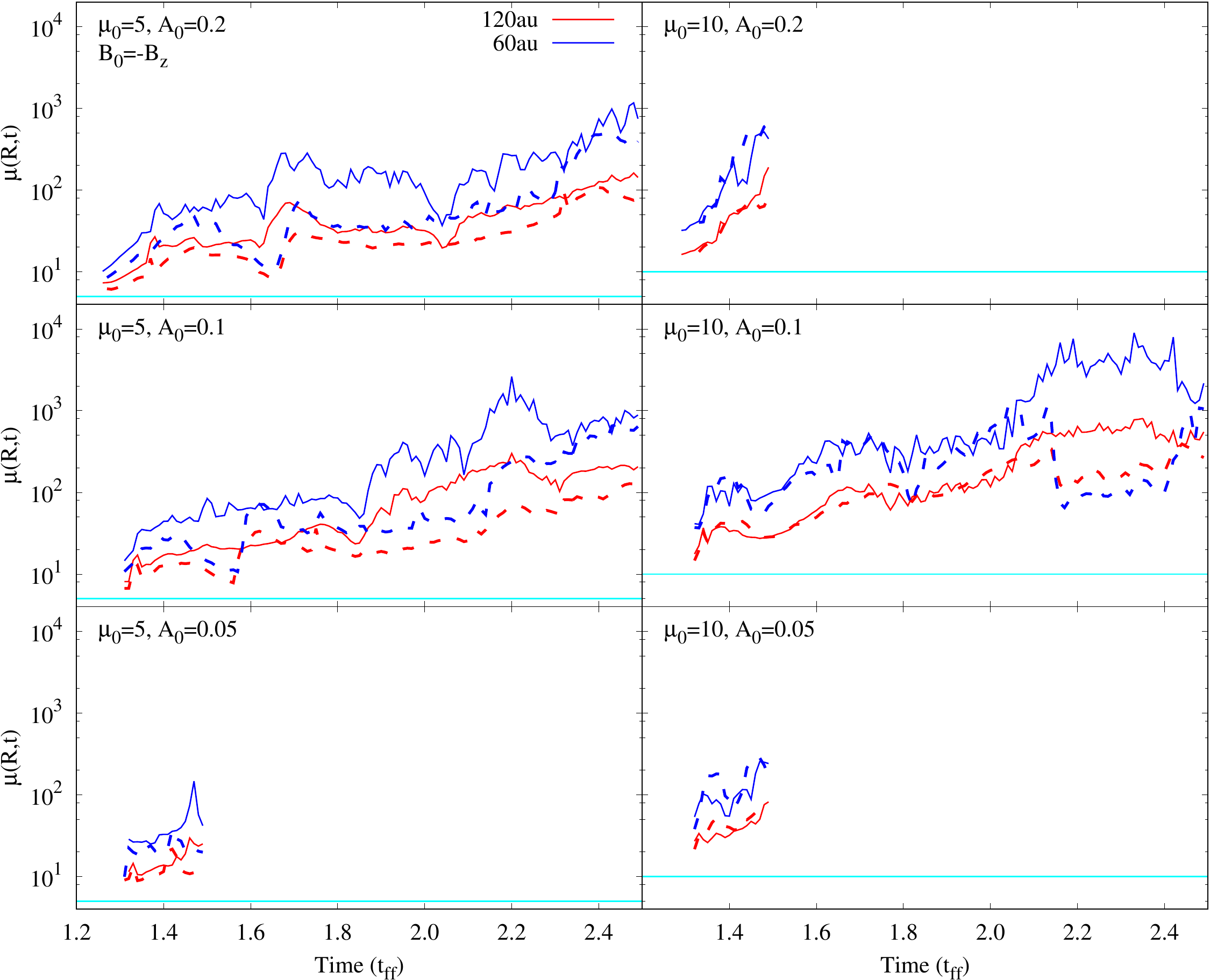}
\includegraphics[width=\columnwidth]{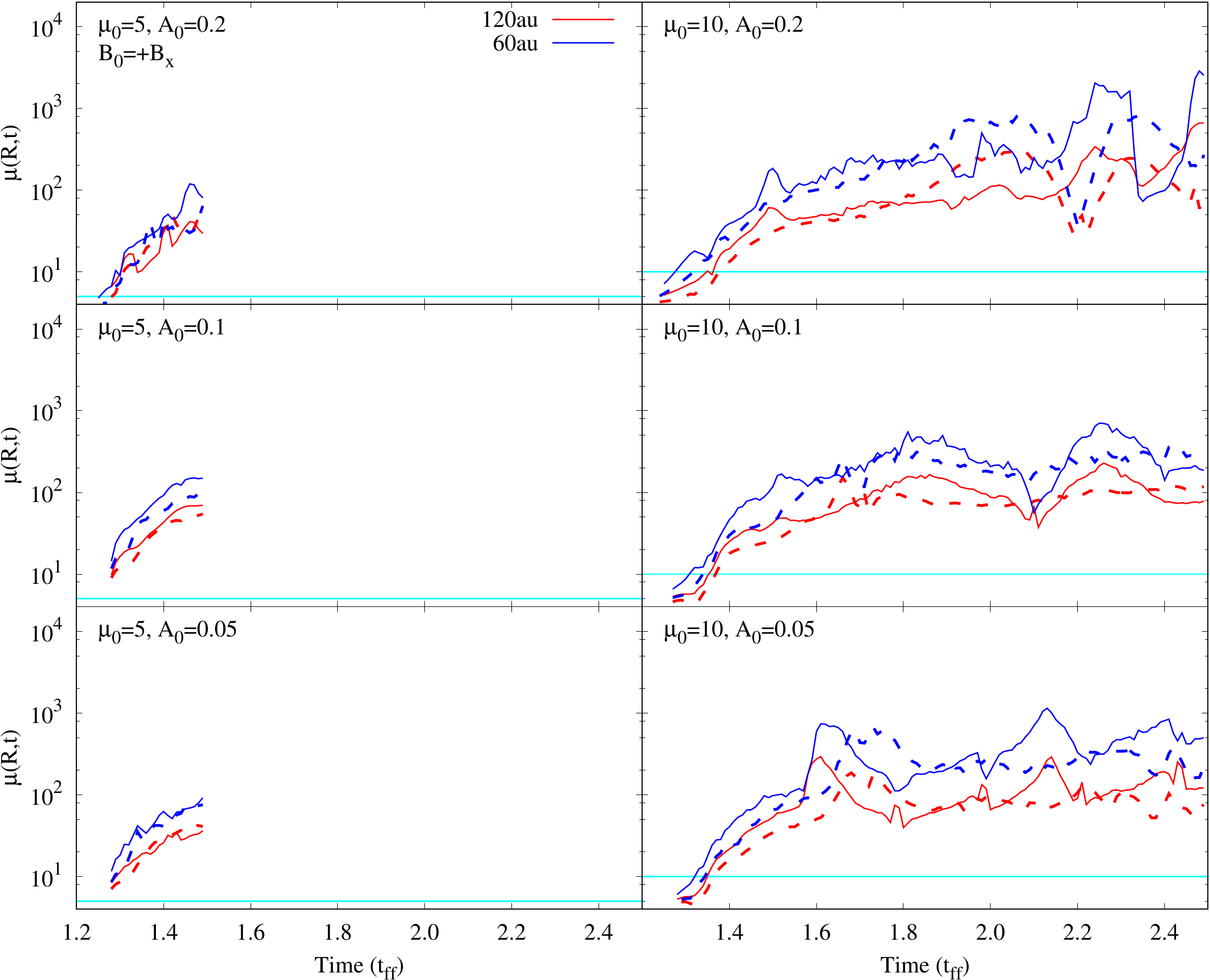}
\caption{Evolution of the mass-to-flux ratio, $\mu(R,t)$ for the ideal (dashed lines) and non-ideal (solid lines) MHD models as in Fig.~\ref{fig:results:mu:global}, but for $R=120$ and $60$ au centred on the first protostar that forms.  The differences in $\mu(R,t)$ between each ideal/non-ideal pair are similar to the differences caused by changing other parameters, suggesting that non-ideal MHD plays a secondary role in binary formation.}
\label{fig:results:mu:disc}
\end{center}
\end{figure*}

The mass-to-flux ratio is spatially dependent, thus we plot four values for selected models:  Fig.~\ref{fig:results:mu:global} shows $\mu(R,t)$ for radii of $R=2680$ au $= 0.013$ pc (i.e. the initial size of the gas cloud) and $500$ au centred on the origin, and Fig.~\ref{fig:results:mu:disc} shows $\mu(R,t)$ for radii of $R=120$ and $60$ au centred on the first protostar that forms.

For each non-ideal model and its ideal MHD counterpart, $\mu(R=2680\text {au},t)$ is similar for all time, while $\mu(R=500\text {au},t)$ begins to diverge at $t \gtrsim 1.5$\tff.  The small differences in $\mu(R=500\text {au},t)$ are a result of the weaker magnetic field and more massive discs in the non-ideal models; the larger differences, including the sharp drops to $\mu(R=500\text {au},t) \ll 1$ are a result of the binary separation surpassing $2R = 1000$ au; thus, these plots also provide insight into the orbital properties of the binaries.

These results are consistent with those found by \citet{TassisMouschovias2007c} and \citet{JoosEtAl2013} who studied the formation of isolated protostars: the mass-to-flux ratio increases around the protostar during its formation, removing memory of its initial value.  In \citet{TassisMouschovias2007c}, their mass-to-flux ratio increases rapidly as the evolution proceeds and density increases.  The increase of the mass-to-flux ratio centred on the first protostar in our models is more gradual after the formation of the protostar since the sink particle removes the central magnetic field upon formation and particle accretion and effectively limits the maximum gas density at $\rho \sim 10^{-10}$ g cm$^{-3}$. 

As shown in Fig.~\ref{fig:results:mu:disc}, the evolution of $\mu(R,t)$ around the protostar is quantitatively different for each model, reflecting the different disc masses and magnetic fields contained within them.  By changing any one parameter, the mass-to-flux ratio either increases or decreases, indicating that no one parameter is dominant in determining its evolution immediately around the protostar.   The change caused by including non-ideal MHD is typically smaller than the change caused by altering another parameter, further suggesting that non-ideal MHD plays a secondary role in binary formation.

\subsection{Influence of the Hall effect}
\label{sec:results_NI:Hall}

In previous studies of collapse to form an isolated first hydrostatic core, the inclusion of non-ideal MHD was found to permit disc formation depending on the direction of the magnetic field with respect to the axis of rotation (\citealt{TsukamotoEtAl2015b}; \citetalias{WPB2016}).  Moreover, the Hall effect was found to influence the formation process even when it was not the dominant non-ideal MHD effect.  

In this study, the faster initial rotation and the initial density perturbations result in discs forming in every super-critical model. Fig.~\ref{fig:results_NI:evol:eta} suggests that the non-ideal MHD effects may influence mainly the inner regions of the discs.  Thus, any changes that switch the direction of the magnetic field in the disc will be amplified by the Hall effect, which may lead to simulations evolving differently.  However, these discs are primarily supported by gas pressure (i.e. $\beta \gg 1$), so the effect of the global change will depend on the amount of modification by the Hall effect and when it occurs.  For example, the initial direction of the magnetic field plays a minimal role in the evolution of the models with \mufive, $A_0 \geq 0.1$ and $\bm{B}_0=\pm\bm{B}_\text{z}$, while it triggers a divergence in the evolutionary paths of \mufive, $A_0=0.1$ and $\bm{B}_0=\pm\bm{B}_\text{x}$.

In the earlier studies, the Hall effect was found to spin up the disc when the magnetic field was initially anti-aligned with the axis of rotation.  To conserve angular momentum, a counter-rotating envelope forms (e.g. \citealp{TsukamotoEtAl2015b}; \citetalias{WPB2016}).  By contrast, the Hall effect does not have a pronounced effect on the rotation of the discs in our binary models, and as a result we see no evidence for counter-rotating envelopes. 

\subsection{Extending the parameter space}
\label{sec:results:altP}

\subsubsection{Sub-critical mass-to-flux ratios}
\label{sec:results:altP:sub}
In the above models, the initial mass-to-flux ratio of $\mu_0 > 1$ means that the magnetic field is not strong enough to prevent gravitational collapse, thus protostar formation is a foregone conclusion.  

\begin{figure}
\begin{center}
\includegraphics[width=0.8\columnwidth]{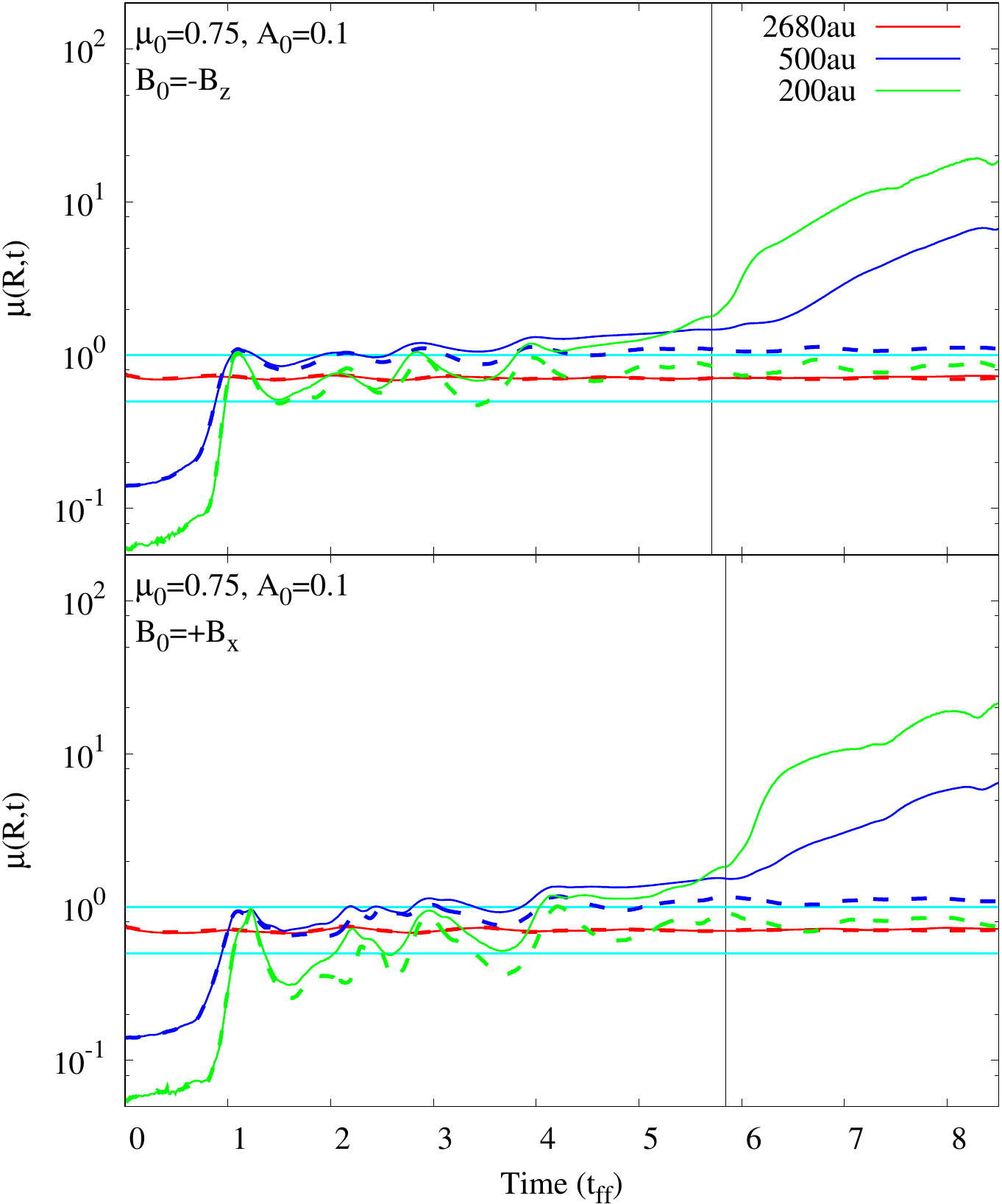}
\caption{Evolution of the mass-to-flux ratio, $\mu(R,t)$, for the ideal (dashed lines) and non-ideal (solid lines) MHD models as in Fig.~\ref{fig:results:mu:global}, but for the sub-critical models with $\mu_0= 0.75$.  The cyan lines are at $\mu=\mu_0$ and $\mu=1$ (i.e. the critical value).  The vertical lines represent when the protostars formed in the non-ideal MHD models.  The ideal MHD models do not collapse during their runtime of $t \approx 17t_\text{ff}$ while the non-ideal MHD models form protostars at $t=5.71$ and $5.84t_\text{ff}$, for $\bm{B}_0=-\bm{B}_\text{z}$ and $\bm{B}_\text{x}$, respectively.}
\label{fig:results:mu:sub}
\end{center}
\end{figure}

A gas cloud with a sub-critical mass-to-flux ratio is magnetically supported and should not collapse when using ideal MHD.  Indeed, our sub-critical models with $\mu_0 = 0.75$ do not collapse during their runtime to $t \approx 17t_\text{ff}$, and their maximum density never surpasses $\rho \approx 6\times 10^{-16}$ g cm$^{-3}$ (recall that the initial density is $\rho_0 = 7.43\times 10^{-18}$ g cm$^{-3}$ and that sink particles are inserted at $\rho \approx 10^{-10}$ g cm$^{-3}$).  As shown by the dashed lines in Fig.~\ref{fig:results:mu:sub}, after $t \gtrsim 4$\tff, the mass-to-flux ratio at $R=2680$, $500$ and $200$ au are approximately constant, with $\mu(R=200\text {au},t) < 1$.  The value of $\left<\bm{B}(R)\right>$ is similar for spheres of both $R=500$ and $200$ au, but the former has more enclosed mass, hence the higher mass-to-flux ratio.

\begin{figure}
\begin{center}
\includegraphics[width=\columnwidth]{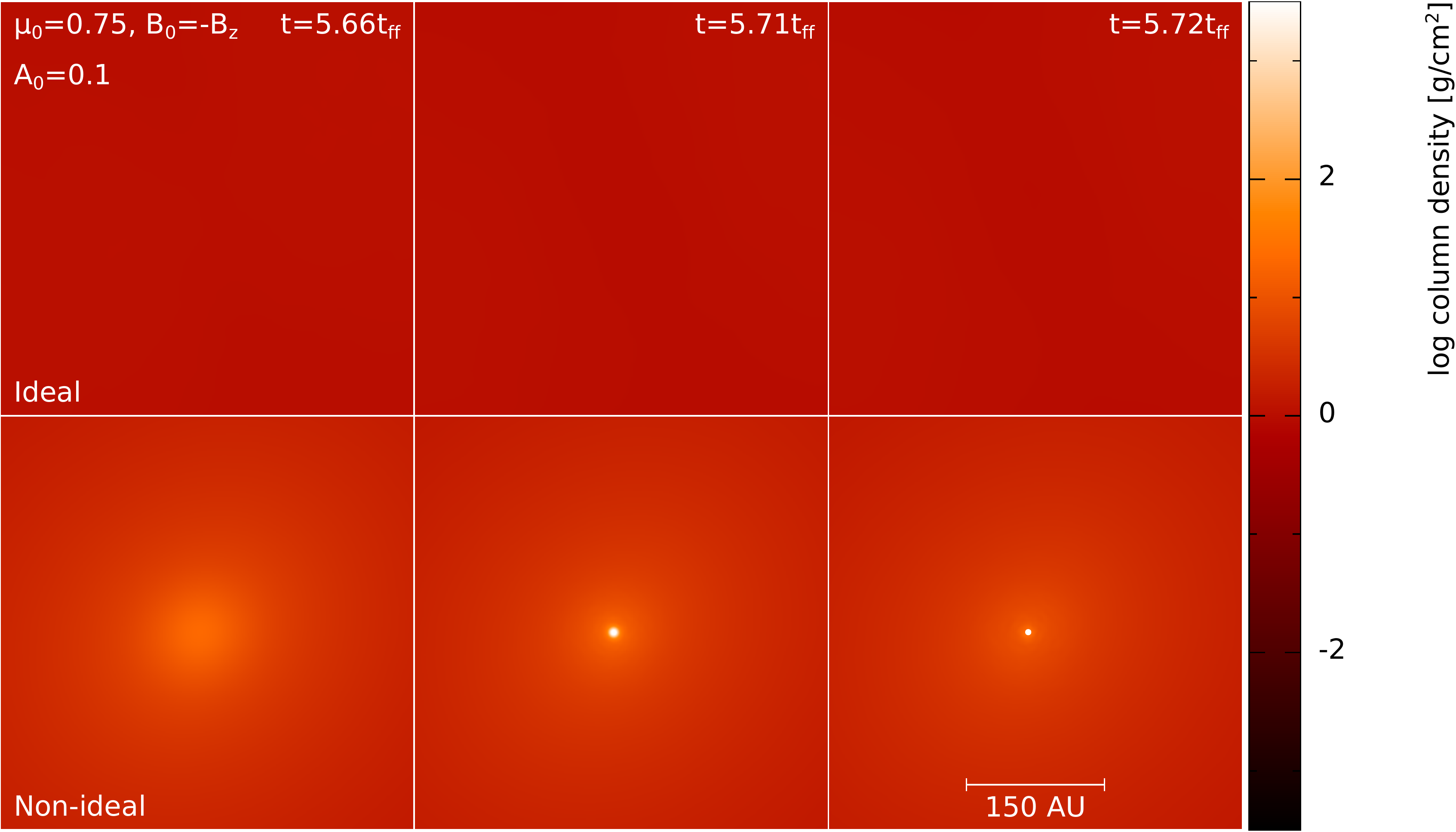}
\includegraphics[width=\columnwidth]{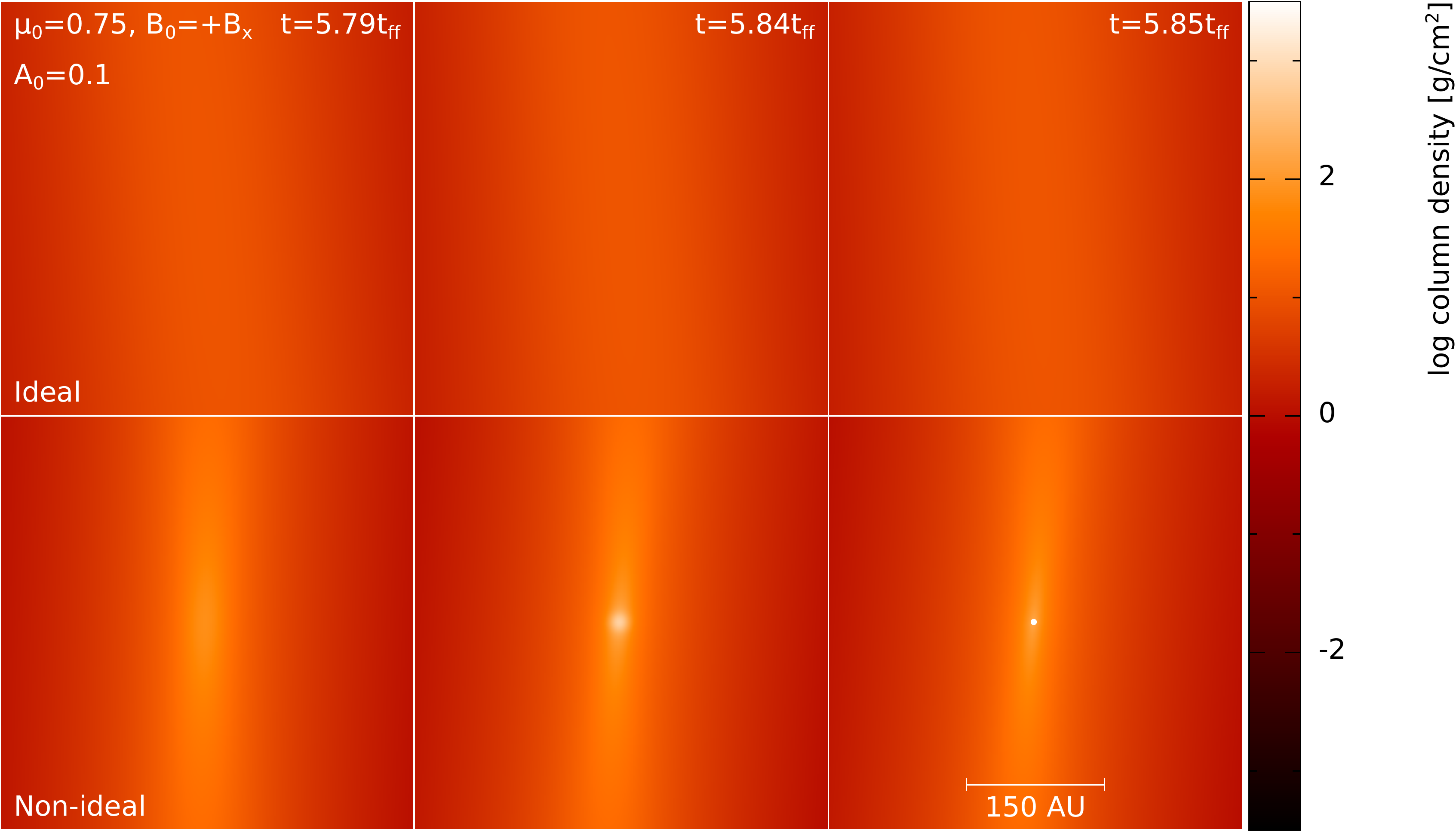}
\caption{Face-on gas column density of the initially sub-critical models ($\mu_0=0.75$) using ideal (top row in each subfigure) and non-ideal (bottom row in each subfigure) MHD.  For both initial magnetic field orientations, the ideal MHD models do not collapse, with their density staying below $\rho \approx 6\times 10^{-16}$ g cm$^{-3}$.  The non-ideal MHD models collapse to form single protostars shortly after $t=5.71$ and $5.84$\tff \ for $\bm{B}_0=-\bm{B}_\text{z}$ (top subfigure) and $\bm{B}_\text{x}$ (bottom subfigure), respectively.}
\label{fig:results:sub:colden}
\end{center}
\end{figure}

For the non-ideal MHD models with $\bm{B}_0=-\bm{B}_\text{z}$ and $\bm{B}_\text{x}$, $\mu(R=200\text {au},t) > 1$ at $t \approx 3.8$ and $4.0$\tff, respectively, as shown by the solid green lines in Fig.~\ref{fig:results:mu:sub}.  After this time, the central regions are no longer magnetically supported, and the clouds collapse to form protostars at $t=5.71$ and $5.84$\tff, respectively.  Fig.~\ref{fig:results:sub:colden} shows the gas column density for the sub-critical models near the time of protostar formation for the non-ideal MHD models.  Only one protostar is formed in each model, and no discs form around them.

Thus, in our models, the non-ideal MHD effects can diffuse enough magnetic field to allow the central regions of initially sub-critical clouds to collapse to form protostars.


\subsubsection{Slower initial rotations}
\label{sec:results:altP:slow}

\begin{figure}
\begin{center}
\includegraphics[width=\columnwidth]{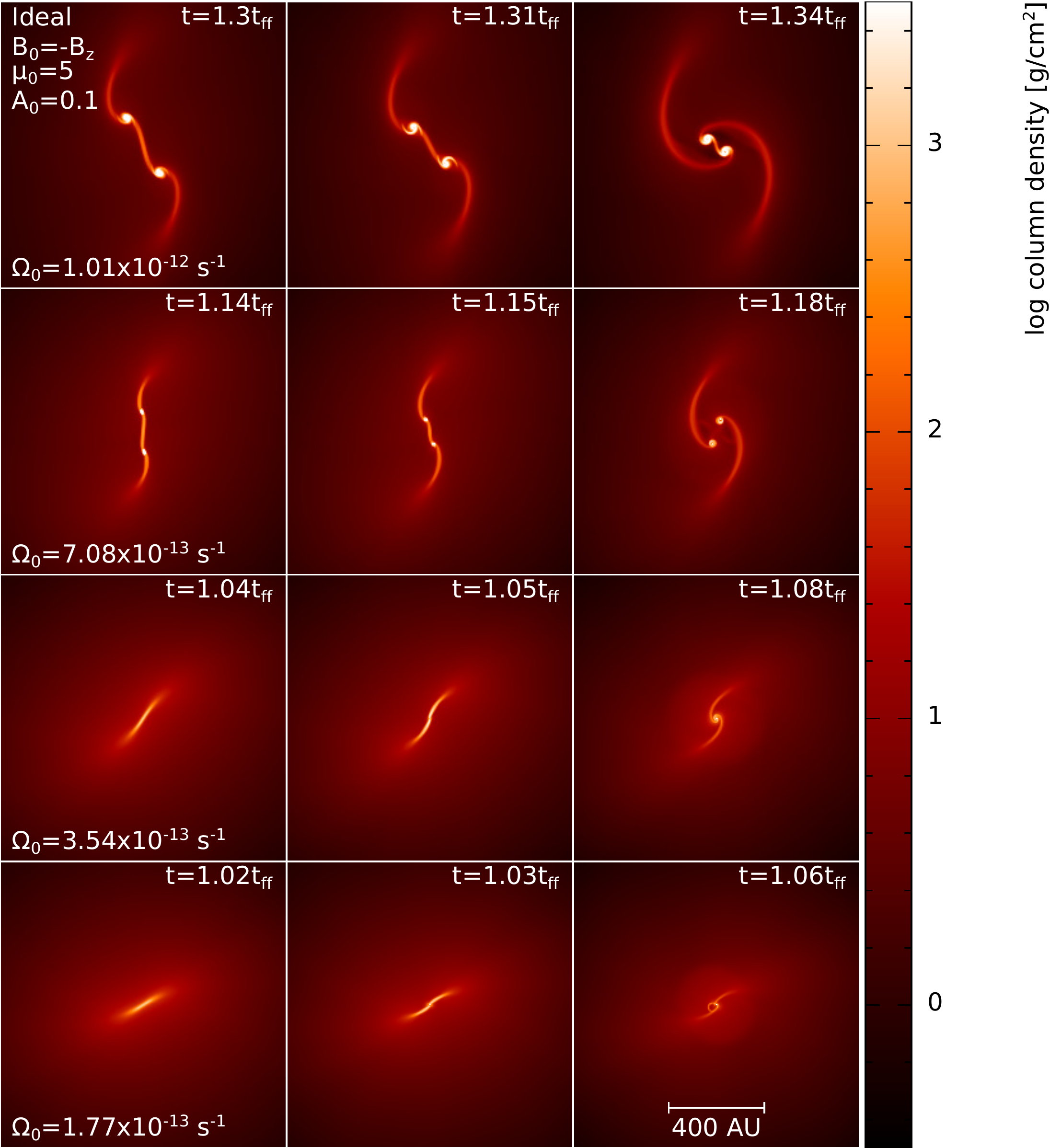}
\includegraphics[width=\columnwidth]{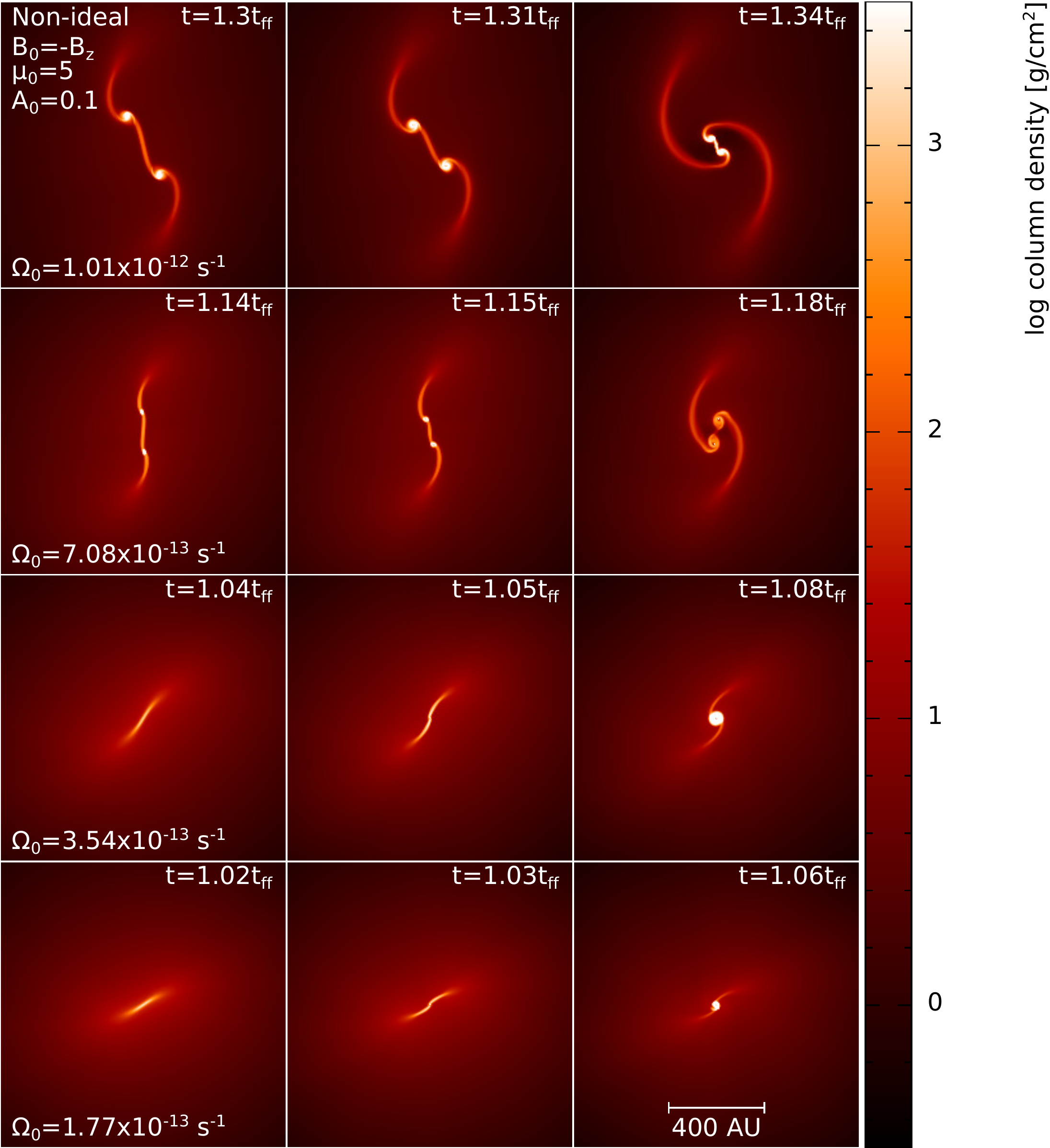}
\caption{Face-on gas column density of models using the fiducial initial rotation of $\Omega_0 = 1.006\times 10^{-12}$ rad s$^{-1}$ and the slower rotations of $\Omega_0 = 7.08\times 10^{-13}$,  $3.54\times 10^{-13}$ and $1.77\times 10^{-13}$ rad s$^{-1}$.  All models use \mufive, $A_0=0.1$, $\bm{B}_0 = -\bm{B}_\text{z}$, and ideal (top subfigure) or non-ideal (bottom subfigure) MHD.  The panels are chosen such that the protostars form between the first two columns, and the third column is d$t \approx 0.03$\tff \ after the protostar's formation.  Decreasing the initial rotation speed decreases the initial binary separation and, if slow enough, prevents the formation of binaries.  Non-ideal MHD has a greater influence on the environment of the initially slower rotating models, forming larger and more massive discs, but has little effect on the large-scale morphology or the number of protostars that form.}
\label{fig:results:omega}
\end{center}
\end{figure}

The models in our primary suite all use an initial rotation of $\Omega_0 = 1.006\times 10^{-12}$ rad s$^{-1}$, and binaries form in all but four of the 36 models.  However, (e.g.) \citet{HennebelleTeyssier2008} and \citet{MachidaEtAl2008} found that initial rotation played an important role in determining the evolution of the system.  Although a full parameter study of the initial rotation is out of the scope of this study, we briefly discuss the early evolution of models with the slower initial rotations of $\Omega_0 = 7.08\times 10^{-13}$,  $3.54\times 10^{-13}$ and $1.77\times 10^{-13}$ rad s$^{-1}$, using \mufive, $A_0=0.1$ and $\bm{B}_0=-\bm{B}_\text{z}$.  Fig.~\ref{fig:results:omega} shows the gas column density of these models at early times.

As the initial rotation speed decreases, the initial separation of the binaries decreases, with only a single protostar forming at the slowest two rotation speeds.  The non-ideal MHD effects become more important as the initial rotation decreases, with larger and more massive discs forming at slower rotation speeds.  This is consistent with previous studies finding larger discs in non-ideal MHD models of isolated protostars than ideal MHD models.  However, the non-ideal MHD effects do not change the global morphology and whether binary or single systems form, which is consistent with our previous results suggesting that non-ideal MHD has a secondary effect on binary formation and evolution.

\subsubsection{Multiple grain populations}
\label{sec:results:altP:grains}
As this study was in progress, Version 1.2.1 of {\sc Nicil} \citep{Wurster2016} was released.  This version differs from v1.1 used here by modelling three grain populations, $n_\text{g}^-$, $n_\text{g}^0$ and $n_\text{g}^+$, with charges $Z=-1,0,+1$, respectively, rather than a single grain population, $n_\text{g}$, with charge $\bar{Z} < 0$.  In v1.2.1, grain number density is conserved, with $n_\text{g} = n_\text{g}^- + n_\text{g}^0 + n_\text{g}^+$, where $n_\text{g}$ is calculated as in v1.1.

\begin{figure}
\begin{center}
\includegraphics[width=\columnwidth]{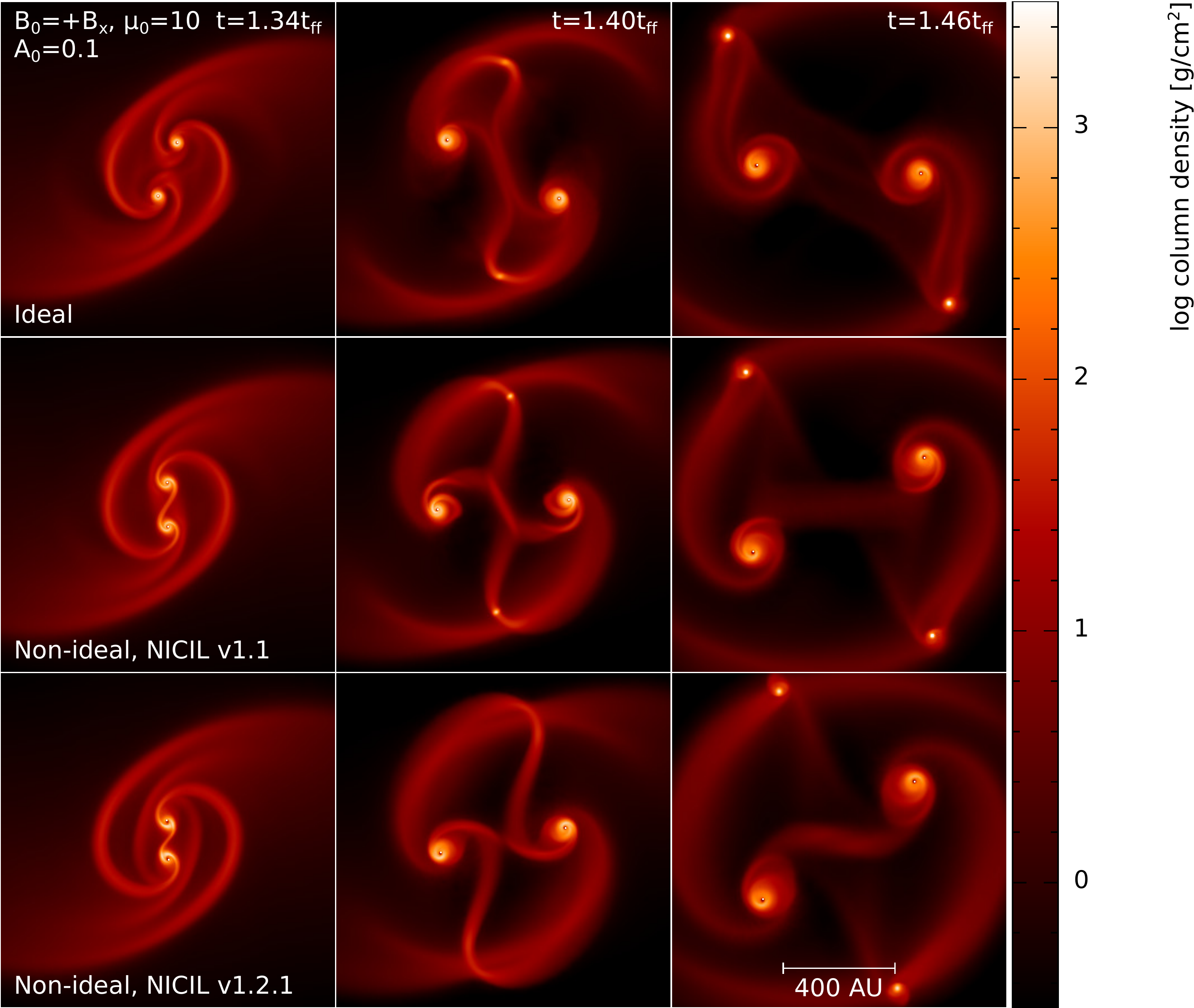}
\includegraphics[width=\columnwidth]{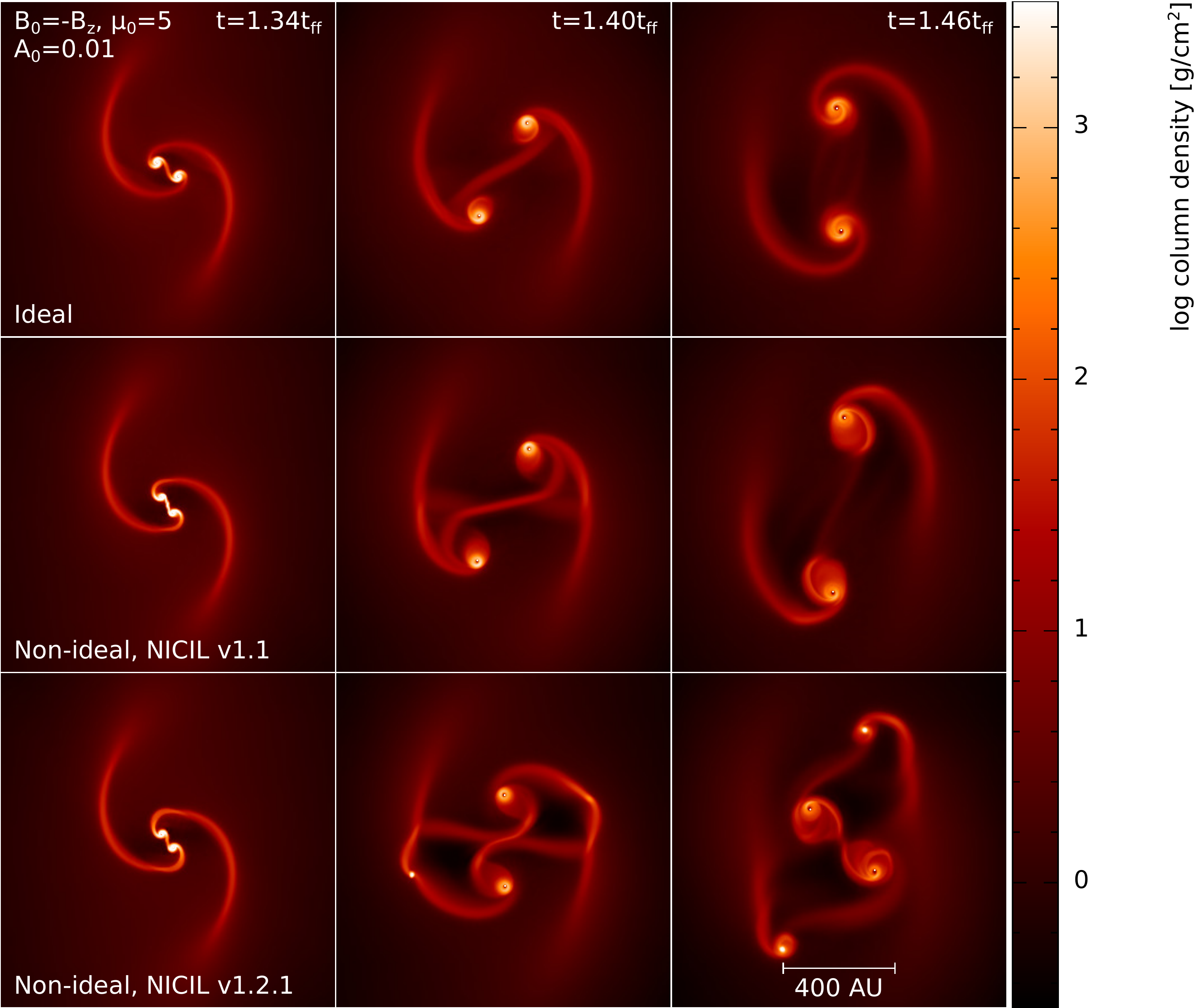}
\caption{Face-on gas column density from selected models using ideal MHD  (top row in each subfigure), {\sc NICIL} Version 1.1 (middle row in each subfigure) and {\sc NICIL} Version 1.2.1 (bottom row in each subfigure).  Version 1.1 uses a single grain population, $n_\text{g}$, with charge $\bar{Z} < 0$ and Version 1.2.1 models three grain populations, $n_\text{g}^-$, $n_\text{g}^0$ and $n_\text{g}^+$, with charges $Z=-1,0,+1$, respectively.   The models with three grain populations yield binaries with smaller first periastron separations, and, for the \mufive \ models, first apoastron and first periods that are $\sim2.5$ times smaller.}
\label{fig:results:nicil:colden}
\end{center}
\end{figure}

\begin{figure}[!h]
\begin{center}
\includegraphics[width=0.8\columnwidth]{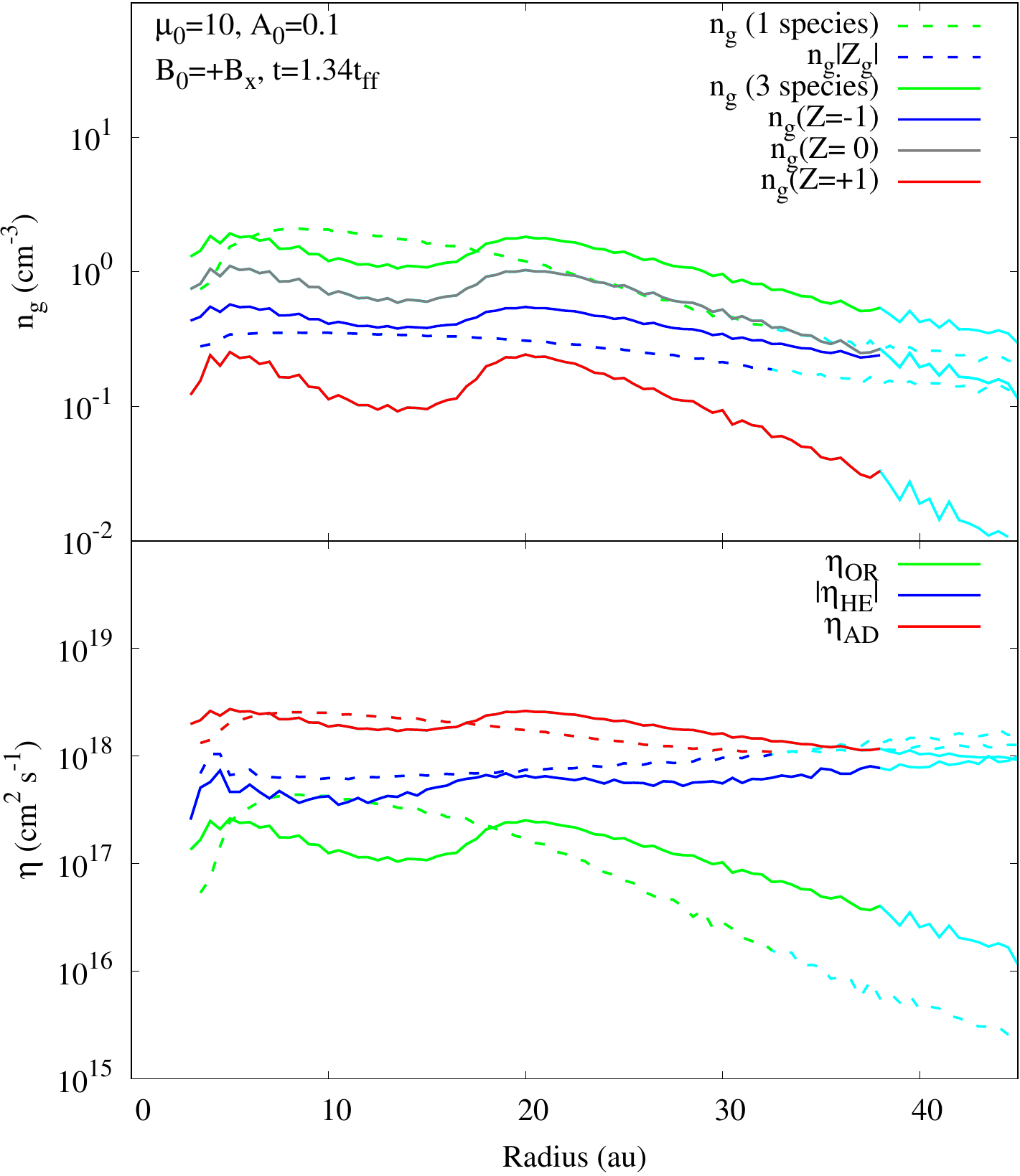}
\includegraphics[width=0.8\columnwidth]{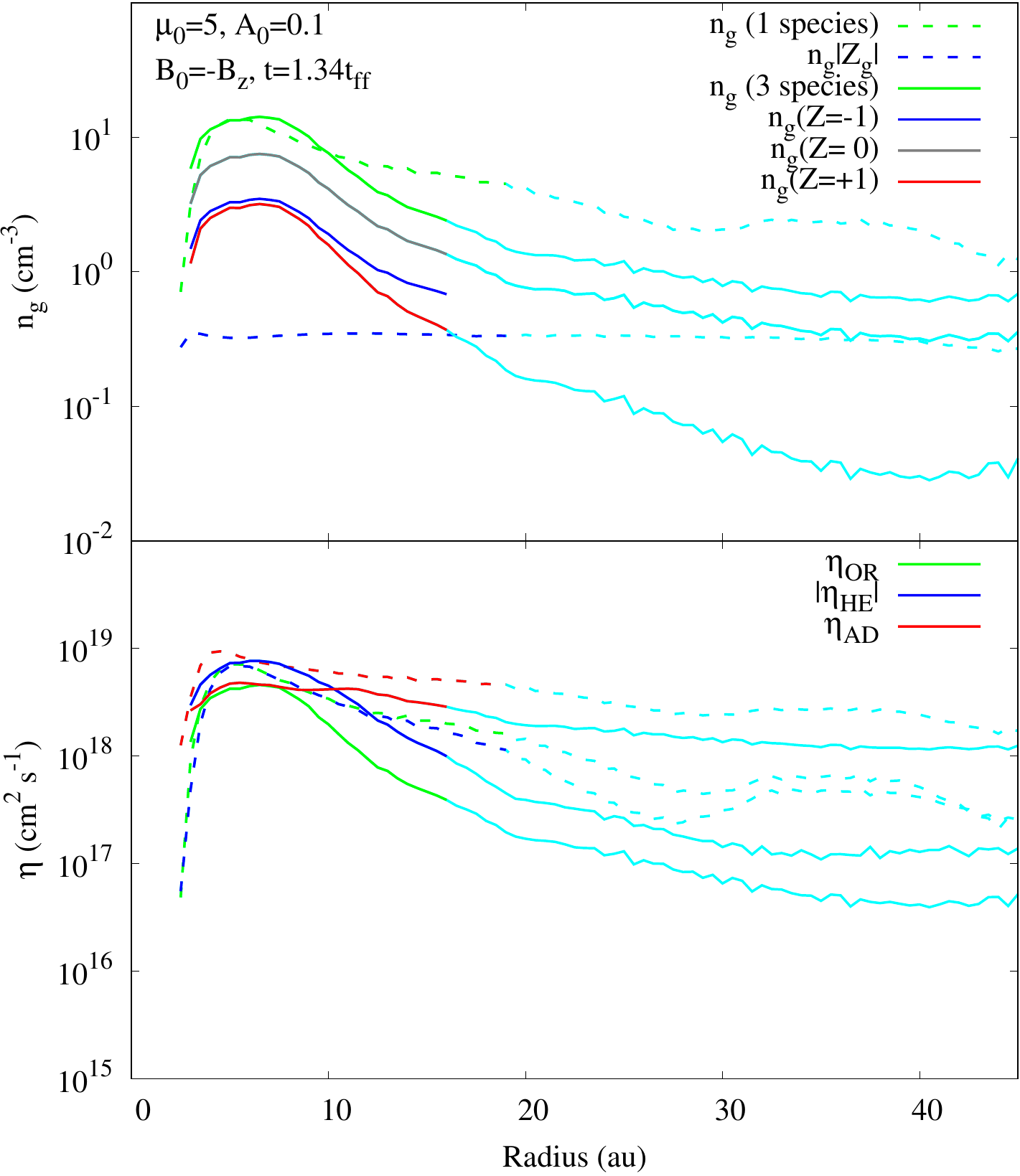}
\caption{\emph{Top panel in each subfigure}:  The grain number density in the disc around one protostar at $t=1.34$\tff \ using v1.1 of {\sc NICIL} (dashed lines) and v1.2.1 (solid lines).  For v1.1, there is a single grain population with an average negative charge,  thus the dashed blue line is $n_\text{g}|Z_\text{g}|$, which effectively represents $n_\text{g}^-$.  For v1.2.1, all three grain populations are self-consistently calculated, and $n_\text{g} = n_\text{g}^- + n_\text{g}^0 + n_\text{g}^+$.  For reference, the neutral grain number density is $n_\text{n} \sim \mathcal{O}(10^{12})$ cm$^{-3}$.  While the total grain number density, $n_\text{g}$, is only weakly dependent on grain model, the effective, $n_\text{g}|Z_\text{g}|$, and real,  $n_\text{g}^-$, number densities of negatively charged grains can differ by an order of magnitude.  \emph{Bottom panel in each subfigure}:  As in Fig.~\ref{fig:results_NI:evol:eta}, average values of Ohmic, ambipolar and Hall diffusion coefficients for the gas.   The grain model only weakly affects the non-ideal MHD coefficients for weak magnetic fields, but decreases the values of ambipolar diffusion and Ohmic resistivity in the high density regions of the disc in the strong field models.}
\label{fig:results:nicil:radprof}
\end{center}
\end{figure}

To test the effect of the grain model, we run two additional models using v1.2.1:  \muten, $A_0=0.1$, $\bm{B}_0 = +\bm{B}_\text{x}$ and \mufive, $A_0=0.1$, $\bm{B}_0 = -\bm{B}_\text{z}$.  Fig.~\ref{fig:results:nicil:colden} shows the gas column density at selected times, and Fig.~\ref{fig:results:nicil:radprof} shows the radial profile of the grain populations and non-ideal MHD coefficients in the disc around one protostar at \otftff.

Modelling three grain populations yields binaries with smaller first periastron separations than the single grain model.  The first apoastron separation and first period are similar for the \muten \ models, but for the \mufive \ models, they are $\sim2.5$ times larger when using v1.1.

Once the discs form, the neutral number density is only weakly dependent on the grain model, and is $n_\text{n} \sim \mathcal{O}(10^{12})$ cm$^{-3}$ for the duration of the simulation.  Averaged over the entire disc, the total grain number density differs by $\lesssim2$ between the two grain models, with the largest differences occurring at larger radii and at later times during the evolution.  At \otftff \ and during the evolution of the disc, $n_\text{g}^0 > n_\text{g}^- > n_\text{g}^+$.  To approximate $n_\text{g}^-$ in the single grain model, we use $n_\text{g}^- \approx n_\text{g}\left|Z_\text{g}\right|$, which varies only slightly with both radius and time.  This value is $\sim$$2$ times smaller for our \muten \ model, and up to $\sim$$11$ times smaller for the \mufive \ model.

The different grain models affect the calculation of the non-ideal MHD coefficients, $\eta$, since we are improving the calculation of $n_\text{g}^-$ and adding a charged species,  $n_\text{g}^+$.   When comparing $\eta$ for the different grain models (bottom panel of each subfigure in Fig.~\ref{fig:results:nicil:radprof}) in the weak field model (\muten), ambipolar diffusion is the largest term for both grain models, followed by the Hall effect.  At this time in the \mufive \ model, all values are similar in the inner disc ($R \sim 7$ au, $n_\text{n} \sim 5\times 10^{12}$ cm$^{-3}$), however, the order of the terms differs depending on the grain model; ambipolar diffusion is reduced in strength to be similar to Ohmic resistivity at this radius.  This is consistent with \citet{TassisMouschovias2007c} and \citet{KunzMouschovias2010}, who find that Ohmic resistivity becomes more important than ambipolar diffusion at $n_\text{n} \sim \mathcal{O}(10^{13})$ cm$^{-3}$.  

When considering the average values of $\eta$ over the entire disc, we find that $\eta_\text{AD} > \left|\eta_\text{HE}\right| > \eta_\text{OR}$, and that $n_\text{n} \lesssim 10^{12}$ cm$^{-3}$.  Thus, we expect ambipolar diffusion to dominate in our models.

\subsection{Resolution}
\label{sec:results:res}

The calculations presented above used $3\times 10^5$ particles in the initial sphere.  This number satisfies the Jeans criteria (c.f. Section \ref{sec:ic}), while allowing us to perform a large suite of simulations, even with the small timesteps required to properly evolve the non-ideal MHD terms.  To test the effect of resolution, we ran selected models initialised with $10^6$ particles in the sphere.

\begin{figure}
\begin{center}
\includegraphics[height=0.6\columnwidth]{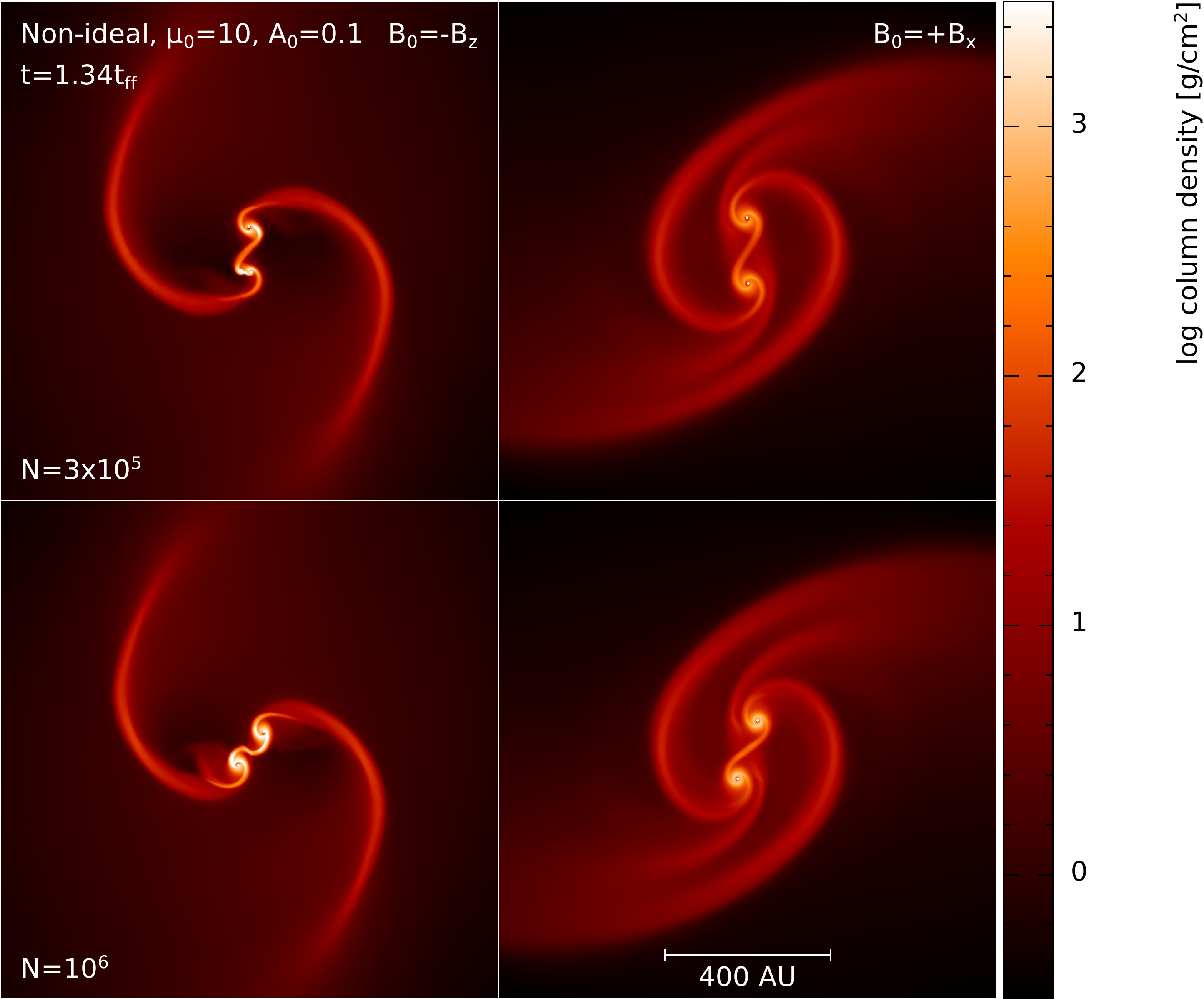}
\caption{Gas column density for four non-ideal MHD models with \muten \ and $A_0=0.1$ at \otftff.  The left- (right-) hand column shows the models with \Bizm \ (\Bixp), and the models in the top and bottom rows are initialised with $3\times 10^5$ and $10^6$ particles, respectively, in the sphere.  The higher resolution models produce somewhat larger and more massive discs that are more susceptible to fragmentation.}
\label{fig:results:resolution}
\end{center}
\end{figure}

Fig.~\ref{fig:results:resolution} shows the non-ideal MHD models at \otftff \ with \muten \ and $A_0=0.1$ using both $3\times 10^5$ and $10^6$ particles in the initial sphere.  
The $10^6$ particle models yield discs that are somewhat larger and more massive than their lower resolution counterparts.   This results in greater interaction at first periastron and shorter periods.  Although the quantitative values change with resolution, the trends previously discussed are independent of resolution.

The higher resolution discs are more massive and thus more susceptible to fragmentation.  We note this fragmentation would likely not be a problem if we included a proper treatment of radiation rather than the barotropic equation of state used here.

\section{Summary and conclusion}
\label{sec:discussion}

We have presented a suite of simulations studying the effect of non-ideal MHD on the formation and early evolution of binary stars.   Our models were initialised as a $1$M$_\odot$ rotating, uniform density sphere, which was given an $m=2$ density perturbation with amplitudes of $A_0 = 0.2, \ 0.1$ and $0.05$.  We threaded the sphere with an initially uniform magnetic field.  We tested our suite of simulations using both ideal MHD and non-ideal MHD at initial mass-to-flux ratios of 10, 5 and 0.75 times the critical value. The ideal MHD models were run using $\bm{B}_0=-\bm{B}_\text{z}$ and $+\bm{B}_\text{x}$, while the non-ideal MHD models were run using $\bm{B}_0=\pm\bm{B}_\text{z}$ and $\pm\bm{B}_\text{x}$ since the Hall effect depends on the direction of the magnetic field with respect to the axis of rotation.  In the models that formed binaries, we followed the gravitational collapse until at least first apoastron.  All of the simulations were performed using the SPMHD code {\sc Phantom}.  Our key results are as follows:
\begin{enumerate}

\item \emph{Sub-critical cores}:  Using ideal MHD, the sub-critical cores did not collapse during their runtime of $t\approx17$\tff, and their maximum density never surpassed $\rho\approx100\rho_0$.  When using non-ideal MHD, the cores collapsed to form single protostars at $t\lesssim5.84$\tff.

\item \emph{Ideal MHD}: $\bm{B}_0=-\bm{B}_\text{z}$:
Decreasing the amplitude of the initial density perturbation yields earlier times of first periastron, and smaller separations.  Decreasing the magnetic field strength (i.e. increasing $\mu_0$) increases first periastron separation and discs sizes.

\item \emph{Ideal MHD}: $\bm{B}_0=+\bm{B}_\text{x}$:  Strong magnetic fields suppress the formation of binaries, as found by previous authors, with a binary only forming for $A_0=0.2$; a tight binary with a common disc forms for $A_0=0.1$ and a single protostar and disc forms for $A_0=0.05$.  Binaries form in all the weak field models, with larger first periastron separations and smaller discs masses and radii than in their \Bizm \ counterparts.  The magnetic fields in the disc are $\sim$$10$ times stronger than in their \Bizm \ counterparts, despite the same initial strength in the initial cloud.

\item \emph{Non-ideal MHD}: $\bm{B}_0=\pm\bm{B}_\text{z}$:  
The time of first periastron is not affected by the inclusion of non-ideal MHD, however, at later times, the non-ideal MHD models tend to have longer periods and larger apoastron separations than the ideal MHD models, as well as larger and more massive discs.  When discs become massive enough to fragment, the fragmentation occurs in the non-ideal MHD models more easily.  The evolution of the \Bzm \ and \Bz \ models tends to diverge between first apoastron and second periastron; the differences are initially small, but are enhanced by the dynamics and subsequent interactions, rather than influences of non-ideal MHD.

\item \emph{Non-ideal MHD}: $\bm{B}_0=\pm\bm{B}_\text{x}$:  
The non-ideal MHD models that form binaries yield smaller first periastron separations and larger disc radii compared to their ideal MHD counterparts.  Divergence in periods and periastron and apoastron separations between ideal and non-ideal MHD models occurs shortly after first periastron, while the divergence between the two non-ideal MHD models with the same $\mu_0$ and $A_0$ occurs later, once local changes have modified the vertical component of the magnetic field such that the Hall effect produces a different evolution in each model.

\item \emph{The Hall effect}:  Unlike models that form an isolated protostar, the Hall effect does not have a global impact on the evolution of the binaries.  Rather, local changes to the magnetic field will be enhanced by the Hall effect causing small modification to the evolution of the model.  These small modifications are further enhanced at periastron, causing the evolutionary paths to slowly diverge.

\item \emph{Rotation speeds}:  Decreasing the initial rotation speed hindered binary formation.  At the slowest two speeds tested, only a single protostar formed.  The inclusion of non-ideal MHD did not affect the global morphology or the number of protostars that formed.

\item \emph{Grain  model}:  The same qualitative conclusions are reached if the non-ideal MHD algorithm used a single population with average charge or three separate grain populations.  The first periastrons were smaller in the models that used the three grain populations.

\end{enumerate}

In the formation of binary systems, the initial parameters --- amplitude of the initial density perturbation, magnetic field strength and orientation, and rotation --- determine their evolution.  The main effect of non-ideal MHD is to enable the formation of larger and more massive discs form around the protostars, and produce binaries that have larger separations and longer periods.  

\section*{Acknowledgements}

JW and MRB acknowledge support from the European Research Council under the European Community's Seventh Framework Programme (FP7/2007- 2013 grant agreement no. 339248).  JW also acknowledges support from the Australian Research Council (ARC) Discovery Projects Grant DP130102078.  
DJP is funded by ARC Future Fellowship FT130100034.  This work was supported by resources on the gSTAR national facility at Swinburne University of Technology and by Zen. gSTAR is funded by Swinburne and the Australian Government's Education Investment Fund.  Several calculations for this paper were performed on the University of Exeter Supercomputer, a DiRAC Facility jointly funded by STFC, the Large Facilities Capital Fund of BIS, and the University of Exeter.  For the column density figures, we used {\sc splash} \citep{Price2007}. 

\bibliography{Wurster_Bib.bib}
\appendix
\section{Models with initial conditions and selected results}
\label{app:results:all}
Table~\ref{table:results:all} summaries the initial parameters of all our models, along with the time of first periastron $t_\text{peri}$, the initial period $T_0$, and the first periastron $R_\text{peri}$ and first apoastron $R_\text{apo}$ separations.  
\begin{table*}
\begin{center}
\begin{tabular}{c c c c c c c c c}
\hline
$\mu_0$  &  $A_0$ & $\bm{B}_0$   & MHD      & Alternate        & $t_\text{peri}$& $T_0$         &    $R_\text{peri}$  & $R_\text{apo}$    \\
                 &              &                     &                       & parameter      & (\tff)               &  ( \tff )          &    (au)                    & (au)                     \\
\hline
 5               & 0.2       &   \Bzm          &  ideal             &                       & 1.370             &  0.28           &   110                      &     530                 \\ 
 5               & 0.1       &   \Bzm          &  ideal             &                       & 1.333             &  0.25           &    68                       &     440                 \\ 
 5               & 0.05     &   \Bzm          &  ideal             &                       & 1.318             &  0.097         &    49                       &     210                 \\ 
 10             & 0.2       &   \Bzm          &  ideal             &                       & 1.369             &  n/a:P          &  140                       &   n/a:P                \\ 
 10             & 0.1       &   \Bzm          &  ideal             &                       & 1.340             &  1.7             &  100                       &   2800                 \\ 
 10             & 0.05     &   \Bzm          &  ideal             &                       & 1.324             &  0.67           &    54                       &   1100                 \\ 
\\
  5              & 0.2         &  \Bx            &  ideal             &                       & 1.287             &   0.033        &    39                      &      100                 \\ 
  5              & 0.1         &  \Bx            &  ideal             &                       & 0                    &   0               &      0                      &          0                 \\ 
  5              & 0.05       &  \Bx            &  ideal             &                       & 0                    &    0              &      0                      &          0                 \\ 
  10            & 0.2         &  \Bx            &  ideal             &                       & 1.320             &   0.89          &  210                      &    1200                 \\  
  10            & 0.1         &  \Bx            &  ideal             &                       & 1.330             &   0.39          &  190                      &      650                 \\  
  10            & 0.05       &  \Bx            &  ideal             &                       & 1.345             &   0.39          &  190                      &      570                 \\  
\\
 5               & 0.2       &   \Bzm         & non-ideal        &                       &     1.372         &    0.29         &   95                       &      500                 \\ 
 5               & 0.1       &   \Bzm         & non-ideal        &                       &     1.337         &    0.52         &   61                       &      820                 \\  
 5               & 0.05     &   \Bzm         & non-ideal        &                       &     1.320         &    0.12         &   43                       &      200                 \\  
 5               & 0.2       &   \Bz            & non-ideal        &                       &     1.373         &    0.24         &   98                       &      460                 \\  
 5               & 0.1       &   \Bz            & non-ideal        &                       &     1.337         &    0.53         &   63                       &      800                  \\  
 5               & 0.05     &   \Bz            & non-ideal        &                       &     1.320         &    0.12         &   43                       &      200                  \\  
 10             & 0.2       &   \Bzm         & non-ideal        &                       &      1.374        &    n/a:P        & 120                       &      n/a:P               \\ 
 10             & 0.1       &   \Bzm         & non-ideal        &                       &      1.341        &    n/a:A        & 110                       &      n/a:A               \\ 
 10             & 0.05     &   \Bzm         & non-ideal        &                       &      1.337        &    n/a:A        &   62                       &      n/a:A               \\ 
 10             & 0.2       &   \Bz            & non-ideal        &                       &      1.378        &    n/a:P        & 120                       &      n/a:P               \\ 
 10             & 0.2       &   \Bz            & non-ideal        &                       &      1.340        &    n/a:A        & 110                       &      n/a:A                \\ 
 10             & 0.05     &   \Bz            & non-ideal        &                       &      1.325        &    n/a:A        &   58                       &      n/a:A                \\ 
\\
 5               & 0.2       &   \Bx            & non-ideal        &                       &     1.288         &    0.036       &   14                       &      130                  \\ 
 5               & 0.1       &   \Bx            & non-ideal        &                       &     0                &    0              &     0                       &          0                  \\  
 5               & 0.05     &   \Bx            & non-ideal        &                       &     n/a:P          &    n/a:P       &  n/a:P                    &      n/a:P                \\  
 5               & 0.2       &   \Bxm         & non-ideal        &                       &     1.288         &    0.035       &   14                       &      120                  \\  
 5               & 0.1       &   \Bxm         & non-ideal        &                       &     0                &     0             &     0                       &          0                  \\  
 5               & 0.05     &   \Bxm         & non-ideal        &                       &     n/a:P          &    n/a:P       &  n/a:P                    &      n/a:P                \\  
 10             & 0.2       &   \Bx            & non-ideal        &                       &      1.347        &    n/a:A       & 120                        &      n/a:A                \\ 
 10             & 0.1       &   \Bx            & non-ideal        &                       &      1.332        &    0.79        & 150                        &    1100                   \\ 
 10             & 0.05     &   \Bx            & non-ideal        &                       &      1.324        &    0.25        & 190                        &      470                   \\ 
 10             & 0.2       &   \Bxm         & non-ideal        &                       &      1.347        &   n/a:A        & 120                        &      n/a:A                 \\ 
 10             & 0.1       &   \Bxm         & non-ideal        &                       &      1.332        &   0.79         & 150                        &     1100                   \\ 
 10             & 0.05     &   \Bxm         & non-ideal        &                       &      1.324        &    0.29        & 190                        &       480                   \\ 
 \\
 0.75          & 0.1       &   \Bzm         &  ideal              & sub-critical $\mu_0$                                       &     n/a:NC      &    n/a:NC     &  n/a:NC                &      n/a:NC             \\
  0.75          & 0.1       &   \Bx           &  ideal              & sub-critical $\mu_0$                                      &     n/a:NC      &    n/a:NC     &  n/a:NC                &      n/a:NC             \\
  0.75          & 0.1       &   \Bzm        & non-ideal       & sub-critical $\mu_0$                                      &     0                &    0              &     0                       &          0                  \\  
 0.75          & 0.1       &   \Bx            & non-ideal        & sub-critical $\mu_0$                                      &     0                &    0              &     0                       &          0                  \\  
\\
 5               & 0.1       &   \Bzm         & ideal              & $\Omega_0 = 7.08\times10^{-13}$ s$^{-1}$  &     1.161         &    0.039       &     19                     &         110               \\ 
 5               & 0.1       &   \Bzm         & ideal              & $\Omega_0 = 3.54\times10^{-13}$ s$^{-1}$  &     0                &    0              &     0                       &          0                  \\ 
 5               & 0.1       &   \Bzm         & ideal              & $\Omega_0 = 1.77\times10^{-13}$ s$^{-1}$  &     0                &    0              &     0                       &          0                  \\ 
 5               & 0.1       &   \Bzm         & non-ideal       & $\Omega_0 = 7.08\times10^{-13}$ s$^{-1}$  &     1.161         &    0.045       &     17                     &        110                 \\ 
 5               & 0.1       &   \Bzm         & non-ideal       & $\Omega_0 = 3.54\times10^{-13}$ s$^{-1}$  &     0                &    0              &     0                       &          0                  \\ 
 5               & 0.1       &   \Bzm         & non-ideal       & $\Omega_0 = 1.77\times10^{-13}$ s$^{-1}$  &     0                &    0              &     0                       &          0                  \\ 
\\
  5              & 0.1       &   \Bzm         & non-ideal       &   {\sc NICIL} v1.2.1                                        &     1.336         &    0.17         &     56                     &        350                \\
  10            & 0.1       &   \Bx            & non-ideal       &   {\sc NICIL} v1.2.1                                        &     1.332         &    0.79         &     130                   &      1100                 \\ 
\hline
\end{tabular}
\caption{The initial parameters and early results of our suite of models.  The first four columns are the initial conditions: the initial mass-to-flux ratio $\mu_0$ in units of the critical mass-to-flux ratio, the amplitude of the initial density perturbation $A_0$, the initial orientation of the magnetic field $\bm{B}_0$, and whether the model uses ideal or non-ideal MHD.  The fifth column lists deviations from the parameters used in the primary suite (see Section~\ref{sec:results:altP}).  The remaining columns are  the time of first periastron $t_\text{peri}$, the initial period $T_0$ calculated using first periastron and first apoastron, and the separations at first periastron $R_\text{peri}$ and first apoastron $R_\text{apo}$. Entires with zeros indicate that only one disc formed, thus  $T_0$, $R_\text{peri}$ and $R_\text{apo}$ do not exists.  Entires with n/a:P indicate that one or both discs fragmented near first periastron and became unstable, entires with n/a:A indicate that one or both discs interacted with younger protostars near first apoastron which modified the primary binary's orbit, and entries with n/a:NC indicate that the cloud did not collapse to form protostars; in these three cases, separations and periods have no useful meaning.}
\label{table:results:all} 
\end{center}
\end{table*}


\label{lastpage}
\end{document}